\documentclass[11pt]{article}
\pdfoutput=1

\RemoveFromHook{package/amsthm/after}[firstaid/aliascounter]
\usepackage{amssymb,amsmath,amsthm,amsfonts}
\usepackage[utf8]{inputenc}
\usepackage[T1]{fontenc}
\usepackage[tt=false, type1=true]{libertine}
\usepackage[varqu]{zi4}
\usepackage[libertine]{newtxmath}
\usepackage[margin=1in]{geometry}
\usepackage{hyperref}
\hypersetup{
    colorlinks,
    allcolors=teal
}

\usepackage[sortcites,sorting=nyt,style=alphabetic,maxbibnames=7]{biblatex}
\usepackage[ruled]{algorithm2e}
    
    \SetAlFnt{\small}
    \SetAlCapFnt{\small}
    \SetAlCapNameFnt{\small}
    \SetAlCapHSkip{0pt}
    \IncMargin{-\parindent}

\usepackage{tikz, xifthen, physics, xcolor, suffix, nicefrac, enumitem, subcaption, float, thm-restate, wrapfig, mathtools}
\usetikzlibrary{shapes, arrows, positioning}
\usepackage[bb=dsserif]{mathalpha}
\usepackage[capitalize]{cleveref}

\renewcommand{\paragraph}[1]{\smallskip\noindent\textbf{#1.}}
\newcommand{\itparagraph}[1]{\textit{#1.}}

\newenvironment{optimization}{\crefalias{equation}{optimization}\equation\aligned}{\endaligned\endequation}
\newenvironment{optimization*}{\displaymath\aligned}{\endaligned\enddisplaymath}

\crefname{optimization}{Optimization Program}{Optimization Programs}
\creflabelformat{optimization}{(#2#1#3)}

\DeclareMathOperator*{\E}{\mathbb{E}}

\newcommand{\Ex}[2][]{\E_{\substack{#1}}\left[\mathchoice{\big.}{}{}{}#2\right]}

\newcommand{\ExC}[3][]{\E_{#1}\left[\mathchoice{\big.}{}{}{}#2\,\middle|\,#3\right]}

\newcommand{\One}[1]{\mathbb{1}_{#1}}
\newcommand{\orderT}[1]{\tilde{\mathcal{O}}\qty(#1)}
\WithSuffix\newcommand\orderT*[1]{\tilde{\mathcal O}(#1)}
\DeclareMathOperator*{\argmax}{arg\;\!max\,}
\DeclareMathOperator*{\argmin}{arg\;\!min\,}
\newcommand{\m}[1]{\begin{bmatrix}#1\end{bmatrix}}

\newcommand{\R}{\mathbb{R}}
\newcommand{\N}{\mathbb{N}}

\newcommand{\calUC}{\mathcal{UC}}
\newcommand{\floor}[1]{\left\lfloor{#1}\right\rfloor}
\WithSuffix\newcommand\floor*[1]{\lfloor{#1}\rfloor}
\newcommand{\ceil}[1]{\left\lceil{#1}\right\rceil}
\WithSuffix\newcommand\ceil*[1]{\lceil{#1}\rceil}

\let\e\varepsilon
\let\l\lambda
\let\d\delta

\newcommand{\SE}{\mathtt{SE}}
\newcommand{\BSE}{\mathtt{BSE}}
\newcommand{\WBSE}{\mathtt{WBSE}}

\newcommand{\UO}{U_{\smash{\mathtt{O}}}}
\newcommand{\UL}{U_{\smash{\mathtt{L}}}}

\newcommand{\rO}{\rho_{\smash{\mathtt{O}}}}
\newcommand{\rL}{\rho_{\smash{\mathtt{L}}}}

\newcommand{\BRL}{\textrm{BR}_{\smash{\mathtt{L}}}}

\newcommand{\CO}{C_{\smash{\mathtt{O}}}}
\newcommand{\CL}{C_{\smash{\mathtt{L}}}}
\newcommand{\nL}{n_{{\mathtt{L}}}}
\newcommand{\nO}{n_{\smash{\mathtt{O}}}}

\newcommand{\hatnO}{\hat{n}_{\mathtt{O}}}

\newcommand{\Line}[4]{%
    #1&%
    \ifthenelse{\isempty{#2}}{\phantom{=}}{#2}%
    #3%
    \ifthenelse{\isempty{#4}}{}{&&\qquad\left(\big.\substack{#4}\right)}%
}

\newtheorem{theorem}{Theorem}[section]

\newtheorem{lemma}[theorem]{Lemma}
\newtheorem{claim}[theorem]{Claim}
\newtheorem{proposition}[theorem]{Proposition}

\theoremstyle{definition}
\newtheorem{remark}[theorem]{Remark}

\crefname{property}{Property}{Properties}

\newtheorem*{fact*}{Fact}
\crefname{fact}{Fact}{Facts}

\crefname{assumption}{Assumption}{Assumptions}

\newcommand{\citet}[1]{\textcite{#1}}

\title{Learning vs. Optimizing Players in Budgeted Games}

\author{
Giannis Fikioris\\
Google Research%
\footnote{Part of this work was done while a PhD student at Cornell, supported by the ONR MURI grant N000142412742.}\\
\texttt{giannisf@google.com}
\and
Balasubramanian Sivan\\
Google Research\\
\texttt{balusivan@google.com}
\and
\'Eva Tardos%
\thanks{Supported in part by AFOSR grant FA9550-23-1-0410, AFOSR grant FA9550-231-0068, ONR MURI grant N000142412742, and through a SPROUT Award from Cornell Engineering.}\\
Cornell University \& Google Research\\
\texttt{eva.tardos@cornell.edu}
}

\date{}

\begin{document}

\pagenumbering{roman}
\maketitle{}
\thispagestyle{empty}

\begin{abstract}
    We study repeated general games over $T$ rounds between a utility-maximizing optimizer and a learning agent, where budget constraints couple the players' decisions across rounds. Without budgets, low swap regret simultaneously guarantees the learner sublinear regret and caps the optimizer at its Stackelberg value. Budgets break this picture: adversarial no-regret becomes impossible, so the natural goals are \emph{non-manipulability}---no adaptive optimizer earns more than $\mathtt{BSE} \cdot T + o(T)$, where $\mathtt{BSE}$ is its \emph{Budgeted Stackelberg Equilibrium} value---and \emph{low stochastic regret} when the optimizer plays a fixed mixed strategy.

We give an efficient, budget-feasible learning algorithm that achieves both goals in every budgeted game: any adaptive optimizer, budgeted or not, earns at most $\mathtt{BSE} \cdot T + O(\sqrt{T})$ in expectation, and if the optimizer plays a fixed mixed strategy on each of a constant number of intervals, the learner has $O(\sqrt{T})$ regret on each such interval. Our learner best responds to forecasts of the optimizer's play that are calibrated in two novel senses, since under budgets the calibration that prevents manipulation no longer yields low stochastic regret. \emph{Response calibration}, which conditions on the learner's own mixed response by embedding its continuous mixing weight as a step function in $L^2([-1,1])$, prevents manipulation without discretizing responses; dual-weighted \emph{Lagrangian calibration} ensures low regret. Although low calibration error requires randomized forecasts---distributions over a continuous space that need not have a succinct representation or be directly computable---we construct such forecasts in polynomial time by simulating online gradient ascent.

\end{abstract}

\clearpage
\pagenumbering{arabic}
\section{Introduction}\label{sec:intro}

Learning algorithms increasingly act on behalf of agents in repeated strategic interactions: they bid in ad auctions, set prices, and allocate computing resources. The party on the other side is often not another learner but a forward-looking \emph{optimizer} that knows how the learner behaves and plans against it. The learner's predictability then becomes a liability, which raises two basic questions~\cite{BMSW18,DSS19,MMSS22,ACMMSS25}: \emph{How much can an optimizer extract from a learner? And which learning algorithms are non-manipulable, in the sense that their predictability gives the optimizer no advantage?}

In standard (unbudgeted) repeated games, these questions have a clean answer. The natural benchmark is the optimizer's \emph{Stackelberg Equilibrium value} $\SE$: the best payoff it can obtain by committing to a mixed strategy to which the learner best responds. Against any no-regret learner, the optimizer can secure (essentially) $\SE$ per round by committing to its Stackelberg strategy, and against common no-regret algorithms, such as multiplicative weights, it can, in some games, earn strictly more~\cite{BMSW18,DSS19}. A learner is protected when it has low \emph{swap regret}: low swap regret caps the optimizer at $\SE\cdot T+o(T)$~\cite{DSS19}, and it is, in a precise sense, also necessary~\cite{MMSS22,ACMMSS25}. Since best responding to calibrated forecasts guarantees low swap regret~\cite{FV97}, a single principle---\emph{calibration}---delivers both robustness to manipulation and good learning performance.

This theory treats the rounds as decoupled. Yet in many of the interactions that motivate it---such as \emph{autobidding} in online advertising~\cite{AggarwalBBBDFGLLMMMMLPP24,BG19,DBLP:journals/ior/ConitzerKSM22,DBLP:journals/mansci/ConitzerKPSMSW22}, but also cloud resource allocation and repeated procurement---agents operate under long-term \emph{budgets} that couple their decisions across rounds. This motivates the study of \emph{repeated budgeted games}. For $T$ rounds, an optimizer with $\nO$ actions and a learner with $\nL$ actions play a stage game with utility matrices $\UO,\UL\in[0,1]^{\nO\times\nL}$ and cost matrices $\CO,\CL\in[0,1]^{\nO\times\nL}$: in round $t$ the optimizer plays $i_t\in[\nO]$, the learner plays $j_t\in[\nL]$, and each player wants to maximize its total utility while keeping its total cost within its budget---$\rO T$ for the optimizer and $\rL T$ for the learner. Budgets change both sides of the interaction:
\begin{itemize}
    \item \emph{The optimizer may want to switch strategies.} A budgeted optimizer can do strictly better than commit to a single mixed strategy: because its value need not be concave in its spending, it may profit from \emph{time-multiplexing}---for instance, by spending aggressively for $qT$ rounds and frugally for the remaining $(1-q)T$ rounds. The resulting benchmark is the \emph{Budgeted Stackelberg Equilibrium} (BSE) (defined in our recent work~\cite{FST26a}): the optimizer commits to (at most) two phases, each met by the learner's budgeted best response. When the optimizer is unbudgeted, one phase suffices, and the BSE value is simply the value of the stage game in which the follower best responds subject to its budget.
    
    \item \emph{The learner can no longer guarantee low regret.} Against an adversarial sequence of actions, no budgeted learner can guarantee sublinear regret~\cite{MTY09,ISSS22}, let alone low swap regret, due to facing the spend or save dilemma. So the swap-regret characterization above has no direct budgeted analog.
\end{itemize}

What, then, should a budgeted learner aim for? We ask for two properties:
\begin{enumerate}
    \item \emph{Non-manipulability:} no adaptive optimizer earns more than $\BSE\cdot T+o(T)$ in expectation.\footnote{Our learner aims to spend at most $\rL$ per round in every phase. Learners that are also willing to spend accumulated surplus can only increase the optimizer's equilibrium value (\cref{sec:app:bse_defs}), so the BSE is the most demanding benchmark for non-manipulability.}
    \item \emph{Low interval stochastic regret\footnote{Note that in a stochastic environment regret and swap regret are the same.}:} if the optimizer's play is piecewise stationary---i.i.d.\ from a fixed mixed strategy within each of a few intervals, unknown to the learner---then on every interval, the learner's utility is within $o(T)$ of the utility from the best budget-feasible response to the corresponding mixed strategy.
\end{enumerate}

\begin{table}[t]
    \centering
    \small
    \renewcommand{\arraystretch}{1.15}
    \begin{tabular}{lcc}
        \hline
        Guarantee for the learner & Without budgets & With budgets \\
        \hline
        Low regret against every sequence & \checkmark & $\times$~\cite{MTY09,ISSS22} \\
        Low swap regret & \checkmark~\cite{BM05} & $\times$ \\
        Low regret against a stationary optimizer & \checkmark & \checkmark~\cite{BKS18,AD16} \\
        Non-manipulability, plus the previous row & \checkmark\ via swap regret~\cite{DSS19} & \checkmark\ \textbf{this paper} \\
        \hline
    \end{tabular}
    \caption{What a learner can guarantee, without and with budgets.}
    \label{tab:achievable}
\end{table}

Since adversarial no-regret is out of reach, the second property is the natural learning goal: it is the standard benchmark for learning under budgets in stochastic environments~\cite{BKS18,AD16}, applied to every stationary stretch of play and it also provides a sanity check, that the learner is not just trying to harm the optimizer.
Without budgets, a single property---low swap regret---yields both non-manipulability and low regret (\cref{tab:achievable}). With budgets, the calibration that prevents manipulation no longer yields low regret (Challenge~\hyperlink{chal:3}{3} in \cref{sec:intro:challenges}), so each property needs its own tool. The central question of this paper is:
\begin{quote}
    \itshape Is there an efficient learning algorithm that, in every repeated budgeted game, is non-manipulable and has low interval stochastic regret?
\end{quote}

\paragraph{Our results}
We answer this question in the affirmative. We assume full-information feedback about the learner's own payoffs and costs: after round, the learner observes the utility and cost of each of its actions, i.e., the vector $(\UL[i_t,:],\CL[i_t,:])\in[0,1]^{2\nL}$.
It observes neither $i_t$ nor the optimizer's payoffs, and it does not know in advance which utility-cost vectors can arise. Each player has a \emph{null action} that incurs no cost.

\begin{theorem}[Informal version of \cref{thm:cal_min:main,thm:cal_min:main_budgeted}]\label{thm:intro_main}
    There is a learning algorithm that never exceeds the learner's budget, runs in $\mathrm{poly}(T,\nL)$ time per round, and in every repeated budgeted game simultaneously guarantees:
    \begin{enumerate}
        \item \emph{Non-manipulability:} against every adaptive optimizer,
        \[
            \E\Big[\sum_{t=1}^T \UO[i_t,j_t]\Big] \;\le\; \BSE\cdot T + \order{ \sqrt{T} },
        \]
        where $O(\cdot)$ hides factors polynomial in $\nO$, $\nL$, and $1/\rL$.\footnote{Precisely, the additive error is $O\big((\sqrt{\nO}\,\nL+\sqrt{\nL}/\rL)(\sqrt{T}+\nO)\big)$.} If the optimizer is also budgeted, the same holds relative to its two-phase BSE, with an additional factor of $1/\rO$ in the error term.
        
        \item \emph{Low interval stochastic regret:} if $[T]$ is partitioned into $S$ intervals, unknown to the learner, such that within each interval $\mathcal{I}$ the optimizer draws its actions i.i.d.\ from a mixed strategy $x_{\mathcal I}$, then on every interval,
        \[
            \Ex{
                \sum_{t\in\mathcal{I}} \UL[i_t,j_t]
            }
            \;\ge\; |\mathcal{I}|\cdot x_{\mathcal I}^\top \UL\, y_{\mathcal I}^\star - \order{ \frac{\sqrt{S\,\nL\,T}}{\rL} },
        \]
        where $y_{\mathcal I}^\star$ is a best response to $x_{\mathcal I}$ subject to the learner's per-round budget.
    \end{enumerate}
\end{theorem}

Three features of \cref{thm:intro_main} are worth highlighting. First, the regret bound is independent of the number of the optimizer's actions: it holds even if no utility-cost vector is ever observed twice. Second, the first guarantee is essentially tight: against our learner, an optimizer that plays a fixed mixed strategy in each of (at most) two phases secures the weak BSE value (\cref{thm:app:budgeted_optimizers:wbse}). Thus, up to the usual gap between weak and strong Stackelberg values, the BSE is the value of the interaction. Third, the algorithm needs no prior knowledge of the game.

\subsection{Comparison With Budgeted Auctions}\label{sec:intro:comparison}

\paragraph{Budgeted auctions: simplicity from structure}
In earlier work~\cite{FST26a}, we studied these questions for repeated budgeted auctions in a Bayesian setting, where in each round the two bidders' values are drawn from a joint distribution. Here, the algorithm used in practice is remarkably simple. In a second-price auction, the learner's best response to any fixed competition is to bid its value scaled by a single \emph{pacing multiplier}, so the learner only needs to track one number, and large advertising platforms adjust it with simple feedback (PID-style) controllers: raise the multiplier after underspending, lower it after overspending~\cite{GooglePacing,TwitterPacing,DBLP:journals/mansci/ConitzerKPSMSW22}. \cite{FST26a} shows that the simplest such controller---a deterministic, one-line proportional update---is non-manipulable: under a mild assumption on the value distribution (some such assumption is necessary), the optimizer earns at most $\BSE\cdot T + \order*{T^{2/3}}$.

\paragraph{General games: universality without structure}
Given the previous results, a natural question arises. Is robustness to manipulation an artifact of the special structure of auctions, or can it be achieved in \emph{every} budgeted game? In a general game, no scalar parameterizes the learner's best responses: a budget-constrained best response is the solution of a linear program---often a strict mixture of two actions---and it jumps discontinuously as the environment changes. The learner does not know the optimizer's payoffs. And there is no Bayesian structure to lean on: everything the learner faces is chosen adaptively by the optimizer. In fact, no deterministic learner is non-manipulable in every game: in matching pennies (with zero costs), an optimizer that simulates the learner always mismatches it, earning $1$ per round, twice its Stackelberg value of $1/2$. In auctions, structure makes determinism harmless; general games have no such structure. 

\paragraph{Why both settings}
Our work on budgeted auctions~\cite{FST26a} and this paper answer on general games complementary questions---\emph{is the algorithm used in practice robust?} and \emph{is robustness achievable in every budgeted game?}---and neither result implies the other (\cref{tab:comparison} in \cref{sec:app:comparison} compares them in detail). The controller of~\cite{FST26a} has no counterpart in general games, where no deterministic learner is robust.
The same pattern is familiar without budgets: against a revenue-maximizing seller, simple no-regret bidding algorithms (in the right parameterization) are already robust~\cite{KSS24,cai2026onlinelinearoptimizationsufficient,DBLP:journals/corr/abs-2601-03853}, whereas in general games, robustness requires low swap regret~\cite{MMSS22}. As often in algorithm design, the special case admits a strikingly simple algorithm, while the general case requires more complex algorithms.

\paragraph{Why the techniques differ}
Beyond the BSE benchmark and Lagrangian duality, the two analyses share very little. In~\cite{FST26a}, the learner is a fixed, deterministic dynamical system with a one-dimensional state, and the proof bounds the value of the optimizer's control problem over this state with a potential function built from Lagrangian duality. Here, there is no small state to control, and the learner must randomize: our learner best responds to forecasts that are calibrated in two new senses---conditionally on its own mixed response (which prevents manipulation), and weighted by its dual variable (which yields low regret)---and computes them by simulating a no-regret dynamic (see challenges and techniques in \cref{sec:intro:challenges,sec:intro:techniques}).

\subsection{Why General Budgeted Games Are Hard: Three Challenges}\label{sec:intro:challenges}

Three challenges rule out natural approaches and shape our solution.

\paragraph{1.\ Interval no-regret does not prevent manipulation}
Since adversarial regret is impossible, one might combine steady spending of budget with an online learning algorithm that has low regret on every interval. This fails even without budgets: projected online gradient ascent has $O(\sqrt{T})$ regret on every interval against every sequence, yet in a simple $3\times 3$ game, an optimizer that switches actions whenever the learner's mixed strategy reaches a corner of the simplex earns $T/2-o(T)$, compared with $\SE\cdot T=T/3$ (see~\cite{FST26a}, also included for completeness in \cref{sec:app:OGA_manipulable}).
Robustness in general games requires guarantees conditioned on the learner's own responses, as in swap regret and calibration~\cite{MMSS22}.

\paragraph{2.\ Budgeted best responses are mixed and discontinuous}
Following Foster and Vohra~\cite{FV97}, the learner could forecast the utility-cost vector that the optimizer will induce, and best respond to the forecast. Merely unbiased forecasts are not enough: since best responses are discontinuous, forecasts with zero cumulative error can lie on opposite sides of a discontinuity and induce opposite actions, which the optimizer can exploit (see \cref{sec:app:resp_cal_example} for an example). Full calibration, which conditions on the forecast itself, avoids this but needs exponentially (in the dimension of forecasts) many buckets. Without budgets, the efficient remedy is to condition only on the learner's response, and since best responses can be taken to be pure, $\nL$ ``buckets'' suffice. With a budget, however, the best response solves a linear program with two constraints (the simplex and the budget), so when the budget binds, it is generically a \emph{strict mixture} of two actions $(j_1,j_2)$, with a mixing weight $q\in[0,1]$ that varies \emph{continuously}. Conditioning on the exact response would require infinitely many buckets, while coarser conditioning---ignoring the weight, or tracking the two actions separately---can be exploited; the latter because several optimizer actions may induce the same feedback for the learner but have different value for the optimizer (see \cref{sec:app:resp_cal_pairs} for an example).

\hypertarget{chal:3}{}
\paragraph{3.\ Non-manipulability no longer implies low regret}
Unlike in unbudgeted games, response calibration does not yield low regret: there are forecasters with \emph{zero} response-conditioned calibration error---so the learner cannot be manipulated---under which the learner suffers $\Omega(T)$ regret against an i.i.d.\ optimizer (\cref{sec:app:resp_cal:not_enough}). The reason is that two different forecasts can induce the same conservative action, so their errors cancel, and the learner never discovers an action that is affordable only on average.

\subsection{Our Techniques}\label{sec:intro:techniques}

\paragraph{Forecast, then best respond---safely}
Our learner keeps a small reserve: in any round $t$ in which its past spending exceeds $t\rL-1$, it plays its null action, so it never exceeds its budget. In every other (\emph{active}) round, it draws a forecast $f$---a mixture of the utility-cost vectors observed so far---from a distribution $F_t$ of such forecasts, and plays a budget-constrained best response $y(f)$ to the forecast, supported on (at most) two actions $\vec{J}(f)=(j_1,j_2)$ with probability $Q(f)$ on $j_1$. Everything then hinges on the quality of the forecasts, which we measure with two new calibration errors.

\paragraph{Response calibration in a function space}
To bound the optimizer's gain over the BSE, forecasts must be calibrated conditionally on the learner's response $(\vec{J}(f),Q(f))$. Instead of discretizing the weight $Q(f)$, we represent it by the step function $q\mapsto\One{q\le Q(f)}$ in the Hilbert space $L^2([-1,1])$. The \emph{response-calibration error} $R_t$ has one coordinate for each outcome $k$ (a possible utility-cost vector) and each pair $(j_1,j_2)$; in each round, coordinate $(k,\vec{J}(f))$ accumulates the forecast error $\One{k_t=k}-f(k)$ times this step function. Two facts make this representation work. First, step functions have the closed-form inner product $\langle\One{\cdot\le q},\One{\cdot\le q'}\rangle=1+\min\{q,q'\}$, so $R_t$ can be stored and manipulated efficiently. Second, the optimizer's best payoff against the mixed response $(j_1,j_2,q)$, among its actions that induce outcome $k$, is a $1$-Lipschitz piecewise-linear function $g(q)$, which can be written as the inner product $\langle g',\One{\cdot\le q}\rangle$ with its derivative (working on $[-1,1]$ rather than $[0,1]$ to accommodate affine functions). By linearity and the Cauchy--Schwarz inequality, the optimizer's total gain over the BSE in active rounds is at most $\langle g',R_T\rangle\le\|g'\|\,\|R_T\|$, simultaneously for all optimizer payoff matrices (which the learner does not know).

\paragraph{Lagrangian calibration}
To obtain low stochastic regret and keep spending on track, we maintain a second error $L_t$. The best response $y(f)$ maximizes the Lagrangian $u(f)^\top y+\lambda(f)\big(\rL-c(f)^\top y\big)$, where $u(f)$ and $c(f)$ are the forecast utility and cost vectors and $\lambda(f)\in[0,1/\rL]$ is the optimal dual multiplier of the learner's budget constraint. When comparing the learner with a budget-feasible comparator $y^\star$, the troublesome term is the cost misprediction \emph{weighted by the dual variable}, $\lambda(f)\,(c_t-c(f))^\top y^\star$, where $c_t$ is the realized cost vector: a forecast that overestimates the cost of $y^\star$ makes it look unaffordable exactly when the budget is tight, as in Challenge~\hyperlink{chal:3}{3} in \cref{sec:intro:challenges} (see example in \cref{sec:app:resp_cal:not_enough}). The \emph{Lagrangian-calibration error} $L_t$ tracks this dual-weighted cost error, together with the utility error and the gaps between the learner's realized and forecast utility and cost. A small $\|L_t\|$ bounds both the number of non-active rounds and the learner's regret on every stationary interval.

\paragraph{Efficient forecasting by simulating the adversary}
We combine the two errors into $Z_t=(R_t,L_t)$ and keep the potential $\|Z_t\|^2$ growing by $O(1)$ per round, in the spirit of Blackwell approachability and forecast hedging~\cite{Blackwell56,FV97,FH21}. This reduces each round to a zero-sum game: the forecaster picks a forecast $f$, the environment picks an outcome $k$, and the payoff is the increase in the potential. If the environment commits to a mixed strategy $w$, the forecaster can simply predict that exact environment $f=w$ and pay exactly zero. However, the payoff is discontinuous in $f$ (both $y(f)$ and $\lambda(f)$ jump), so minimax theorems do not apply directly; an equilibrium forecast distribution, even when it exists, need not have a succinct representation; and discretizing the simplex requires exponentially many points. We sidestep these issues constructively: we simulate projected online gradient ascent for the environment player, and the forecaster answers each iterate $f^{(m)}$ with the forecast $f^{(m)}$ itself. The uniform distribution over the iterates is our forecast distribution $F_t$, and the regret bound of gradient ascent shows that it loses at most $O(1)$ against every previously observed outcome. This yields $\|Z_T\|=O(\sqrt{T}+\nO)$ against any optimizer---each of the at most $\nO$ distinct utility-cost vectors may cost $O(1)$ when it first appears---in time $\mathrm{poly}(T,\nL)$. Against a piecewise-stationary optimizer, a finer argument that compares the forecasts with the empirical distribution of outcomes within each interval gives $\E[\max_t\|Z_t\|]=O(\sqrt{ST})$, independently of~$\nO$.

\subsection{Related Work}\label{sec:intro:related}

\paragraph{Strategizing against learners}
How a utility-maximizing optimizer should play against a learning algorithm, and which algorithms resist such manipulation, was first studied by~\cite{BMSW18} for selling to a no-regret buyer. The question has since been studied in normal-form games~\cite{DSS19,BSV23,ACS24}, Bayesian games~\cite{MMSS22}, polytope games~\cite{ACMMSS25}, mechanism design and contracting with learning agents~\cite{CHJ20,GKSTVWW24}, Stackelberg games against calibrated agents~\cite{HPY24}, and repeated auctions~\cite{BMSW18,DSS19a,KSS24,RZ24,cai2026onlinelinearoptimizationsufficient,DBLP:journals/corr/abs-2601-03853}. The swap-regret characterization recalled above is due to~\cite{DSS19,MMSS22} for normal-form games, and to~\cite{ACMMSS25} for polytope games (via profile swap regret). In auctions, recent work designs bidding algorithms that are both no-regret and robust to a manipulative seller~\cite{KSS24,cai2026onlinelinearoptimizationsufficient,DBLP:journals/corr/abs-2601-03853}. All of these works consider rounds that are not linked by budgets.

\paragraph{Calibration, approachability, and swap regret}
Foster and Vohra~\cite{FV97} showed that best responding to calibrated forecasts yields no internal regret, so that such dynamics converge to correlated equilibria; see also~\cite{HM00,BM05,Blackwell56}. Since calibrating high-dimensional forecasts by discretization is exponentially costly, a recent line of work calibrates only what downstream decisions depend on: decision calibration~\cite{ZKSME21}, response-based approachability~\cite{bernstein-shimkin-2015,anagnostides-et-al-2026}, pseudonorm approachability~\cite{DMMSS23}, and forecasters that guarantee low regret or swap regret to downstream agents~\cite{KLPST23,NRRX25,RS24,HW24,rivkin-et-al-2026}. Our response-calibration error is in the same spirit, as it conditions on the learner's own response, and in unbudgeted games it reduces to errors that guarantee low swap regret and look like the ones in previous works (\cref{sec:app:resp_cal_unbudgeted}). The budget, however, forces three departures (see issues and techniques in \cref{sec:intro:challenges,sec:intro:techniques}): (i) responses are mixtures with a continuous weight, calibrated through step functions in $L^2([-1,1])$; (ii) response calibration no longer implies low regret,  necessitating the dual-weighted Lagrangian calibration; and (iii) since the set of outcomes is not known in advance and randomized forecasts cannot be represented explicitly, we construct them by simulating online gradient ascent.

\paragraph{Online learning under budget constraints}
Learning under budget constraints has been studied extensively in the frameworks of bandits with knapsacks and online allocation~\cite{BKS18,AD16,ISSS22,KS20,CCK22,CCK24,BCCF24,BLM23,FT23,BLMSX25}.
In stochastic environments, $\tilde{O}(\sqrt{T})$ regret against the budgeted best response is achievable~\cite{BKS18,AD16,BLM23}.
In adversarial environments, the best reward in hindsight is not attainable~\cite{MTY09,ISSS22}: \cite{MTY09} used calibrated forecasts to attain its convex hull, and later work obtained competitive-ratio guarantees~\cite{ISSS22,KS20,CCK22,CCK24}, best-of-both-worlds algorithms~\cite{CCK22,BCCF24,BLM23}, guarantees for approximately stationary environments~\cite{FT23}, and guarantees against weaker benchmarks~\cite{BLMSX25, DBLP:conf/colt/LucierPSZ24}.
In auctions, a large literature studies budget pacing: learning to bid under budget (and return-on-investment) constraints~\cite{BG19,BKMSW23,BLMS23,WYDK23,FPW23,BBFLMSW24,AFZ25,DBLP:conf/sigecom/FikiorisKKKMT25}, pacing equilibria~\cite{BBW15,DBLP:journals/ior/ConitzerKSM22, DBLP:journals/mansci/ConitzerKPSMSW22, BKK23, CKK24}, and the welfare of pacing dynamics~\cite{DBLP:conf/innovations/GaitondeLLLS23,DBLP:journals/mor/FikiorisT25,DBLP:conf/colt/LucierPSZ24}; see~\cite{AggarwalBBBDFGLLMMMMLPP24} for a survey. Our interval stochastic regret guarantee
is the full-information analog of stochastic guarantees for bandits with knapsacks, in piecewise-stationary
environments; what is new is achieving it together with non-manipulability against an arbitrary adaptive
opponent.

\paragraph{Strategizing against budgeted learners}
\cite{fikioris2026robust} study a proportional controller similar to~\cite{FST26a} in repeated auctions where all values equal $1$, and show that each bidder wins a budget-proportional share of the rounds, even against adversarial competitors.
In~\cite{FST26a}, we introduced the BSE and proved that the proportional controller is non-manipulable in repeated auctions with Bayesian values; see \cref{sec:intro:comparison,sec:app:comparison} for a comparison with this paper.
\cite{GJLM23} study an unconstrained seller learning a static posted price via binary search against a budget- and ROI-constrained buyer. Their focus is how the seller can guarantee low regret w.r.t. the Stackelberg value and not whether more revenue than that can be achieved.

\section{Model and Definitions} 
\label{sec:defs}

Two players, called optimizer and learner, are playing a repeated budgeted game over $T$ rounds.
The optimizer has $\nO$ actions and the learner has $\nL$ actions.
In each round $t$, the optimizer picks an action $i_t \in [\nO]$ and the learner picks $j_t \in [\nL]$; the outcome of the round is described by the matrices $\UO, \UL, \CO, \CL \in [0, 1]^{\nO \times \nL}$:
$\UO[i_t, j_t]$ and $\CO[i_t, j_t]$ are the optimizer's utility and cost, respectively, and similarly for $\UL, \CL$ for the learner.
The goal of each player is to maximize their utility, while they also have a budget constraint: the total cost has to be at most $\rL T$ and $\rO T$ respectively, for some $\rL, \rO \in (0, 1]$; we assume $\rL$ and $\rO$ are both $\Omega(1)$ w.r.t. $T$. 
We also assume that each player has an action (called the null action) that guarantees $0$ cost no matter what the other player is doing.
We assume throughout the paper that the learner receives full information feedback: for a chosen action $i_t$ of the optimizer in round $t$, the learner picks action $j_t$ and learns the utilities and costs $\UL[i_t,:]$ and $\CL[i_t,:]$ for each of their actions, not only the cost and utility of their chosen action $j_t$.

For most of the paper we will focus on the version of the problem where only the learner is budget constrained.
Later we show in \cref{sec:budgeted_optimizer} that our non-manipulability result for an unbudgeted optimizer translates to a similar result for a budgeted optimizer.

\paragraph{Budgeted Stackelberg Equilibrium}
In \cite{FST26a} we introduced the \textit{Budgeted Stackelberg Equilibrium} (BSE), that generalizes the concept of Stackelberg Equilibrium to budgeted games.
We first define the best-response function of the learner, given the expected utility and cost vectors $u = x^\top \UL$ and $c = x^\top \CL$ induced by an optimizer's mixed strategy $x \in \Delta_{\nO}$:
\begin{optimization}\label{eq:best_response}
    \BRL(u, c)
    =\quad
    \argmax_{y \in \Delta_{\nL}} \;
    u^\top y
    \qquad\textrm{ such that }\qquad
    c^\top y \le \rL
    .
\end{optimization}

For unbudgeted optimizer, this makes the BSE very similar to a Stackelberg equilibrium here equal to
\begin{optimization} \label{opt:defs:bse}
    \BSE
    =\quad
    \max_{x \in \Delta_{\nO}, \; y \in \Delta_{\nL}} \;
    x^\top \UO y
    \qquad \textrm{\textup{ such that }} \qquad
    y \in \BRL\qty(x^\top \UL, x^\top \CL)
    .
\end{optimization}

Note that there is always a best response $y$ that can be represented with support at most two actions, because it is a solution to a Linear Program with two constraints: the budget and the distribution constraints.

\paragraph{Learner's Desiderata}
Our goal in this paper is to design a learning process to achieve the following.

\itparagraph{Non-manipulability}
No matter what strategy the optimizer follows, the expected utility should be bounded by $\BSE \cdot T + o(T)$, i.e., there is sublinear in $T$ gain to deviate to any strategy other than the one suggested by the BSE.
Concretely, we want this to happen in expectation: for optimizer actions $i_t$ and learner actions $j_t$ we want
$
    \sum_t \Ex{\UO[i_t, j_t]}
    \le
    \BSE \cdot T
    +
    o(T)
$.

Our definition of BSE assumes that the learner's  goal is to best-respond in every round trying to keep her spending at most $\rL$. Alternately, we can consider learners who may be willing to overspend following historical underspending.
We explore these different definitions in \cref{sec:app:bse_defs}, where we show that they can only benefit the optimizer.

\itparagraph{Low stochastic regret}
We want that when the learner is faced with a stationary environment (e.g., the optimizer uses the same mixed strategy in every round), then the learner has no regret.
In fact, we achieve a stronger guarantee: in any interval of rounds $\mathcal I$ the optimizer samples $i_t$ from a fixed distribution of actions $x$, then the learner should have low regret in that interval with respect to the best-response of that interval $y^\star \in \BRL(x^\top\UL, x^\top\CL)$:
$
    \sum_{t \in \mathcal I} \Ex{\UL[i_t, j_t]}
    \ge
    x^\top \UL y^\star \cdot |\mathcal I|
    -
    o(T)
$
    .

Aside from being a strong learning guarantee for the learner, this also offers a lower bound for the optimizer's utility, the \textit{weak Budgeted Stackelberg Equilibrium}.
Specifically, the BSE value we defined in \cref{opt:defs:bse} is analogous to the \textit{strong Stackelberg Equilibrium}, where ties between the learner's best-responses are broken in favor of the optimizer.
In contrast, in a \textit{weak Stackelberg Equilibrium} such ties are broken to punish the optimizer, replacing the maximization over $y$ in \cref{opt:defs:bse} with minimization.
In \cref{ssec:budgeted_optimizer:bse,ssec:budgeted_optimizer:wbse}, we define the strong and weak Budgeted Stackelberg Equilibrium for both budgeted and unbudgeted optimizers, and show why the optimizer can achieve the weak BSE due to the guarantee above.
In \cref{ssec:budgeted_optimizer:generic} we show that in games with generic payoffs the weak and strong version of BSE are the same.

\section{Non-manipulable Algorithm with Low Stochastic Regret} \label{sec:algo}

In this section, we define the first component   of our algorithm that achieves the two objectives of our paper: non-manipulability and low stochastic regret.
This component is based on receiving a forecast predicting the environment induced by the optimizer and then best-responding to this forecast. 
Specifically, when the optimizer picks randomized actions $x$, the learner would like to best respond by knowing $x^\top \UL$ and $x^\top \CL$.
Ideally, this means that the learning algorithm receives a prediction for these two $\nL$-dimensional vectors---which we denote $uc \in [0,1]^{2 \nL}$---and best responds to them.
Our guarantees in \cref{sec:resp_cal,sec:lagr_cal}, then become that as long as these forecasts are calibrated according to the right calibration error, then the optimizer's gain over its BSE and the learner's stochastic regret are proportional to that error.
In \cref{sec:cal_minimization} we show how the learner can generate such calibrated forecasts.

The forecast received is a distribution over utility-cost vectors that the optimizer can generate, that is, a convex combination of such utility-cost vectors.
Let
\begin{equation*}
    \calUC = \qty{ uc \in [0, 1]^{2\nL} : uc = \qty\big(\UL[i, :], \CL[i, :]), i \in [\nO]} = \{uc^{(1)}, uc^{(2)}, uc^{(3)}, \ldots\}
\end{equation*}
and $\hatnO = |\calUC| \le \nO$.
Note that $\hatnO < \nO$ when there are multiple optimizer actions that create the same utility and cost vectors for the learner.
While such actions are identical for the learner, their utility to the optimizer may be different and may depend on which action the learner takes.
We use $i$ to index actions by the optimizer and $k$ to index elements in $[\hatnO]$ that can correspond to multiple different actions by the optimizer.
Therefore, the forecast $f$ is a convex combination of the elements of $\calUC$; the induced expected utility and cost vector for the learner is
$uc(f) =\sum_{k = 1}^{\hatnO} f_k uc^{(k)}$. We similarly define $u(f)$ and $c(f)$.
Then, given this forecast, the learner best responds with mixed strategy $y \in \BRL\qty\big(uc(f))$.
This happens in \textit{active rounds}, i.e., when the historical spending is at most $t \rL - 1$.
The formal description of this algorithm is in \cref{alg:learner}.

\begin{algorithm}[ht]
\DontPrintSemicolon
\SetAlgoNoEnd
\caption{Forecast-Based Best-Response Algorithm}
\label{alg:learner}
\KwIn{Forecaster over utility-cost vectors}
\For{$t = 1, \dots, T$}{
    \uIf(\tcp*[f]{Non-active round}){historical spending $> t \rL - 1$}{
        Play the null action\;
    }
    \Else(\tcp*[f]{Active round}){
        Receive forecast $f^{(t)}$ from the forecaster\;
        Pick best response $y_t \in \BRL\qty\big(uc(f^{(t)}))$ with support at most $2$ (breaking ties consistently)\;
        Sample and play a pure action from $j_t \sim y_t$\;
        Observe realized utility-cost vector $uc_t \in [0,1]^{2 \nL}$ and report it to the forecaster\;
    }
}
\end{algorithm}

In \cref{sec:resp_cal}, we will describe, \textit{response-calibration error}, which, if low, will guarantee that the optimizer's utility is $\BSE \cdot T + o(T)$.
Unfortunately, this non-manipulability property is insufficient to achieve low stochastic regret (see \cref{sec:app:resp_cal:not_enough}).
For this reason, we describe \textit{Lagrangian-calibration error} in \cref{sec:lagr_cal}, which, if low, guarantees low stochastic regret.
Finally, in \cref{sec:cal_minimization} we show how to minimize both calibration errors simultaneously.

\section{Response-calibration Error (Low Manipulability)} \label{sec:resp_cal}

In this section we present the calibration error we minimize that, if small, guarantees that the learner of \cref{alg:learner} is not manipulated by the adversary.
The obvious thing to track is the error between the realized vector $uc_t$ and the one implied by the forecast.
However that doesn't work: a prediction might be very close to its realization, but because of the non-continuous nature of the best-response function, this might not lead to approximately the same action by the learner.
We showcase this with a simple example in \cref{sec:app:resp_cal_example}.
Instead, we introduce a new error, which we call \textit{response-calibration error}.
This tracks the error of the response that the learner takes with respect to the forecast.
Specifically, for a forecast $f$, let $\vec{J}(f) \in [\nL]^2$ be the two pure actions in the learner's best-response (recall the best-response can always be supported on at most two actions) and $Q(f) \in [0, 1]$ be the probability of playing the first of those actions. If the best response to $f$ is a pure action $j$ we assume $\vec J(f) = (j, j)$ and $Q(f) = 0$, while for multiple best responses we assume any consistent tie-breaking rule. 
Ideally, for any triplet $k\in [\hatnO]$ and $j_1,j_2 \in [\nL]$, if we consider rounds when the optimizer chooses an action $i$ that induces a cost-utility vector $k\in [\hatnO]$ and the learner best responded to their forecast by the pair $(j_1,j_2)$ with some weight $q$, the empirical frequency with which the optimizer induces $uc^{(k)}$ is close to the average probability that the forecast assigned to $uc^{(k)}$, and the $q$ values are similar.
As is standard in the calibration literature, low error cannot be achieved unless the forecast is randomized.
Therefore, the forecasting algorithm produces randomized forecasts $F$ over the convex combinations\footnote{Technically, the forecast $f$ can also be seen as a distribution over the utility-cost vectors induced by the optimizer's pure strategies; said distribution corresponds to the expected utility/cost vectors induced by the optimizer's mixed strategies. Then, the randomized forecast $F$ is a distribution over these distributions. However, for exposition we will be calling $f$ a convex combination over $[\hatnO]$ and $F$ a distribution over this convex combination.} $f$ of the possible utility-cost vectors $\calUC$.

\paragraph{The response-calibration error}
We now show how we calculate the response-calibration error of a single round.
For a chosen forecast $f$ we will separately collect the error for each learner response pair $(j_1, j_2)$ and for each possible $k \in [\hatnO]$ that indexes $\calUC$.
The error is proportional to $f_k$ when $k\neq k_t$ and $\qty\big( 1-f_k )$ when $k = k_t$ for the $k_t$ that the optimizer chooses.
To also account for the weight of the two actions in the best response, we define this for each $(k, j_1, j_2)$ as a single parameter step function that is $0$ above $Q(f)$.
Formally, we define the error as a vector indexed by $k \in [\hatnO]$ and two learner actions $j_1, j_2 \in [\nL]$, where each element of the vector is a one dimensional step function (parametrized by $q$):
\begin{equation} \label{eq:algo:delta_R_alt}
    \d^R_{k, j_1, j_2}(f, k_t)(q) =
    \begin{cases}
    0 & \text{if}\; (j_1, j_2) \neq \vec J(f)\; \text{or}\; q>Q(f)\\ 1-f_k & \text{if}\; (j_1, j_2) = \vec J(f),\; q \le Q(f), \; k=k_t\\ -f_k & \text{if}\; (j_1, j_2) = \vec J(f),\; q \le Q(f),\; k \neq k_t
    \end{cases}
    \quad =
    \One{(j_1, j_2) = \vec J(f)}
        \qty\big( \One{k_t = k} - f_k )
        \One{q \le Q(f)}
\end{equation}

Then, the error in round $t$ when the randomized forecast is $F_t$ is $\Ex[f \sim F_t]{ \d^R\qty(f, k_t) }$, an $\hatnO \nL^2$ dimensional vector of functions. We then define the total error up to round $t$ for active rounds to be
\begin{equation}\label{eq:regret_function}
    R_t
    =
    \sum_{t' \le t : t' \in \mathcal T} \Ex[f \sim F_{t'}]{ \d^R\qty(f, k_{t'}) }
    \in
    \qty\big( \R \to \R )^{\hatnO \nL^2}
    .
\end{equation}

Intuitively, \cref{eq:algo:delta_R_alt} measures whether the forecasts are calibrated conditioned on the learner's induced best response.
The middle factor $\One{k_t = k} - f_k$ is the basic forecast error for utility-cost vector $uc^{\qty(k)}$, comparing the indicator that $uc^{\qty(k_t)}$ was realized in round $t$ against its predicted probability $f_k$.
If that was always $0$ (the forecast is always correct), then the error would be $0$.
Because the optimizer's utility in round $t$ depends on which response the forecast induces---and conditioning on the full $\hatnO$-dimensional vector $f$ would require exponentially many buckets---we group these forecast errors by the learner's best response, which is determined by the action pair $\vec J(f) = (j_1, j_2)$ and the mixing probability $Q(f)$ on $j_1$.
Since $\vec J(f)$ ranges over the finite set $[\nL]^2$, the first factor $\One{(j_1, j_2) = \vec J(f)}$ simply dedicates a separate coordinate to each pair $(j_1, j_2)$.
In contrast, the mixing weight $Q(f) \in [0, 1]$ is continuous; to avoid discretizing $[0, 1]$, the third factor represents $Q(f)$ via the step function $q \mapsto \One{q \le Q(f)}$. 

The usage of the step function $q \mapsto \One{q \le q'}$ is more complicated.
Consider a certain $(k, j_1, j_2)$ triplet and the mixing weight $q'$ of a step function.
For the realization of this triplet (the optimizer plays action $i$ that corresponds to $k$ and the learner uses pair $(j_1, j_2)$) the optimizer's expected utility is $q'\UO[i,j_1]+(1-q')\UO[i,j_2]$, which is an affine function in $q'$.
In fact, we can represent this utility via the right inner product between the step function $q \mapsto \One{q \le q'}$ and the derivative of this utility function.
Therefore, if the norm of our calibration error (which is a weighted sum of step functions) is small, this inner product is also going to be small by the Cauchy-Schwarz inequality.
This lets us bound the optimizer's utility for any optimizer, without knowing the optimizer's utility.
We next define this inner product. 

\begin{remark}
    It would be tempting to collect the calibration error for each $(k, j)$ pair rather than 
    $(k,j_1,j_2)$, tracking calibration error when the learner chooses best response 
    $(j_1, j_2)$ separately per action.
    Unfortunately, when multiple actions of the optimizer, say $i_1, i_2$, create the same environment $k$ for the learner, the optimizer can manipulate the learner even if such a calibration error is close to $0$.
    In this case, it is possible that the optimizer would benefit if they can choose $i_1$ when the learner chooses $j_1$ and $i_2$ when the learner chooses $j_2$.
    Specifically, we present an example in \cref{sec:app:resp_cal_pairs}, where the calibration error can be zero for both pairs $(k, j_1)$ and $(k, j_2)$ but by correlating $i_1$ with $j_1$ and $i_2$ with $j_2$ the optimizer achieves $\Omega(T)$ higher utility than $\BSE \cdot T$. As we show, this is not possible when the error is tracked per pair of learner actions.
\end{remark}

\paragraph{Inner product and norm}
The basis of our response-calibration error is a linear combination of step functions (see \eqref{eq:algo:delta_R_alt}), so we define a norm on them to define when this error is close to $0$.
We define this for more general functions, namely the Hilbert space $L^2\qty\big([-1, 1])$.

The space $L^2\qty\big([-1, 1])$ consists of functions $g, \tilde g: [-1, 1] \to \R$ equipped with the inner product and norm
\begin{equation} \label{eq:algo:inner_norm_general}
    \langle g, \tilde g \rangle
    =
    \int_{-1}^1 g(x) \tilde g(x) dx
    \qquad\text{ and }\qquad
    \norm{g}_{L^2}^2
    =
    \langle g, g \rangle
    =
    \int_{-1}^1 g(x)^2 dx < \infty
    .
\end{equation}

Now consider the case of linear combinations of step functions like a coordinate of $\delta^R$ or $R_t$: let $r(\cdot) = \sum_k w_k \One{\cdot \le q_k}$ where $w_k \in \R$, $q_k \in [0, 1]$, and $\One{\cdot \le q_k}$ is the step function.
Similarly let $\tilde r(\cdot) = \sum_\ell \tilde w_\ell \One{\cdot \le \tilde q_\ell}$, then \eqref{eq:algo:inner_norm_general} becomes
\begin{equation} \label{eq:algo:inner_norm_step}
    \langle r, \tilde r \rangle
    =
    \sum_{k, \ell} w_k \tilde w_\ell \qty\big( 1 + \min\{q_k, \tilde q_\ell\} )
    \qquad\text{ and }\qquad
    \norm{r}_{L^2}^2
    =
    \sum_{k, k'} w_k w_{k'} \qty\big( 1 + \min\{q_k, q_{k'}\} )
    .
\end{equation}

We also extend all the above definitions to vectors of such functions, $\qty(L^2\qty\big([-1, 1]))^n$ for some $n \in \N$.
Specifically, if $R = (r_1, \ldots, r_n)$ and $\tilde R$ are vectors of $L^2([-1, 1])$ functions then
\begin{equation*}
    {
    \langle R, \tilde R \rangle
    =
    \sum_{i = 1}^n \langle r_i, \tilde r_i \rangle
    \qquad\text{ and }\qquad
    }
    \norm{R}_{L^2}^2
    =
    \sum_{i=1}^n \norm{r_i}_{L^2}^2
    .
\end{equation*}

We next use this norm for the $n = \hatnO\nL^2$ dimensional function $R_t$ defined in \eqref{eq:regret_function}.
The simplifications of \cref{eq:algo:inner_norm_step} of the general form in \cref{eq:algo:inner_norm_general} imply efficient use of these functions and computation of their norms.
Specifically, we show how to efficiently keep $\norm{R}_{L^2}$ low in \cref{sec:cal_minimization}.

In \cref{ssec:resp_cal:non_manipulability}, we will upper bound the optimizer's utility beyond their BSE value as a product $\langle R_T, g' \rangle$.
In every coordinate $(k, j_1, j_2)$, $g$ has a function that takes as input $q$ and upper bounds the utility of the optimizer with the best response of $(j_1, j_2)$ with mixing weight $q$, assuming the optimizer's action needs to induce $k$.
Then we can bound this additional utility using the Cauchy-Schwarz inequality.
This inner product and norm allows us to do this more efficiently.
In \cref{eq:algo:inner_norm_step} we see how perturbing all the $q_k$'s by a little bit would barely perturb the norm, similar to how the optimizer's utility barely changes.
Instead, an approach that treats all $q_k$'s as distinct (for example via discretization) would yield huge norm increase under such perturbations.
In addition, the $1$ term in \cref{eq:algo:inner_norm_step} is due to making this definition over $L^2([-1, 1])$ instead of $L^2([0, 1])$.
Without this term, always using $q_k = 0$ would imply a $0$ norm, which is not meaningful.
In fact, this $1$ term is key to re-writing the optimizer's utility as an inner product between the step functions and the affine function $g$---without the $1$ term we could do this only for linear functions (functions whose value is $0$ at $0$).

\subsection{Response-calibration Error Implies Non-manipulability}
\label{ssec:resp_cal:non_manipulability}

In this section, we prove that low response-calibration error implies that the optimizer cannot get utility substantially more than the BSE value during active rounds.
Along with the fact the number of non-active rounds is small (see \cref{lem:lagr_cal:null_rounds}), we get our non-manipulability result.

\begin{theorem} \label{thm:resp_cal:manipulability}
    Let $\mathcal T$ be the set of active rounds of \cref{alg:learner}.
    Then the optimizer's utility is
    \begin{equation*}
        \sum_{t \in \mathcal T} \Ex[f \sim F_t]{ e_{i_t}^\top \UO y(f) }
        \le
        \BSE \cdot |\mathcal T|
        +
        \sqrt{2 \nO} \nL \norm{R_T}_{L^2}
        .
    \end{equation*}
\end{theorem}

The proof's first step is to observe that $\BSE$ upper-bounds the optimizer's value if the forecast is correct, for any forecast.
In other words, for a forecast $f$, if the optimizer plays a mixed strategy $x^\star(f)$ that induces $f$ and the learner uses $y(f)$, then $\qty(x^\star(f))^\top \UO y(f) \le \BSE$.
This implies that the optimizer's benefit in round $t$ over $\BSE$ can be expressed via $\qty( e_{i_t} - x^\star(f) )^\top \UO y(f)$, which measures the difference between the optimizer's true utility and the one that would maximize her utility while consistent with forecast $f$. 

The next step is to relate this to our response-calibration error.
We do this via a function $g_{k, j_1, j_2}(q)$, defined for every triplet $(k, j_1, j_2)$ as the maximum reward the optimizer can get when the learner uses $(j_1, j_2)$ with mixing probability $q$ and the optimizer uses an action consistent with $uc^{(k)}$.
This makes the quantity we cared about above to be upper-bounded by a linear combination of terms $\langle g_{k, j_1, j_2}' , \One{\cdot \le Q(f)} \rangle$.
Summing over active rounds, yields the calibration error via $\langle g' , R_T \rangle$.
Cauchy-Schwarz yields the bound.

\begin{proof}
    We first show that if the learner's forecasts are correct, then the optimizer can get at most $\BSE$ utility per round. 
    For a forecast $f$, let $x^\star(f)$ be a mixed strategy of the optimizer that aligns with the forecast, while maximizing utility under that forecast:
    \begin{equation*}
        x^\star(f) \in \argmax_{x \in \Delta_{\nO}} \; x^\top \UO y(f)
        \qquad \text{ such that } \qquad
        u(f) = x^\top \UL
        \quad \text{ and } \quad
        c(f) = x^\top \CL
        .
    \end{equation*}

    By definition of the Budgeted Stackelberg Equilibrium, the optimizer's utility above is at most $\BSE$ for any mixed strategy of the optimizer
    $
        x^\star(f)$, i.e., $\qty(x^\star(f))^\top \UO y(f)
        \le \BSE
        .
    $
    
    So if the learner always responds to a correct forecast, the optimizer can get utility at most $\BSE$.

    Instead, the expected utility of the optimizer when they take action $i$ and the forecast is $f$ is 
    $
        e_i^\top \UO y(f)
        .
    $ 
    Combining these two expressions, we can transform the quantity that we bound. 
    Specifically, because
    \begin{equation} \label{eq:resp_cal:11}
        e_{i_t}^\top \UO y(f) - \BSE
        \le
        \qty( e_{i_t} - x^\star(f) )^\top  \UO y(f)
        ,
    \end{equation}
    if we take expectation $f \sim F_t$, sum over $t \in \mathcal T$, and upper bound this by $\sqrt{2 \nO} \nL \norm{R_T}_{L^2}$ we get the theorem.

    Recall that the response-calibration error is a vector indexed by the learner's action pairs $(j_1, j_2)$ and the environment $k$ induced by the optimizer (while $Q(f)$ appears inside the terms of each index).
    For each such $(k, j_1, j_2)$, the optimizer may choose different actions (among those inducing the environment $uc^{(k)}$) depending on the $Q$ value of the learner's response.
    Let $g$ be a $\hatnO\nL^2$ dimensional vector of functions $[0, 1] \to \R$, defined as the maximum optimizer utility over all pure actions $i \in [\nO]$ that induce $uc^{(k)} \in \calUC$:
    \begin{equation*} 
        g_{k, j_1, j_2}(q)
        =
        \max_{i \in [\nO] \,:\, \qty(\UL[i, :], \CL[i, :]) = uc^{(k)}} \qty\Big(
            q \UO[i, j_1] + (1 - q) \UO[i, j_2]
        )
        .
    \end{equation*}
    
    Using this function we bound the difference in \cref{eq:resp_cal:11} above as
    \begin{alignat}{2} \label{eq:resp_cal:12}
        \Line{
            \qty( e_{i_t} - x^\star(f) )^\top  \UO y(f)
        }{\le}{
            \sum_{k \in [\hatnO]}
            \qty( \One{k = k_t} - f_k ) \cdot g_{k, J_1(f), J_2(f)} \qty\big(Q(f))
        }{}
        \nonumber\\
        \Line{}{=}{
            \sum_{k \in [\hatnO]}
            \sum_{j_1, j_2}
            \One{(j_1, j_2) = \vec J(f)}
            \qty( \One{k = k_t} - f_k ) \cdot g_{k, j_1, j_2}\qty\big(Q(f))
            .
        }{}
    \end{alignat}
    
    The inequality above holds because $i_t$ induces $uc^{(k_t)}$, so $e_{i_t}^\top \UO y(f) \le g_{k_t, J_1(f), J_2(f)}(Q(f))$,
    and because
    \begin{equation*}
        x^\star(f)^\top \UO y(f) \ge \sum_{k \in [\hatnO]} f_k g_{k, J_1(f), J_2(f)}(Q(f))
    \end{equation*}
    since if we consider the $i$ that is the maximizer in the definition of $g_{k, J_1(f), J_2(f)}(Q(f))$ for the $Q(f)$ value and for each $k$ and take expectation over $k \sim f$, we would get expected utility-cost vectors that match $uc(f)$ (which is what the definition of $x^\star(f)$ requires).
    
    We proceed to express $g_{k, j_1, j_2}$ as the integral of its derivative.
    To make this a clean, we also define for $q < 0$, $g_{k, j_1, j_2}(q) = (1 + q) g_{k, j_1, j_2}(0)$, which makes $g_{k, j_1, j_2}$ have domain $[-1, 1]$.
    Overall $g_{k, j_1, j_2}$ is piecewise linear, with slope at most $\qty\big| g_{k, j_1, j_2}'(q) | \le 1$, and $g_{k, j_1, j_2}(-1) = 0$; these ensure that 
    it is an element of $L^2([-1, 1])$ and
    \begin{equation*}
        g_{k, j_1, j_2}\qty\big(q)
        =
        \int_{-1}^q g_{k, j_1, j_2}'(x) dx
        =
        \int_{-1}^1 g_{k, j_1, j_2}'(x) \One{x \le q} dx
        =
        \left\langle
            g_{k, j_1, j_2}', \One{\cdot \le q}
        \right\rangle
        .
    \end{equation*}
    
    Applying this equality to \cref{eq:resp_cal:12} we get
    \begin{equation*}
        \qty( e_{i_t} - x^\star(f) )^\top  \UO y(f)
        \le
        \sum_{k \in [\hatnO]}
            \sum_{j_1, j_2}
            \One{(j_1, j_2) = \vec J(f)}
            \qty( \One{k = k_t} - f_k )
            \left\langle
                g_{k, j_1, j_2}', \One{\cdot \le Q(f)}
            \right\rangle
        =
        \left\langle
            g', \d^R(f, k_t)
        \right\rangle
    \end{equation*}

    Using \cref{eq:resp_cal:11}, taking expectation $f \sim F_t$, and summing over $t \in \mathcal T$ we get
    \begin{equation*}
        \sum_{t \in \mathcal T} \Ex[f \sim F_t]{ e_{i_t}^\top \UO y(f) }
        -
        \BSE \cdot |\mathcal T|
        \le
       \left\langle
            g', R_T
        \right\rangle
        \le
        \norm{g'}_{L^2}
        \norm{R_T}_{L^2}
        \le
        \sqrt{2 \nO} \nL \norm{R_T}_{L^2}
    \end{equation*}
    where in the second to last inequality we used Cauchy-Schwarz and in the last inequality we used
    \begin{equation*}
        \norm{g'}_{L^2}^2
        =
        \sum_{k \in [\hatnO]}
        \sum_{j_1, j_2}
        \norm{ g_{k, j_1, j_2}' }_{L^2}^2
        =
        \sum_{k \in [\hatnO]}
        \sum_{j_1, j_2}
        \int_{-1}^1 g_{k, j_1, j_2}'(x)^2 dx
        \le
        \sum_{k \in [\hatnO]}
        \sum_{j_1, j_2}
        2
        =
        2 \hatnO \nL^2
    \end{equation*}
    which uses that the slope of $g$ is between $-1$ and $1$.
\end{proof}

\paragraph{Why this norm is not enough for stochastic regret}
Unfortunately, even if $R_T = 0$, the learner may suffer $\Omega(T)$ regret in a stochastic environment by best-responding to each of the optimizer's pure actions instead of the mixed ones.
Best-responding to a mixed action allows the learner to overspend for some optimizer actions and underspend on others, which can lead to more learner utility.
If the forecast is correlated with the realization of the optimizer's action, then the learner is forced to best-respond every round, which forces not overspending in every optimizer action.
The example in \cref{sec:app:resp_cal:not_enough} shows that this can happen, even if the forecast is not at all correlated with the optimizer's random choice.
If distinct forecasts induce the \emph{same} best response, the errors in forecasts can cancel each other in $R_T$, even though the true average environment admits a strictly better feasible response.

\section{Lagrangian-calibration Error (Low Stochastic Regret)} \label{sec:lagr_cal}

As the example of \cref{sec:app:resp_cal:not_enough} illustrates, keeping $\norm{R_t}_{L^2}$ small is not enough to guarantee low regret in stochastic environments.
Moreover, bounding utility and cost errors through $R_t$ alone would introduce a dependence on $\hatnO$, whereas the stochastic regret bound we show will be independent of $\hatnO$, similar to classic results in online learning.
To resolve both issues, we introduce a second calibration error, which we call \textit{Lagrangian-calibration error} and denote by $L_t$.
This error allows us to get low stochastic interval regret, i.e., the learner's regret in every interval where the optimizer uses a fixed distribution of actions is $o(T)$.
As we mentioned before, this also implies a lower bound for a budgeted optimizer's utility, see \cref{ssec:budgeted_optimizer:wbse} for details.

\paragraph{Lagrangian dual of the learner's best response}
Recall that a single forecast $f \in \Delta_{\hatnO}$ induces the predicted utility and cost vectors $uc(f) = \qty\big(u(f), c(f)) = \sum_{k=1}^{\hatnO} f_k uc^{(k)}$, and that the learner plays $y(f) \in \BRL\qty\big(uc(f))$.
Because $y(f)$ is the solution to the linear program $\max_{y \in \Delta_{\nL} : c(f)^\top y \le \rL} u(f)^\top y$, we can associate with each forecast $f$ an optimal dual (Lagrange) multiplier $\l(f) \ge 0$ for the budget constraint $c(f)^\top y \le \rL$:
\begin{equation} \label{eq:algo:dual_price}
    \l(f)
    \in
    \argmin_{\l \ge 0} \qty{ \l \rL + \max_{j \in [\nL]} \qty\big( u(f)[j] - \l c(f)[j] ) }
    .
\end{equation}
Intuitively, $\l(f)$ measures the marginal value of budget under forecast $f$ (how tight the budget constraint is).
At $\l = 0$ the dual objective in \eqref{eq:algo:dual_price} is at most $1$, and because the null action has cost $0$ and utility at least $0$, the objective at any $\l \ge 0$ is at least $\l \rL$; hence we can always pick $\l(f) \in [0, 1/\rL]$.

\paragraph{The Lagrangian-calibration error}
Fix a round $t$ with forecast distribution $F_t$ and realized utility-cost vector $uc^{(k_t)} = \qty\big(u^{(k_t)}, c^{(k_t)})$.
For a single forecast $f$, we define the single-round Lagrangian-calibration error $\d^L(f, k_t) = \qty\big( \d^{L, u}(f, k_t),\, \d^{L, c}(f, k_t),\, \d^{L, uy}(f, k_t),\, \d^{L, cy}(f, k_t) ) \in \R^{2\nL + 2}$ via the four components
\begin{equation}\label{eq:algo:delta_L}
\begin{aligned}
    \d^{L, u}(f, k_t)
    &\coloneq
    \frac{1}{2\sqrt{\nL}} \qty\big( u^{(k_t)} - u(f) ) \in \R^{\nL}
    , &
    \d^{L, c}(f, k_t)
    &\coloneq
    \frac{\rL \l(f)}{2\sqrt{\nL}} \qty\big( c^{(k_t)} - c(f) ) \in \R^{\nL}
    , \\
    \d^{L, uy}(f, k_t)
    &\coloneq
    \frac{1}{2\sqrt{\nL}} \qty\big( u^{(k_t)} - u(f) )^\top y(f) \in \R
    , &
    \d^{L, cy}(f, k_t)
    &\coloneq
    \frac{1}{2\sqrt{\nL}} \qty\big( c^{(k_t)} - c(f) )^\top y(f) \in \R
    . 
\end{aligned}
\end{equation}

The total Lagrangian-calibration error up to round $t$ is then the sum of their expectations over $f \sim F_{t'}$:
\begin{equation*}
    L_t
    =
    \qty\big( L_t^u, L_t^c, L_t^{uy}, L_t^{cy} )
    =
    \sum_{t' \le t: t' \in \mathcal T} \Ex[f \sim F_{t'}]{ \d^L(f, k_{t'}) }
    \in
    \R^{2\nL + 2}
    .
\end{equation*}

Because $L_t$ is a standard vector in $\R^{2\nL + 2}$, we measure its magnitude using the ordinary Euclidean norm $\norm{L_t}_2^2 = \norm{L_t^u}_2^2 + \norm{L_t^c}_2^2 + (L_t^{uy})^2 + (L_t^{cy})^2$.
Since all entries of $u^{(k_t)} - u(f)$ and $\rL \l(f)\qty(c^{(k_t)} - c(f))$ lie in $[-1, 1]$, the normalization factor $\frac{1}{2\sqrt{\nL}}$ in \eqref{eq:algo:delta_L} guarantees that every single-round error vector satisfies $\norm{\d^L(f, k_t)}_2^2 \le \frac{\nL + \nL + 1 + 1}{4\nL} \le 1$.

In \cref{ssec:lagr_cal:stochastic_regret}, and ignoring some technical details and terms not depending on $T$, we show how each of the four components of $L_t$ controls a distinct step in bounding the learner's regret against an optimal budget-feasible strategy $y^\star \in \Delta_{\nL}$:
\begin{itemize}
    \item $\abs{L_t^{cy}}$ bounds the learner's cumulative overspending relative to $\rL t$, which bounds the expected number of non-active rounds by $\order*{\max_{t \le T}\abs{L_t^{cy}} + \sqrt{T}}$.
    \item $\abs{L_T^{uy}}$ bounds the gap between the learner's \emph{realized} and \emph{predicted} utilities for active rounds by $\order*{\abs{L_T^{uy}}}$.
    \item $\norm{L_T^c}_2$ bounds the gap between the learner's \emph{predicted} utility and any $y^\star$'s \emph{predicted} utility by $\order*{\norm{L_T^c}_2}$: although $y^\star$ is budget-feasible under the true environment, a forecast $f$ may overestimate $y^\star$'s cost when $\l(f) > 0$ (making $y^\star$ appear infeasible under $f$, as in the example of \cref{sec:app:resp_cal:not_enough}), and the resulting Lagrangian cost penalty is controlled by $\norm{L_T^c}_2$.
    \item $\norm{L_T^u}_2$ bounds the gap between $y^\star$'s \emph{predicted} and $y^\star$'s \emph{realized} utilities across active rounds by $\order*{\norm{L_T^u}_2}$.
\end{itemize}

Combined, a small norm $\norm{L_t}_2$ guarantees that the learner's stochastic regret is at most $\order{\norm{L_T}_2 + \sqrt{T}}$.

\subsection{Lagrangian-calibration Error Implies Low Stochastic Regret}
\label{ssec:lagr_cal:stochastic_regret}

We now formalize how a small Lagrangian-calibration error $L_t$ bounds both the number of non-active rounds and the learner's regret.
The first part also completes the proof that the optimizer's benefit over her $\BSE$ value is small by showing there are only a few non-active rounds and then using \cref{thm:resp_cal:manipulability}. 
For the learner's regret, we will focus on a piecewise-stationary stochastic optimizer that switches mixed strategies across $S$ intervals (with $S = 1$ recovering the standard i.i.d. stochastic setting).

\paragraph{Bounding the number of non-active rounds}
We first bound the expected number of non-active rounds, using $L_t^{cy}$.
Our bound holds against any (even adversarial) optimizer.

\begin{lemma} \label{lem:lagr_cal:null_rounds}
    Let $\mathcal T \subseteq [T]$ be the rounds in which \cref{alg:learner} follows the forecast.
    Against any optimizer, the expected number of non-active rounds satisfies
    \begin{equation*}
        \E\qty\big[ T - |\mathcal T| ]
        \le
        \frac{2\sqrt{\nL}\, \E\qty[ \max_{t \le T} L_t^{cy} ] + 2\sqrt{T} + 1}{\rL}
        .
    \end{equation*}
\end{lemma}

Due to the definition of $L_T^{cy}$, the proof is quite simple.
The expected overspending of the learner above $\rL$ in round $t$ is bounded by $\Ex{\d^{L, cy} (f, k_t)}$ since there is no overspending w.r.t. the forecast, $c(f)^\top y(f) \le \rL$.
This makes the total expected overspending roughly at most $L_T^{cy}$.
Relating the expected spending with the actual spending via concentration inequalities implies the lemma.

\begin{proof}[Proof of \cref{lem:lagr_cal:null_rounds}]
    First let $\mathcal T_t = \mathcal T \cap [t]$ be the set of active rounds up to $t$.
    Also, let $\tau \in [T] \setminus \mathcal T$ be the last non-active round of \cref{alg:learner} (if $\mathcal T = [T]$, then $T - |\mathcal T| = 0$).
    Because round $\tau$ triggered the spending threshold in \cref{alg:learner} and the null action has cost $0$, the cumulative realized cost up to step $\tau$, $C_{\tau-1} = C_\tau = \sum_{r=1}^\tau \CL[i_r, j_r]$ satisfies $C_\tau > \tau \rL - 1$.
    Since all rounds after $\tau$ belong to $\mathcal T$, the total number of non-active rounds is $T - |\mathcal T| = \tau - |\mathcal T_\tau|$.
    Rearranging $C_\tau > \tau \rL - 1$ therefore gives
    \begin{equation} \label{eq:lagr_cal:null_pathwise}
        T - |\mathcal T|
        =
        \tau - |\mathcal T_\tau|
        <
        \frac{C_\tau - \rL |\mathcal T_\tau| + 1}{\rL}
        \le
        \frac{\max_{t \in [T]} \qty\big( C_t - \rL |\mathcal T_t| ) + 1}{\rL}
        .
    \end{equation}

    Since the forecast's mixed strategy always has spending at most $\rL$ w.r.t. the forecast, we can write
    \begin{alignat*}{3}
        \Line{
            C_t - \rL |\mathcal T_t|
        }{\le}{
            \sum_{t' \in \mathcal T_t}\qty(
                e_{i_{t'}}^\top \CL e_{j_{t'}}
                -
                \Ex[f \sim F_{t'}]{ c(f)^\top y(f) }
            )
        }{}
        \\
        \Line{}{=}{
            \sum_{t' \in \mathcal T_t}\qty(
                e_{i_{t'}}^\top \CL e_{j_{t'}}
                -
                \Ex[f \sim F_{t'}]{ e_{i_{t'}}^\top \CL y(f) }
            )
            -
            \sum_{t' \in \mathcal T_t}\Ex[f \sim F_{t'}]{ c(f)^\top y(f) - e_{i_{t'}}^\top \CL y(f) }
        }{}
        \\
        \Line{}{=}{
            \sum_{t' \in \mathcal T_t}\qty(
                e_{i_{t'}}^\top \CL e_{j_{t'}}
                -
                \Ex[f \sim F_{t'}]{ e_{i_{t'}}^\top \CL y(f) }
            )
            +
            2 \sqrt{\nL} L_t^{cy}
        }{}
    \end{alignat*}

    Given the above and \eqref{eq:lagr_cal:null_pathwise}, to bound the lemma we have to bound the expected maximum of the above sum.
    We notice that it is a martingale $M_t$: let
    \begin{equation*}
        m_t
        =
        \One{t \in \mathcal T}
        \qty(
            e_{i_t}^\top \CL e_{j_t}
            -
            \Ex[f \sim F_t]{ e_{i_t}^\top \CL y(f) }
        )
        \qquad \text{ and } \qquad
        M_t = \sum_{t' \le t} m_{t'}
    \end{equation*}

    The expectation of $m_t$ conditioned on the history up to round $t-1$ is $0$, making $M_t$ a martingale.
    Given that $|m_t| \le 1$, Doob's $L^2$ maximal inequality \cite[Theorem 4.4.4]{durrett2019probability} proves that
    \begin{equation*}
        \Ex{\qty( \max_t |M_t| )^2}
        \le
        4
        \Ex{M_T^2}
        \quad\overset{\text{Jensen's}}{\implies}\quad
        \Ex{\max_t |M_t|}
        \le
        2
        \sqrt{\Ex{M_T^2}}
    \end{equation*}

    For $\Ex{M_T^2}$ we notice that by expanding the sum, cross terms have expectation $0$, i.e., $\Ex{m_t m_{t'}} = 0$ for $t \ne t'$.
    This implies
    \begin{equation*}
        \Ex{M_T^2}
        =
        \Ex{ \qty(
            \sum_{t' \le T} m_{t'}
        )^2 }
        =
        \Ex{ \sum_{t' \le T} m_{t'}^2 }
        \le T
    \end{equation*}

    This completes the proof of the lemma.
\end{proof}

We now prove that the learner enjoys low regret.
Specifically, we show that if the $T$ rounds are split into $S \ge 1$ phases, then the learner will approximately best respond in each phase, assuming it has low Lagrangian-calibration error.

\begin{theorem} \label{thm:lagr_cal:stochastic_regret}   
    Let $\mathcal I = \{t_1 + 1, \dots, t_2\} \subseteq [T]$ be any contiguous interval of rounds in which the optimizer draws $i_t \sim x$ independently in each round $t \in \mathcal I$ for some mixed strategy $x \in \Delta_{\nO}$, and let $y^\star \in \BRL(x)$ be the learner's optimal budget-feasible response to $x$.
    Then, regardless of the optimizer's play outside $\mathcal I$, the learner's expected utility in interval $\mathcal I$ satisfies
    \begin{equation*}
        \E\qty[ \sum_{t \in \mathcal I} \UL[i_t, j_t] ]
        \ge
        |\mathcal I| \cdot x^\top \UL y^\star
        -
        \frac{14 \sqrt{\nL}}{\rL} \E\qty[ \max_{t \le T} \norm{L_t}_2 ]
        -
        \frac{2\sqrt{T} + 1}{\rL}
        .
    \end{equation*}
\end{theorem}

\begin{proof}
    By strong duality for the learner's best-response linear program $\max_{y \in \Delta_{\nL} : c(f)^\top y \le \rL} u(f)^\top y$ and its dual \eqref{eq:algo:dual_price}, for any forecast $f$, the best response $y(f)$ maximizes the Lagrangian $y \mapsto \qty\big(u(f) - \l(f) c(f))^\top y$ and satisfies complementary slackness $\l(f)\qty\big(\rL - c(f)^\top y(f)) = 0$.
    Consequently, for every $y \in \Delta_{\nL}$,
    \begin{equation} \label{eq:lagr_cal:kkt}
        u(f)^\top y(f)
        =
        \qty\big( u(f) - \l(f) c(f) )^\top y(f) + \l(f) \rL
        \ge
        u(f)^\top y + \l(f)\qty\big( \rL - c(f)^\top y )
        .
    \end{equation}

    Now consider the suboptimality gap at round $t \in \mathcal I \cap \mathcal T$ with respect to the best response of the learner $y = y^\star \in \BRL(x)$, some forecast $f$, and the $k_t$ induced by the optimizer's action
    \begin{alignat*}{3}
        &
        \qty( u^{(k_t)} )^\top \qty( y(f) - y^\star )
        \\
        \Line{}{=}{
            \qty( u^{(k_t)} - u(f) )^\top y(f)
            +
            u(f)^\top y(f)
            -
            \qty( u^{(k_t)} )^\top y^\star
        }{}
        \\
        \Line{}{\ge}{
            \qty(u^{(k_t)} - u(f))^\top y(f)
            +
            \qty( u(f) - u^{(k_t)} )^\top y^\star
            +
            \lambda(f) \qty(\rL - c(f)^\top y^\star)
        }{}
        \\
        \Line{}{=}{
            \qty(u^{(k_t)} - u(f) )^\top y(f)
            +
            \qty( u(f) - u^{(k_t)} )^\top y^\star
            +
            \lambda(f) \qty( c^{(k_t)} - c(f) )^\top y^\star
            +
            \lambda(f) \qty(\rL - \qty( c^{(k_t)} )^\top y^\star)
        }{}
    \end{alignat*}

    By taking expectation $f \sim F_t$ and summing over $t \in \mathcal I \cap \mathcal T$ we bound each of the terms.

    For the first term,
    \begin{equation*}
        \sum_{t \in \mathcal I \cap \mathcal T} \Ex[f\sim F_t]{ \qty(u^{(k_t)} - u(f) )^\top y(f) }
        =
        2 \sqrt{\nL} \qty( L^{uy}_{t_2} - L^{uy}_{t_1} )
        \ge
        - 4 \sqrt{\nL} \max_{t \in [T]} |L_t^{uy}|
        .
    \end{equation*}

    For the second term,
    \begin{equation*}
        \sum_{t \in \mathcal I \cap \mathcal T} \Ex[f\sim F_t]{
            \qty( u(f) - u^{(k_t)} )^\top y^\star
        }
        =
        - 2 \sqrt{\nL} \qty( L^{u}_{t_2} - L^{u}_{t_1} )^\top y^\star
        \ge
        - 4 \sqrt{\nL} \max_{t \in [T]} \norm{ L_t^u }_2
        .
    \end{equation*}

    For the third term,
    \begin{equation*}
        \sum_{t \in \mathcal I \cap \mathcal T} \Ex[f\sim F_t]{
            \lambda(f) \qty( c^{(k_t)} - c(f) )^\top y^\star
        }
        =
        \frac{2 \sqrt{\nL}}{\rL} \qty( L_{t_2}^c - L_{t_1}^c )^\top y^\star
        \ge
        - \frac{4 \sqrt{\nL}}{\rL} \max_{t \in [T]} \norm{ L_t^c }_2
        .
    \end{equation*}

    Adding up the first three terms makes them lower-bounded by $- \frac{12 \sqrt{\nL}}{\rL} \max_{t \in [T]} \norm{ L_t }_2$.

    For the fourth term we instead take expectation over everything (and especially over the draw of $k_t$),
    \begin{equation*}
        \sum_{t \in \mathcal I} \Ex{
            \One{t \in \mathcal T}
            \lambda(f) \qty(\rL - \qty( c^{(k_t)} )^\top y^\star)
        }
        =
        \sum_{t \in \mathcal I} \Ex{
            \One{t \in \mathcal T}
            \lambda(f) \qty(\rL - x^\top \CL y^\star)
        }
        \ge 0
    \end{equation*}
    where in the inequality we used that $y^\star$ is a best-response to $(x^\top \UL, x^\top \CL)$ and that $\lambda(f) \ge 0$.

    This bounds the error accumulated over active rounds.
    In each non-active round the loss is at most $1$, and the total expected number of non-active rounds is bounded by \cref{lem:lagr_cal:null_rounds}.
    Adding that error completes the proof.
\end{proof}

\section{Calibration Error Minimization} \label{sec:cal_minimization}

In \cref{sec:resp_cal,sec:lagr_cal}, we showed that both the optimizer's utility gain over $\BSE \cdot T$ and the learner's stochastic regret are bounded by the norms of the response-calibration error $R_t$ and the Lagrangian-calibration error $L_t$.
In this section, we show how to generate randomized forecasts $F_1, \dots, F_T$ that simultaneously keep $\norm{R_t}_{L^2}$ and $\norm{L_t}_2$ bounded by $\order*{\sqrt{T}}$ across all rounds $t \in [T]$, completing the analysis of our algorithm.
We achieve this without prior knowledge of $\calUC$ or $\hatnO$.

\paragraph{Combining the two calibration errors}
To minimize $\norm{R_t}_{L^2}$ and $\norm{L_t}_2$ simultaneously, we combine them into a single joint error vector.
For a forecast $f \in \Delta_{\hatnO}$ and a realized index $k \in [\hatnO]$ corresponding to $uc^{(k)} \in \calUC$, we define the single-round joint calibration error $\d(f, k) \coloneq \qty\big(\d^R(f, k),\, \d^L(f, k))$ and the cumulative joint error up to round $t$:
\begin{equation*}
    Z_t
    \coloneq
    (R_t, L_t)
    =
    \sum_{t' \le t : t' \in \mathcal{T}} \Ex[f \sim F_{t'}]{ \d(f, k_{t'}) }
    .
\end{equation*}
We equip the joint space with the inner product and norm combining \eqref{eq:algo:inner_norm_general}--\eqref{eq:algo:inner_norm_step} on $R_t$ and the standard Euclidean inner product on $L_t$:
\begin{equation} \label{eq:cal_min:joint_norm}
    \langle Z, Z' \rangle
    =
    \langle R, R' \rangle + \langle L, L' \rangle
    \qquad\text{ and }\qquad
    \norm{Z}
    =
    \sqrt{ \norm{R}_{L^2}^2 + \norm{L}_2^2 }
    .
\end{equation}

For any forecast $f \in \Delta_{\hatnO}$ and any realized index $k \in [\hatnO]$, the single-round error $\d(f, k)$ satisfies two key properties:
\begin{enumerate}
    \item \textit{Bounded norm:}
    In $\d^R(f, k)$, only the coordinates corresponding to the action pair $\vec J(f)$ are nonzero, and $\norm{\One{\cdot \le Q(f)}}_{L^2}^2 = 1 + Q(f) \le 2$ by \eqref{eq:algo:inner_norm_step}.
    Since $\sum_{k' \in [\hatnO]} \qty\big(\One{k = k'} - f_{k'})^2 \le 2$ for any $f \in \Delta_{\hatnO}$, we have $\norm{\d^R(f, k)}_{L^2}^2 \le 4$.
    Combined with $\norm{\d^L(f, k)}_2^2 \le 1$ from \cref{sec:lagr_cal}, every single-round joint error satisfies
    \begin{equation} \label{eq:cal_min:delta_norm}
        \norm{\d(f, k)}^2
        =
        \norm{\d^R(f, k)}_{L^2}^2 + \norm{\d^L(f, k)}_2^2
        \le
        5
        .
    \end{equation}

    \item \textit{Unbiasedness under its own forecast:}
    If the realized index $k$ is drawn from the forecast $f$ itself, then the error is $0$:
    \begin{equation} \label{eq:cal_min:delta_unbiased}
        \Ex[k \sim f]{ \d(f, k) }
        =
        0
        .
    \end{equation}
    
    This follows from $\Ex[k \sim f]{\One{k = k'}} = f_{k'}$ for all $k' \in [\hatnO]$ (which makes $\Ex[k \sim f]{ \d^R(f, k) } = 0$), along with $\Ex[k \sim f]{u^{(k)}} = u(f)$, and $\Ex[k \sim f]{c^{(k)}} = c(f)$ (which makes $\Ex[k \sim f]{ \d^L(f, k) } = 0$).
\end{enumerate}

\paragraph{Potential function and the zero-sum game}
Following the standard Blackwell approachability framework \cite{Blackwell56, FV97, FH21}, we control the magnitude of $Z_t$ by tracking the potential $\norm{Z_t}^2$.
In any active round $t \in \mathcal{T}$, expanding $\norm{Z_t}^2 = \norm{Z_{t-1} + \Ex[f \sim F_t]{\d(f, k_t)}}^2$ and using \eqref{eq:cal_min:delta_norm} yields
\begin{equation} \label{eq:cal_min:potential_step}
    \norm{Z_t}^2 - \norm{Z_{t-1}}^2
    =
    2 \left\langle Z_{t-1}, \Ex[f \sim F_t]{ \d(f, k_t) } \right\rangle
    +
    \norm{ \Ex[f \sim F_t]{ \d(f, k_t) } }^2
    \le
    2 \Ex[f \sim F_t]{ \langle Z_{t-1}, \d(f, k_t) \rangle }
    +
    5
    .
\end{equation}

Consequently, to ensure that $\norm{Z_t}^2$ increases by at most a constant regardless of which $k_t \in [\hatnO]$ the optimizer induces, the forecaster wants to pick $F_t$ so that $\Ex[f \sim F_t]{\langle Z_{t-1}, \d(f, k) \rangle}$ is small for every $k \in [\hatnO]$.
For example, if $\Ex[f \sim F_t]{\langle Z_{t-1}, \d(f, k) \rangle} \le 1$ for every $k \in [\hatnO]$, then \eqref{eq:cal_min:potential_step} implies $\norm{Z_t}^2 - \norm{Z_{t-1}}^2 \le 7$ in every active round, which by telescoping over at most $T$ rounds guarantees $\norm{Z_t}^2 \le 7T$ (and hence $\norm{R_t}_{L^2}, \norm{L_t}_2 \le \sqrt{7T}$) for all $t \in [T]$.

To get intuition why such a distribution $F_t$ exists, fix the cumulative error $Z_{t-1}$ prior to round $t$ and define for each forecast $f \in \Delta_{\hatnO}$ the \textit{potential drift vector} (or \textit{drift vector} for short) $z(f) \in \R^{\hatnO}$ with coordinates
\begin{equation} \label{eq:cal_min:drift}
    z_k(f)
    \coloneq
    \big\langle Z_{t-1}, \d(f, k) \big\rangle
    \qquad
    \forall k \in [\hatnO]
    .
\end{equation}

We can interpret \eqref{eq:cal_min:drift} as the payoff of a two-player zero-sum game in round $t$: the \textit{forecaster} (the minimization player) chooses a forecast $f \in \Delta_{\hatnO}$, and the \textit{environment player} (the maximization player) chooses an outcome $k \in [\hatnO]$.
If the environment player moves first and commits to any mixed strategy $w \in \Delta_{\hatnO}$, its expected payoff against forecast $f$ is $\sum_{k \in [\hatnO]} w_k z_k(f) = w^\top z(f)$.
The forecaster can then simply respond with the identical forecast $f = w$ as the pure action, which satisfies
\begin{equation} \label{eq:cal_min:drift_zero}
    w^\top z(w)
    =
    \sum_k w_k \big\langle Z_{t-1} , \delta(w, k) \big\rangle
    =
    \left\langle Z_{t-1} , \sum_k w_k \delta(w, k) \right\rangle
    =
    \left\langle Z_{t-1} , \Ex[k \sim w]{ \delta(w, k) }  \right\rangle
    =
    0
\end{equation}
by \cref{eq:cal_min:delta_unbiased}.
Thus, the value of the game when the environment player moves first is at most $0$:
\begin{equation*}
    \max_{w \in \Delta_{\hatnO}} \min_{f \in \Delta_{\hatnO}} w^\top z(f)
    \le
    0
    .
\end{equation*}

If a minimax theorem applied directly to this game, the Nash equilibrium would provide the forecaster with a randomized forecast $F_t$ over $\Delta_{\hatnO}$ guaranteeing $\max_{k \in [\hatnO]} \Ex[f \sim F_t]{z_k(f)} \le 0$.
However, because the learner's best response $y(f)$ and dual multiplier $\l(f)$ can change discontinuously with $f$, the payoff $f \mapsto z_k(f)$ is discontinuous, standard minimax theorems for infinite games do not apply directly.
Moreover, even if we use an extension like Sion's minimax theorem to prove the existence of such a randomized forecast $F_t$, there is no guarantee that this $F_t$ can be represented efficiently.
The obvious solution is to discretize the forecast space to restore a finite zero-sum game, but discretizing $\Delta_{\hatnO}$ into a fine grid would require exponential in $\hatnO$ number of points.

\paragraph{Forecasting without knowing $\calUC$ in advance}
In addition to the continuity and discretization issues above, the learner does not know the matrices $(\UL, \CL)$ or the number of distinct utility-cost vectors $\hatnO = |\calUC|$ in advance, aside from the all-zeros vector $uc^{(1)} = 0 \in \calUC$ corresponding to the null action.
Therefore, the forecaster can only use forecasts that predict utility-cost vectors that have already been observed: let $\calUC_t \subseteq \calUC$ be the set of distinct utility-cost vectors observed in rounds $1, \dots, t$ (initialized at $\calUC_0 = \{0\}$).
Without loss of generality, we index the elements of $\calUC = \{uc^{(1)}, \dots, uc^{(\hatnO)}\}$ in the order they are first observed, so that prior to round $t$ the observed dictionary is $\calUC_{t-1} = \{uc^{(1)}, \dots, uc^{(\hat{n}_{t-1})}\}$ of size $\hat{n}_{t-1} \coloneq |\calUC_{t-1}| \in \qty[1, \min\{\hatnO, t\}]$.
In every active round $t \in \mathcal{T}$, the forecaster is restricted to the known indices $[\hat{n}_{t-1}]$: it outputs a distribution $F_t$ over forecasts $f \in \Delta_{\hat{n}_{t-1}}$ (viewed as elements of $\Delta_{\hatnO}$ by setting $f_k = 0$ for all unobserved $k > \hat{n}_{t-1}$) against outcomes $k \in [\hat{n}_{t-1}]$, with drift vector $z(f) = \qty\big(z_1(f), \dots, z_{\hat{n}_{t-1}}(f)) \in \R^{\hat{n}_{t-1}}$.
Because $\operatorname{supp}(w) \subseteq [\hat{n}_{t-1}]$ for any $w \in \Delta_{\hat{n}_{t-1}}$, the identity $w^\top z(w) = 0$ in \eqref{eq:cal_min:drift_zero} continues to hold over $\Delta_{\hat{n}_{t-1}}$.
For the rounds where the optimizer induces a vector that has not been seen before, the potential $Z$ can increase by more than the above bound.
However, the number of rounds this can happen is bounded by $\hatnO$, which introduces an additional error of that size.

\paragraph{Approximating the Nash equilibrium via projected gradient ascent}
Here we show how to get an \textit{approximate} Nash equilibrium strategy $F_t$ over $\Delta_{\hat{n}_{t-1}}$ that guarantees expected payoff at most $1$ (instead of $0$) against every known $k \in [\hat{n}_{t-1}]$.
We compute $F_t$ constructively by simulating a learning process for the above zero-sum game.
Specifically, the environment player runs \textit{Online Gradient Ascent}~\cite{Hazan16} over $M_t = 50 \hat{n}_{t-1} T^2$ iterations, to which the forecaster responds every time with the strategy we used in \cref{eq:cal_min:drift_zero}.
Thus, in iteration $m \in [M_t]$ the maximization player maintains a distribution over its $\hat{n}_{t-1}$ actions $f^{(m)} \in \Delta_{\hat{n}_{t-1}}$ (recall the environment player's pure actions are the outcomes $k \in [\hat{n}_{t-1}]$) and the minimization player responds with the forecast $f^{(m)}$ (recall the forecaster's pure actions are forecasts) that guarantees at most $0$ payoff by \cref{eq:cal_min:drift_zero}.
The final randomized forecast $F_t$ is the uniform distribution over the forecasts $f^{(m)}$, which as we show next guarantees expected payoff at most $1$ against any $k \in [\hat{n}_{t-1}]$.

\begin{algorithm}[ht]
\DontPrintSemicolon
\SetAlgoNoEnd
\caption{Projected Gradient Ascent Forecaster for Active Round $t \in \mathcal{T}$}
\label{alg:forecaster}
\KwIn{Observed dictionary $\calUC_{t-1}$ of size $\hat{n}_{t-1} \ge 1$ and cumulative calibration error $Z_{t-1} = (R_{t-1}, L_{t-1})$}
Set $M_t = 50 \hat{n}_{t-1} T^2$, step size $\eta_t = \frac{2}{M_t}$, and initial forecast $f^{(1)} = \qty(1/\hat{n}_{t-1}, \dots, 1/\hat{n}_{t-1}) \in \Delta_{\hat{n}_{t-1}}$\;
\For{$m = 1, \dots, M_t$}{
    Compute drift vector $z\qty(f^{(m)}) \in \R^{\hat{n}_{t-1}}$ with $z_k\qty(f^{(m)}) = \left\langle Z_{t-1}, \d\qty(f^{(m)}, k) \right\rangle$ for each $k \in [\hat{n}_{t-1}]$\;
    Update $f^{(m+1)} = \Pi_{\Delta_{\hat{n}_{t-1}}}\qty\big( f^{(m)} + \eta_t z\qty(f^{(m)}) )$, where $\Pi_{\Delta_{\hat{n}_{t-1}}}(\cdot)$ is the Euclidian projection\;
}
\KwOut{Uniform distribution $F_t = \mathrm{Unif}\qty\big(\{f^{(1)}, \dots, f^{(M_t)}\})$}
\end{algorithm}

\begin{lemma} \label{lem:cal_min:oga}
    In every active round $t \in \mathcal{T}$, the forecast distribution $F_t$ returned by \cref{alg:forecaster} satisfies
    \begin{equation} \label{eq:cal_min:oracle_guarantee}
        \Ex[f \sim F_t]{ \langle Z_{t-1}, \d(f, k) \rangle }
        \le
        1
        \qquad
        \forall k \in [\hat{n}_{t-1}]
        .
    \end{equation}
\end{lemma}

\begin{proof}
    Fix a round $t$.
    We start with a crude bound on $Z_{t-1}$: since $Z_{t-1}$ is the sum of at most $t - 1 < T$ single-round errors of norm at most $\sqrt{5}$ by \eqref{eq:cal_min:delta_norm}, the triangle inequality gives $\norm{Z_{t-1}} \le \sqrt{5}T$.
    By Cauchy--Schwarz and \eqref{eq:cal_min:delta_norm}, each coordinate of the drift vector is bounded by $\abs{z_k(f)} \le \norm{Z_{t-1}} \norm{\d(f, k)} \le 5T$ for all $k \in [\hat{n}_{t-1}]$, and thus $\norm{z(f)}_2^2 \le 25 \hat{n}_{t-1} T^2$ for every $f \in \Delta_{\hat{n}_{t-1}}$.
    This bound explains the number of iterations $M_t = 50 \hat{n}_{t-1} T^2 \ge 2 \norm{z(f)}_2^2$.
    Note that a larger $M_t$ would have led to less error (specifically we will get linear in $M_t$ regret) but that is sufficient for the result we are looking for.

    We now follow the standard analysis of Projected Gradient Descent, which we include here for completeness.
    Fix any pure action of the environment player $k \in [\hat{n}_{t-1}]$ and its forecast $e_k \in \Delta_{\hat{n}_{t-1}}$.
    By definition of the Euclidean projection we have for each iteration $m \in [M_t]$,
    \begin{equation*}
        \norm{f^{(m+1)} - e_k}_2^2
        \le
        \norm{f^{(m)} + \eta_t z\qty(f^{(m)}) - e_k}_2^2
        =
        \norm{f^{(m)} - e_k}_2^2
        +
        2\eta_t \qty(f^{(m)} - e_k)^\top z\qty(f^{(m)})
        +
        \eta_t^2 \norm{z\qty(f^{(m)})}_2^2
        .
    \end{equation*}
    Rearranging, summing over $m = 1, \dots, M_t$, telescoping, and using $\norm{f^{(M_t+1)} - e_k}_2^2 \ge 0$ yields the regret of this process:
    \begin{equation*}
        \sum_{m=1}^{M_t} \qty(e_k - f^{(m)})^\top z\qty(f^{(m)})
        \le
        \frac{\norm{f^{(1)} - e_k}_2^2}{2\eta_t}
        +
        \frac{\eta_t}{2} \sum_{m=1}^{M_t} \norm{z\qty(f^{(m)})}_2^2
        \le
        \frac{1}{\eta_t}
        +
        \frac{\eta_t}{2} M_t \frac{M_t}{2}
        =
        M_t
    \end{equation*}
    where we used $\norm{ f^{(1)} - e_k }_2^2 \le 2$, $\norm{z\qty(f^{(m)})}_2^2 \le M_t / 2 = 1 / \eta_t$.
    
    By \eqref{eq:cal_min:drift_zero}, the forecaster's choice $f^{(m)}$ satisfies $\qty(f^{(m)})^\top z\qty(f^{(m)}) = 0$ for every $m \in [M_t]$.
    Dividing by $M_t$ makes the above inequality
    \begin{equation*}
        \Ex[f \sim F_t]{ \langle Z_{t-1}, \d(f, k) \rangle }
        =
        \frac{1}{M_t} \sum_{m=1}^{M_t} z_k\qty(f^{(m)})
        \le
        1
        ,
    \end{equation*}
    proving the lemma.
\end{proof}

\subsection{Bounding the Cumulative Calibration Error}
\label{ssec:cal_min:error_bounds}

We now use \cref{lem:cal_min:oga} to bound the cumulative error $\norm{Z_t}$ in both the adversarial and stochastic settings. By examining the stochastic setting more carefully, we get a bound that does not depend on the number of actions that the optimizer has, which matches the bounds in Stochastic BwK.

\paragraph{Bound against any optimizer}
The bound in this case follows from \cref{lem:cal_min:oga}.
For rounds where the realized vector $uc^{(k_t)}$ has already been observed, the strategy produced by \cref{alg:forecaster} guarantees low norm increase.
On the other hand, there are at most $\hatnO$ rounds where a new vector can be observed.
This makes the total error $\order*{\sqrt{T} + \hatnO}$.

\begin{lemma} \label{lem:cal_min:adversarial}
    Let $\mathcal T_t$ be the active rounds up to round $t$.
    Against any optimizer and in every round $t \in [T]$, it holds
    \begin{equation} \label{eq:cal_min:norm_bound}
        \sqrt{ \norm{R_t}_{L^2}^2 + \norm{L_t}_2^2 }
        =
        \norm{Z_t}
        \le
        \sqrt{7 |\mathcal{T}_t|} + \sqrt{5}\hatnO
        \le
        \sqrt{7T} + \sqrt{5}\nO
        .
    \end{equation}
\end{lemma}

\begin{proof}
    We show by induction on $t$ that $\norm{Z_t} \le \sqrt{7|\mathcal{T}_t|} + \sqrt{5}\hat{n}_t$.
    The claim trivially holds for $t = 0$, since $\norm{Z_0} = 0$.

    We now prove the lemma assuming it is true for $t-1$.
    For non-active rounds $t \notin \mathcal{T}$, $Z_t = Z_{t-1}$.
    For active rounds $t \in \mathcal{T}$, we distinguish two cases.
    First, if $k_t \in [\hat{n}_{t-1}]$ has already been seen, then $\hat{n}_t = \hat{n}_{t-1}$ and it holds 
    \begin{equation*}
        \norm{Z_t}^2
        =
        \norm{Z_{t-1} + \Ex[f \sim F_t]{\delta(f, k_t)} }^2
        =
        \norm{Z_{t-1}}^2
        +
        2 \left\langle Z_{t-1} , \Ex[f \sim F_t]{\delta(f, k_t)} \right\rangle
        +
        \norm{ \Ex[f \sim F_t]{\delta(f, k_t)} }^2
        \le
        \norm{Z_{t-1}}^2
        +
        2
        +
        5
    \end{equation*}
    where in the inequality we used \cref{lem:cal_min:oga} and \cref{eq:cal_min:delta_norm}.
    Using the inductive hypothesis and some algebra proves the desired bound for $\norm{Z_t}$:
    \begin{equation*}
        \norm{Z_t}^2
        \le
        \qty(\sqrt{7|\mathcal{T}_{t-1}|} + \sqrt{5}\hat{n}_{t-1} )^2
        +
        7
       =
        \qty(\sqrt{7(|\mathcal{T}_t|-1)} + \sqrt{5}\hat{n}_t )^2
        +
        7
        \le
        \qty(\sqrt{7|\mathcal{T}_t|} + \sqrt{5}\hat{n}_t )^2
    \end{equation*}
    where the last inequality holds because $|\mathcal{T}_t| \ge 1$ and $\hat{n}_t \ge 0$.

    The second case is when $k_t \notin [\hat{n}_{t-1}]$, i.e., the utility-cost vector has not been observed before.
    In that case, $\hat{n}_t = \hat{n}_{t-1} + 1$, which makes $\norm{Z_t} \le \norm{Z_{t-1}} + \sqrt{5} \le \sqrt{7|\mathcal{T}_t|} + \sqrt{5} \hat{n}_t$ by \cref{eq:cal_min:delta_norm}.
\end{proof}

\paragraph{Dimension-free bound in stochastic environments}
While \cref{lem:cal_min:adversarial} suffices to bound the optimizer's utility we prove in our main \cref{thm:cal_min:main} using \cref{thm:resp_cal:manipulability}, plugging it into \cref{thm:lagr_cal:stochastic_regret} would add an $\order{\hatnO}$ term to the learner's stochastic regret.
When the optimizer plays i.i.d.\ strategies across $S \ge 1$ intervals, we can remove this dependence on $\hatnO$ completely.
This dependence is also more natural given existing bounds for Stochastic BwK, where the learner's regret does not depend on the number of different environments it will face.

The key intuition is simple: although the forecaster does not know the true distribution of $k_t$ in interval $\mathcal{I}_s$, the \textit{empirical distribution} of the vectors observed in the earlier rounds of $\mathcal{I}_s$ is supported entirely on the known vectors $\calUC_{t-1}$. 
Hence in our analysis we can use \cref{lem:cal_min:oga} assuming the optimizer uses the empirical distribution instead of the true distribution.
However, this introduces an error proportional to the error between the true and empirical distributions.
We show this error is small by standard concentration bounds between the two quantities, i.e., the variance after $\tau$ rounds of the interval has passed is $\order*{1/\sqrt \tau}$ in expectation, which is independent of the support size $\hatnO$.

\begin{lemma} \label{lem:cal_min:stochastic_norm}
    Suppose the horizon $[T]$ is partitioned into $S \ge 1$ contiguous intervals $\mathcal{I}_1, \dots, \mathcal{I}_S$, where in each interval $\mathcal{I}_s$ the optimizer draws $i_t \sim x^{(s)}$ independently for some $x^{(s)} \in \Delta_{\nO}$.
    Then the expected maximum calibration error satisfies
    \begin{equation} \label{eq:cal_min:stochastic_norm_bound}
        \E\qty[ \max_{t \le T} \norm{Z_t} ]
        \le
        18 \sqrt{S T}
        .
    \end{equation}
\end{lemma}

\begin{proof}
    For each interval $\mathcal{I}_s = \{t_{s-1} + 1, \dots, t_s\}$, let $f^{\star(s)} \in \Delta_{\hatnO}$ be the distribution of indices $k_t$ of $\calUC$ induced by $i_t \sim x^{(s)}$\, i.e., what would be the ``correct'' forecast.
    Also let $\hat{f}_{t-1} \in \Delta_{\hatnO}$ be the empirical distribution of environments of rounds $\mathcal I_s \cap [t-1]$ (for the first round in an interval define empirical distribution as $e_1$); even though $\hat{f}_{t-1}$ is not a forecast, it is convenient to see it as one, since it is the optimal prediction with the samples available, assuming knowledge of when interval started.
    
    Define $\bar{\d}_t(k) \coloneq \Ex[f \sim F_t]{\d(f, k)}$ to be the expected calibration error of outcome $k$ in round $t$.
    We decompose the potential drift of an active round $t$:
    \begin{equation} \label{eq:cal_min:drift_decomposition}
    \begin{split}
        \left\langle  Z_{t-1}, \bar \delta_t(k_t) \right\rangle
        & =
        \left\langle
            Z_{t-1}
            ,\;
            \bar\delta_t(k_t)
            -
            \sum_{k \in [\hatnO]} f^{\star(s)}_k \bar{\d}_t(k)
        \right\rangle
        \\
        &\;+
        \left\langle
            Z_{t-1}
            ,
            \sum_{k \in [\hatnO]} \qty( f^{\star(s)}_k - \hat{f}_{t-1, k} ) \bar{\d}_t(k)
        \right\rangle
        \\
        &\;+
        \sum_{k \in [\hatnO]} \hat{f}_{t-1, k} \left\langle Z_{t-1}, \bar{\d}_t(k) \right\rangle
        .
    \end{split}
    \end{equation}

    The error in the first line corresponds to the difference in the optimizer's realized behavior and expected behavior, the second to the difference in the optimizer's expected behavior 
    and empirical one, observed so far from rounds in this interval, and the third line to the calibration error with respect to the empirical behavior of the optimizer of each interval.
    We proceed to bound each term separately.
    
    \begin{claim} \label{cl:cal_min:first_term}
        For the first term in \cref{eq:cal_min:drift_decomposition} it holds
        \begin{equation*}
            \Ex{
                \max_{t \le T}
                \sum_{\tau \in \mathcal{T}_t}
                \left\langle
                    Z_{\tau-1}
                    ,\;
                    \bar\delta_\tau(k_\tau)
                    -
                    \sum_{k \in [\hatnO]} f^{\star(s)}_k \bar{\d}_\tau(k)
                \right\rangle
            }
            \le
            2\sqrt{5 T}\, \sqrt{\E\qty[\max_{t \le T} \norm{Z_t}^2]}
            .
        \end{equation*}
    \end{claim}

    The proof of the claim is pretty straightforward: the right part of the inner product has expectation $0$ so the sum is a martingale and we can bound it using Doob's $L^2$ maximal inequality.
    
    \begin{proof}[Proof of \cref{cl:cal_min:first_term}]
        Define the sequences
        \begin{equation*}
            \xi_t
            \coloneq
            \One{t \in \mathcal{T}}
            \left\langle
                Z_{t-1}
                ,\;
                \bar\delta_t(k_t)
                -
                \sum_{k \in [\hatnO]} f^{\star(s)}_k \bar{\d}_t(k)
            \right\rangle
            \qquad\text{ and }\qquad
            \Xi_t
            \coloneq
            \sum_{\tau \le t} \xi_\tau
            .
        \end{equation*}
        Conditioned on the history $\mathcal{H}_{t-1}$ up to round $t - 1$, the quantities $Z_{t-1}$, $\One{t \in \mathcal{T}}$, and $F_t$ (hence $\bar{\d}_t$) are fixed, while $k_t \sim f^{\star(s)}$ is drawn independently.
        Thus $\ExC{\xi_t}{\mathcal{H}_{t-1}} = 0$, making $\Xi_t$ a martingale.
        Moreover, since the variance of a random variable is bounded by its second moment, Cauchy--Schwarz and \eqref{eq:cal_min:delta_norm} give
        \begin{equation*}
            \ExC{\xi_t^2}{\mathcal{H}_{t-1}}
            \le
            \Ex[k \sim f^{\star(s)}]{ \left\langle Z_{t-1}, \bar{\d}_t(k) \right\rangle^2 }
            \le
            \norm{Z_{t-1}}^2 \Ex[k \sim f^{\star(s)}]{ \norm{\bar{\d}_t(k)}^2 }
            \le
            5 \norm{Z_{t-1}}^2
            .
        \end{equation*}
        
        By Doob's $L^2$ maximal inequality~\cite[Theorem 4.4.4]{durrett2019probability}, Jensen's inequality, and the orthogonality of martingale differences ($\Ex{\xi_t \xi_{t'}} = 0$ for $t \ne t'$), we obtain
        \begin{equation*}
            \E\qty[\max_{t \le T} \Xi_t]
            \le
            2\sqrt{\E\qty[\Xi_T^2]}
            =
            2\sqrt{\sum_{t=1}^T \E\qty[\xi_t^2]}
            \le
            2\sqrt{5 \sum_{t=1}^T \E\qty[\norm{Z_{t-1}}^2]}
            \le
            2\sqrt{5 T}\, \sqrt{\E\qty[\max_{t \le T} \norm{Z_t}^2]}
            .
            \qedhere
        \end{equation*}
    \end{proof}

    \begin{claim} \label{cl:cal_min:second_term}
        For the second term in \cref{eq:cal_min:drift_decomposition} it holds
        \begin{equation*}
            \E\qty[
                \max_{t \le T}
                \sum_{\tau \in \mathcal{T}_t}
                \left\langle
                    Z_{\tau-1}
                    ,
                    \sum_{k \in [\hatnO]} \qty( f^{\star(s)}_k - \hat{f}_{\tau-1, k} ) \bar{\d}_\tau(k)
                \right\rangle
            ]
            \le
            4\sqrt{S T}\, \sqrt{\E\qty[\max_{t \le T} \norm{Z_t}^2]}
            .
        \end{equation*}
    \end{claim}

    This is the most complicated quantity to bound.
    On a high level, it is bounded because in expectation, $\hat f_{\tau-1}$ is $f^{\star(s)}$ for all but the first round of an interval.
    However, a naive bound would depend on the dimension of $\bar \delta_\tau(k)$, which can have $\nL^2 \min\{\tau , \hatnO \}$ non-zero coordinates.
    To achieve the improved bound, we decompose the right part of the inner product into three separate errors: the variance between $\hat f_{\tau-1}$ and $f^{\star(s)}$, along with the variance of the induced utilities and costs.
    Since $\hat f_{\tau-1}$ and $f^{\star(s)}$ are distributions, while the utilities and costs have dimension $\nL$, we can bound their variance via standard concentration of averages.

    \begin{proof}[Proof of \cref{cl:cal_min:second_term}]
        For each round $t \in \mathcal{I}_s$, let $\Gamma_t \coloneq \sum_{k \in [\hatnO]} \qty( f^{\star(s)}_k - \hat{f}_{t-1, k} ) \bar{\d}_t(k)$.
        Applying Cauchy--Schwarz first to the inner product $\langle Z_{\tau-1}, \Gamma_\tau \rangle \le \norm{Z_{\tau-1}} \norm{\Gamma_\tau}$ and then to the expectation $\E\qty[\norm{Z_{\tau-1}} \norm{\Gamma_\tau}] \le \sqrt{\E\qty[\norm{Z_{\tau-1}}^2]} \sqrt{\E\qty[\norm{\Gamma_\tau}^2]}$ yields
        \begin{equation} \label{eq:cal_min:second_term_cs}
        \begin{split}
            \E\qty[
                \max_{t \le T}
                \sum_{\tau \in \mathcal{T}_t} \left\langle Z_{\tau-1}, \Gamma_\tau \right\rangle
            ]
            & \le
            \max_{t \le T}\sum_{\tau= 1}^t \E\qty[ \norm{Z_{\tau - 1}} \norm{\Gamma_\tau} ]
            =
            \sum_{t=1}^T \E\qty[ \norm{Z_{t-1}} \norm{\Gamma_{t}} ]
            \\ & \le
            \sum_{t=1}^T \sqrt{\E \norm{Z_{t-1}}^2} \sqrt{\E \norm{\Gamma_t}^2}
            \le
            \sqrt{\E\qty[\max_{t \le T} \norm{Z_t}^2]}
            \sum_{t=1}^T \sqrt{\E\qty[\norm{\Gamma_t}^2]}
            .
        \end{split}
        \end{equation}
        
        We now bound $\E\qty[\norm{\Gamma_t}^2]$ for a fixed $t \in \mathcal I_s$.
        Recall the definition of $\Gamma_t$ and $\bar \delta_t(k)$ inside it:
        \begin{equation} \label{eq:cal_min:1}
            \norm{ \Gamma_t }^2
            =
            \norm{ \sum_{k \in [\hatnO]} \qty( f^{\star(s)}_k - \hat{f}_{t-1, k} ) \Ex[f \sim F_t]{\d(f, k)} }^2
            \le
            \Ex[f \sim F_t]{ \norm{ \sum_{k \in [\hatnO]} \qty( f^{\star(s)}_k - \hat{f}_{t-1, k} ) \d(f, k) }^2 }
        \end{equation}
        where the inequality uses triangle inequality of the norm $\norm{\cdot}$ and then Jensen's inequality on the square function.
        Since $f^{\star(s)}, \hat{f}_{t-1} \in \Delta_{\hatnO}$ are distributions, we have $\sum_{k \in [\hatnO]} \qty\big(f^{\star(s)}_k - \hat{f}_{t-1, k}) = 0$;  this makes many terms of $\delta(f, k)$ to cancel out.
        Let $\e_k = f^{\star(s)}_k - \hat{f}_{t-1, k}$; for $\delta^R(f, k)$ we have
        \begin{alignat*}{3}
            \Line{
                \norm{ \sum_{k \in [\hatnO]} \e_k \d^R(f, k) }^2_{L^2}
            }{=}{
                \sum_{k' \in [\hatnO] }
                \norm{ \sum_{k \in [\hatnO]} \e_k \qty\big( \One{k' = k} - f_{k'} ) \One{\cdot \le Q(f)} }^2_{L^2}
            }{\text{def of } \delta^R\text{, \eqref{eq:algo:delta_R_alt}}}
            \\
            \Line{}{=}{
                \sum_{k' \in [\hatnO] }
                \norm{ \e_{k'} \One{\cdot \le Q(f)} }^2_{L^2}
            }{\sum_k \e_k = 0}
            \\
            \Line{}{=}{
                \sum_{k' \in [\hatnO] }
                \e_{k'}^2
                \norm{ \One{\cdot \le Q(f)} }^2_{L^2}
                =
                \qty\big( 1 + Q(f) )
                \sum_{k' \in [\hatnO] }
                \e_{k'}^2
            }{}
            \\
            \Line{}{\le}{
                2 \sum_{k' \in [\hatnO] } \e_{k'}^2
                =
                2 \norm{ f^{\star(s)} - \hat f_{t-1} }_2^2
            }{}
        \end{alignat*}

        Similarly, using the definition of $\delta^L$ we get
        \begin{alignat*}{3}
            \Line{
                \norm{ \sum_{k \in [\hatnO]} \e_k \d^L(f, k) }_2^2
            }{=}{
                \frac{1}{4 \nL} \left(
                    \norm{ \sum_{k \in [\hatnO]} \e_k \qty( u^{(k)} - u(f) ) }_2^2
                    +
                    \norm{ \sum_{k \in [\hatnO]} \e_k \rL \lambda(f) \qty( c^{(k)} - c(f) ) }_2^2
                \right.
            }{}
            \\
            \Line{}{}{
                \qquad\;\;\left.
                    +
                    \qty( \sum_{k \in [\hatnO]} \e_k \qty( u^{(k)} - u(f) )^\top y(f) )^2
                    +
                    \qty( \sum_{k \in [\hatnO]} \e_k \qty( c^{(k)} - c(f) )^\top y(f) )^2
                \right)
            }{}
            \\
            \Line{}{=}{
                \frac{1}{4 \nL} \left(
                    \norm{ \sum_{k \in [\hatnO]} \e_k u^{(k)} }_2^2
                    +
                    \qty\big( \rL \lambda(f) )^2\norm{ \sum_{k \in [\hatnO]} \e_k c^{(k)} }_2^2
                \right.
            }{}
            \\
            \Line{}{}{
                \qquad\;\;\left.
                    +  
                    \qty( \sum_{k \in [\hatnO]} \e_k \qty( u^{(k)} )^\top y(f) )^2
                    +
                    \qty( \sum_{k \in [\hatnO]} \e_k \qty( c^{(k)} )^\top y(f) )^2
                \right)
            }{}
            \\
            \Line{}{\le}{
                \frac{1}{4 \nL} \left(
                    2 \norm{ \sum_{k \in [\hatnO]} \e_k u^{(k)} }_2^2
                    +
                    \qty( \qty\big( \rL \lambda(f) )^2 + 1) \norm{ \sum_{k \in [\hatnO]} \e_k c^{(k)} }_2^2
                \right)
            }{}
            \\
            \Line{}{=}{
                \frac{1}{4 \nL} \left(
                    2 \norm{ u\qty(f^{\star(s)}) - u\qty(\hat{f}_{t-1}) }_2^2
                    +
                    \qty( \qty\big( \rL \lambda(f) )^2 + 1) \norm{ c\qty(f^{\star(s)}) - c\qty(\hat{f}_{t-1}) }_2^2
                \right)
            }{}
        \end{alignat*}
        where in the inequality we used Cauchy-Schwarz to bound the last two terms and then used $\norm{y(f)}_2^2 \le 1$.
        Also using $\lambda(f) \le 1/\rL$ and plugging in the above two bounds into \cref{eq:cal_min:1} we get
        \begin{equation} \label{eq:cal_min:11}
            \norm{\Gamma_t}^2
            \le
            2 \norm{ f^{\star(s)} - \hat f_{t-1} }_2^2
            +
            \frac{1}{2\nL}
            \norm{ u\qty(f^{\star(s)}) - u\qty(\hat{f}_{t-1}) }_2^2
            +
            \frac{1}{2\nL}
            \norm{ c\qty(f^{\star(s)}) - c\qty(\hat{f}_{t-1}) }_2^2
        \end{equation}

        We now prove that the above quantities are small in expectation via standard concentration of the empirical average of a quantity, i.e., $\Ex{\hat f_{t-1}} = f^{\star(s)}$, $\Ex{u(\hat f_{t-1})} = u(f^{\star(s)})$, and $\Ex{c(\hat f_{t-1})} = c(f^{\star(s)})$ (recall $u(f)$ and $c(f)$ are linear in the mass of $f$).
        Consider an interval $\mathcal I_s$.
        In the first round $t$ of $\mathcal I_s$, $\norm{\Gamma_t}^2 \le 5$ by the boundedness of each of the norms.
        For every subsequent round $t$ where $L$ rounds of interval $\mathcal I_s$ have elapsed we have
        \begin{equation*}
            \Ex{ \norm{ f^{\star(s)} - \hat f_{t-1} }_2^2 }
            =
            \Ex{ \norm{ \frac{1}{L} \sum_{\tau \in \mathcal I_s \cap [t-1]} \qty( f^{\star(s)} - e_{k_\tau} ) }_2^2 }
            \le
            \frac{1}{L}
            ,
        \end{equation*}
        \begin{equation*}
            \Ex{ \norm{ u\qty(f^{\star(s)}) - u\qty(\hat{f}_{t-1}) }_2^2 }
            =
            \Ex{ \norm{ \frac{1}{L} \sum_{\tau \in \mathcal I_s \cap [t-1]} \qty( u(f^{\star(s)}) - u(e_{k_\tau}) ) }_2^2 }
            \le
            \frac{\nL}{L},
        \end{equation*}
        and similarly for the costs.
        Plugging this in \cref{eq:cal_min:11} implies that $\Ex{\norm{\Gamma_t}^2} \le \frac{3}{L}$, which in turn yields
        \begin{equation*}
            \sum_{t=1}^T \sqrt{ \Ex{\norm{\Gamma_t}^2} }
            \le
            \sum_{s = 1}^S \qty( \sqrt{5} + \sum_{L = 1}^{|\mathcal I_s|-1} \sqrt{ \frac{3}{L} } )
            \le
            \sum_{s = 1}^S 2 \sqrt{3 |\mathcal I_s|}
            \le
            4 \sqrt{S T}
            .
        \end{equation*}

        Plugging the above into \cref{eq:cal_min:second_term_cs} completes the proof of the claim.
    \end{proof}

    \begin{claim} \label{cl:cal_min:third_term}
        For the third term in \cref{eq:cal_min:drift_decomposition} it holds, in every active round $t \in \mathcal{T}$,
        \begin{equation*}
            \sum_{k \in [\hatnO]} \hat{f}_{t-1, k} \left\langle Z_{t-1}, \bar{\d}_t(k) \right\rangle
            \le
            1
            .
        \end{equation*}
    \end{claim}

    The proof follows by \cref{lem:cal_min:oga} since $\hat{f}_{t-1}$ is supported on environments seen prior to round $t$.

    \begin{proof}[Proof of \cref{cl:cal_min:third_term}]
        In any round $t \in \mathcal{I}_s$, every outcome $k_{t'}$ from the preceding rounds $t' \in \mathcal{I}_s \cap [t-1]$ has already been observed prior to round $t$ (and for the first round of $\mathcal{I}_s$, $\hat{f}_{t-1} = e_1$ puts all mass on $uc^{(1)} = 0 \in \calUC_0$).
        Hence $\operatorname{supp}(\hat{f}_{t-1}) \subseteq [\hat{n}_{t-1}]$.
        By \cref{lem:cal_min:oga}, $\langle Z_{t-1}, \bar{\d}_t(k) \rangle = \Ex[f \sim F_t]{\langle Z_{t-1}, \d(f, k) \rangle} \le 1$ for every $k \in [\hat{n}_{t-1}]$, and taking the convex combination over $k \sim \hat{f}_{t-1}$ proves the claim.
    \end{proof}

    This concludes the proof of \cref{lem:cal_min:stochastic_norm} via self-bounding inequality.
    Let $B \coloneq \sqrt{\E\qty[\max_{t \le T} \norm{Z_t}^2]}$.
    Summing \cref{eq:cal_min:potential_step} over active rounds $t' \in \mathcal{T}_t$, then taking a maximum over $t \in [T]$, then expectation and then using \cref{cl:cal_min:first_term}, \cref{cl:cal_min:second_term}, and \cref{cl:cal_min:third_term} into \cref{eq:cal_min:drift_decomposition} we get
    \begin{equation*}
        B^2
        =
        \E\qty[\max_{t \le T} \norm{Z_t}^2]
        \le
        4 \sqrt{5 T} B
        +
        8 \sqrt{S T} B
        +
        2 T
        +
        5 T
        \le
        17 \sqrt{S T} B
        +
        7 S T
    \end{equation*}
    Solving this quadratic inequality gives $B \le \frac{17 + \sqrt{317}}{2} \sqrt{S T} \le 18\sqrt{S T}$.
    The proof is concluded by Jensen's inequality: $\E\qty[\max_{t \le T} \norm{Z_t}] \le B \le 18\sqrt{S T}$.
\end{proof}

\subsection{Putting Everything Together: Main Guarantees}
\label{ssec:cal_min:main_guarantees}

Combining \cref{lem:cal_min:adversarial,lem:cal_min:stochastic_norm} with \cref{thm:resp_cal:manipulability}, \cref{lem:lagr_cal:null_rounds}, and \cref{thm:lagr_cal:stochastic_regret} yields our main theorem.

\begin{theorem} \label{thm:cal_min:main}
    Running \cref{alg:learner} with the forecaster of \cref{alg:forecaster} runs in time $\operatorname{poly}(T, \nL)$, never runs out of budget, and it guarantees:
    \begin{enumerate}

        \item \textbf{Non-manipulability:} Against any adaptive optimizer,
        \begin{equation*}
            \E\qty[ \sum_{t=1}^T \UO[i_t, j_t] ]
            \le
            \BSE \cdot T
            +
            \order{ \qty( \sqrt{\nO}\nL + \frac{\sqrt{\nL}}{\rL} ) \qty(\sqrt{T} + \hatnO) }
            .
        \end{equation*}

        \item \textbf{Low interval stochastic regret:} If the horizon $[T]$ is partitioned into $S \ge 1$ contiguous intervals $\mathcal{I}_1, \dots, \mathcal{I}_S$ in which the optimizer draws $i_t \sim x^{(s)}$ independently in each round $t \in \mathcal{I}_s$ for some $x^{(s)} \in \Delta_{\nO}$, then in every interval $\mathcal{I}_s$ and for $y_s^\star \in \BRL(x^{(s)})$,
        \begin{equation*}
            \E\qty[ \sum_{t \in \mathcal{I}_s} \UL[i_t, j_t] ]
            \ge
            |\mathcal{I}_s| \cdot \qty(x^{(s)})^\top \UL y_s^\star
            -
            \order{ \frac{\sqrt{S \nL T}}{\rL} }
            .
        \end{equation*}
    \end{enumerate}
\end{theorem}

\begin{proof}
    In any active round $t \in \mathcal{T}$, the historical spending up to round $t - 1$ is at most $t \rL - 1$ and the round-$t$ cost is at most $1$, so total spending up to round $t$ is at most $t \rL$; in non-active rounds $t \notin \mathcal{T}$, the null action has cost $0$, preserving the budget bound by induction.

    Against an arbitrary adaptive optimizer, \cref{lem:cal_min:adversarial} guarantees $\norm{R_T}_{L^2} \le \sqrt{7T} + \sqrt{5}\hatnO$, along with $\max_{t \le T} L_t^{cy} \le \max_{t \le T} \norm{L_t}_2 \le \sqrt{7T} + \sqrt{5}\hatnO$ on every path.
    Plugging this into \cref{lem:lagr_cal:null_rounds} bounds the expected number of non-active rounds by
    \begin{equation*}
        \E\qty\big[ T - |\mathcal{T}| ]
        \le
        \frac{2\sqrt{\nL}\qty(\sqrt{7T} + \sqrt{5}\hatnO) + 2\sqrt{T} + 1}{\rL}
        =
        \order{ \frac{1}{\rL} \sqrt{\nL} \qty( \sqrt{T} + \hatnO ) }
        .
    \end{equation*}
    
    Since the optimizer's utility in each non-active round is at most $1$ and $\BSE \cdot |\mathcal{T}| \le \BSE \cdot T$, combining \cref{thm:resp_cal:manipulability} with $\norm{R_T}_{L^2} \le \sqrt{7T} + \sqrt{5}\hatnO$ and the bound on $\E\qty\big[T - |\mathcal{T}|]$ establishes the non-manipulability guarantee.

    In the piecewise-stationary stochastic setting, \cref{lem:cal_min:stochastic_norm} gives $\E\qty[\max_{t \le T} \norm{L_t}_2] \le \E\qty[\max_{t \le T} \norm{Z_t}] \le 18\sqrt{S T}$, independent of $\hatnO$.
    Substituting this bound directly into \cref{thm:lagr_cal:stochastic_regret} for each interval $\mathcal{I}_s$ gives the interval stochastic regret guarantee.

    Finally, for the running time, since $\hat{n}_{t-1} \le t \le T$, \cref{alg:forecaster} performs $M_t = 50 \hat{n}_{t-1} T^2 \le 50 T^3$ iterations per active round.
    In each iteration $m$, finding $y\qty(f^{(m)})$ and $\l\qty(f^{(m)})$ solves a linear program in $\nL$ variables, and computing $z\qty(f^{(m)})$ evaluates finite step-function inner products via \eqref{eq:algo:inner_norm_step} together with a Euclidean inner product in $\R^{2\nL + 2}$, taking $\operatorname{poly}(T, \nL)$ time in total (independent of $\nO$ and $\hatnO$).
\end{proof}

\section{Budgeted Optimizer and the Value the Optimizer Can Get} 
\label{sec:budgeted_optimizer}

In this section we first show how our non-manipulability results extend to the case when the optimizer is also budgeted.
In \cref{ssec:budgeted_optimizer:bse,ssec:budgeted_optimizer:wbse} we will define the strong and weak versions of BSE, and show that the optimizer can achieve the value of the weak BSE against any learner with low interval stochastic regret.
In \cref{ssec:budgeted_optimizer:generic} we show that when utility and cost values are generic the different definitions of BSE are equal.

Recall from \cref{sec:defs} that the goal of the optimizer is to maximize their utility subject to keeping their cost bounded, $\rO T$ for some $\rO \in (0, 1]$.
We assume that the optimizer also has a null action that guarantees $0$ cost no matter what the learner is doing.

\subsection{Budgeted Stackelberg Equilibrium for a Budgeted Optimizer}
\label{ssec:budgeted_optimizer:bse}

In \cref{opt:defs:bse} we defined \textit{Budgeted Stackelberg Equilibrium} (BSE) for the case when the optimizer is unbudgeted.
In \cite{FST26a} we introduced the \textit{Budgeted Stackelberg Equilibrium} (BSE) more generally. When the optimizer is also budgeted, it is important to note that the utility of the optimizer from the best response of the learner is not a concave function of their cost. As a result, a strategic optimizer may want to use multiple strategies, maybe overspending for a while, and then underspending, effectively ironing their utility function. To define this formally, we use the same best response function as was used in \cref{eq:best_response}.

As shown in \cite{FST26a} the optimizer mixing two distributions is enough to achieve their maximum value, i.e., three or more distributions does not help.
Allowing the optimizer to use two different action distributions in two distinct phases, makes the BSE here equal to
\begin{optimization} \label{opt:app:budgeted_optimizers:bse}
    \BSE =
    \max_{x_1, x_2 \in \Delta_{\nO}, \; y_1, y_2 \in \Delta_{\nL}, \; q \in [0, 1]}
    \quad &
    q \cdot x_1^\top \UO y_1
    +
    (1 - q) \cdot x_2^\top \UO y_2
    \\
    \textrm{\textup{ such that }} \;\;\quad\qquad &
    q \cdot x_1^\top \CO y_1
    +
    (1 - q) \cdot x_2^\top \CO y_2
    \le \rO
    \\
    \textrm{\textup{ and }} \;\;\qquad\qquad &
    y_1 \in \BRL(x_1), \; y_2 \in \BRL(x_2)
\end{optimization}

\subsection{Relaxing the Optimizer to an Unbudgeted Optimizer}

Here we see how we can replace the budget-constraint optimizer with an optimizer who does not have a budget constraint, and extend our results from \cref{ssec:cal_min:main_guarantees} that upper bound the optimizer's utility to budgeted optimizers also.

Consider the Lagrangian relaxation of the optimizer who is a pure utility maximizer and who receives utility $\UO[i, j] - \mu \CO[i, j] + \mu \rO$ when the pure actions are $i, j$, for a Lagrange multiplier $\mu \ge 0$.
The BSE of this optimizer is
\begin{optimization} \label{opt:app:budgeted_optimizers:bse_lagrange}
    \BSE_\mu =
    \max_{x \in \Delta_{\nO}, \; y \in \Delta_{\nL}}
    \quad &
    x^\top \UO y - \mu x^\top \CO y + \mu \rO
    \\
    \textrm{\textup{ such that }} \quad &
    y \in \BRL(x)
\end{optimization}

We notice that for the right $\mu$, the two BSEs are equal.

\begin{lemma} \label{lem:app:budgeted_optimizers:lagrangian_equiv}
    It holds $\BSE = \min_{\mu \ge 0} \BSE_\mu$.
    In addition, for the optimal value it holds $\mu^\star \le 1/\rO$.
\end{lemma}

\begin{proof}
    Let $\mathcal{S} = \{ (x^\top \UO y, x^\top \CO y) : x \in \Delta_{\nO}, \, y \in \BRL(x) \} \subseteq [0, 1]^2$ be the set of achievable cost-utility pairs for the optimizer under a single mixed strategy, and let $\bar{\mathcal S} = \operatorname{conv}(\mathcal{S})$ denote its convex hull.
    We can re-write \cref{opt:app:budgeted_optimizers:bse} as
    \begin{equation*}
        \BSE = \max_{(u, c) \in \bar{\mathcal S} \,:\, c \le \rO} u.
    \end{equation*}
    since any extreme point of $\bar{\mathcal S}$ can be written as a convex combination of two points in $\mathcal S$ (see \cite{FST26a} for a more detailed proof).
    
    The optimizer's null action yields a point $(u_0, 0) \in \mathcal{S}$ with cost $c = 0 < \rO$ and utility $u_0 \ge 0$.
    This makes Slater's condition true; by strong Lagrangian duality for the constraint $c \le \rO$ with multiplier $\mu \ge 0$, the dual minimum is attained and
    \begin{equation*}
        \BSE
        = \max_{(u, c) \in \bar{\mathcal S}} \min_{\mu \ge 0} \big( u - \mu c + \mu \rO \big)
        = \min_{\mu \ge 0} \max_{(u, c) \in \bar{\mathcal S}} \big( u - \mu c + \mu \rO \big)
        = \min_{\mu \ge 0} \max_{(u, c) \in \mathcal{S}} \big( u - \mu c + \mu \rO \big)
        = \min_{\mu \ge 0} \BSE_\mu,
    \end{equation*}
    where the third equality holds because the linear in $(u, c)$ objective $u - \mu c + \mu \rO$ achieves its supremum over $\bar{\mathcal S}$ on $\mathcal{S}$ itself.
    Finally, evaluating $\BSE_\mu$ at $(u_0, 0) \in \mathcal{S}$ gives $\BSE_\mu \ge u_0 + \mu \rO \ge \mu \rO$; since $\BSE \le 1$, any minimizer $\mu^\star$ satisfies $\mu^\star \le \BSE / \rO \le 1 / \rO$.
\end{proof}

This allows us to focus on the Lagrangian optimizer instead of the budgeted one for utility upper bounds.
Specifically, the budgeted optimizer's benefit over its BSE is bounded by the benefit of the Lagrangian optimizer's over its BSE.
Consider any sequence of pairs of actions $(i_1, j_1), \ldots, (i_T, j_T)$ such that the optimizer's budget is observed.
Then,
\begin{equation*}
    \sum_{t = 1}^T \UO[i_t, j_t] - T \cdot \BSE
    \le
    \sum_{t = 1}^T \qty( \UO[i_t, j_t]  - \mu^\star \CO[i_t, j_t] + \mu^\star \rO ) - T \cdot \BSE_{\mu^\star}
    .
\end{equation*}
Note that the weaker version $\BSE \le \min_{\mu \ge 0} \BSE_\mu$ that can be trivially proved via weak duality is not enough to prove the above inequality.
This allows us to extend our main result to the case when the optimizer is also budgeted, with an increase  in the error bound since the Lagrangian optimizer's rewards have range $O(1/\rO)$ instead of $1$:

\begin{theorem} \label{thm:cal_min:main_budgeted}
    Running \cref{alg:learner} with the forecaster of \cref{alg:forecaster} ensures the guarantees of \cref{thm:cal_min:main} also against a budgeted optimizer with an additional factor of $1/\rO$ in the error term of the non-manipulability guarantee.
\end{theorem}

\subsection{Achieving the Weak Budgeted Stackelberg Equilibrium}
\label{ssec:budgeted_optimizer:wbse}

We now show that against any learning algorithm with low per-interval stochastic regret and low per-interval overspending (which \cref{alg:learner,alg:forecaster} satisfy via \cref{lem:lagr_cal:null_rounds,thm:lagr_cal:stochastic_regret,thm:cal_min:main}), these two generic properties are sufficient to guarantee that the optimizer can always achieve the value of the \textit{Weak Budgeted Stackelberg Equilibrium} ($\WBSE$), defined by replacing optimistic learner tie-breaking in \eqref{opt:app:budgeted_optimizers:bse} with pessimistic tie-breaking over $\BRL(x)$:
\begin{optimization} \label{opt:app:budgeted_optimizers:wbse}
    \WBSE =
    \sup_{x_1, x_2 \in \Delta_{\nO}, \; q \in [0, 1]}
    \quad &
    q \min_{y_1 \in \BRL(x_1)} x_1^\top \UO y_1
    +
    (1 - q) \min_{y_2 \in \BRL(x_2)} x_2^\top \UO y_2
    \\
    \textrm{\textup{ such that }} \;\qquad &
    q \max_{y_1 \in \BRL(x_1)} x_1^\top \CO y_1
    +
    (1 - q) \max_{y_2 \in \BRL(x_2)} x_2^\top \CO y_2
    \le \rO
    .
\end{optimization}

When the optimizer is unbudgeted ($\rO = 1$), the cost constraint is vacuous and  it is not useful for the optimizer to have two phases, so \eqref{opt:app:budgeted_optimizers:wbse} reduces to $\WBSE = \sup_{x \in \Delta_{\nO}} \min_{y \in \BRL(x)} x^\top \UO y$.
Whenever the learner's best response $\BRL(x)$ is unique at the optimal strategies, $\WBSE$ coincides with $\BSE$.

We first show that in any contiguous interval $\mathcal{I} \subseteq [T]$ where the optimizer commits to a fixed mixed strategy $x \in \Delta_{\nO}$, low stochastic regret and low overspending on $\mathcal{I}$ force the learner's expected empirical play to lie close to the best-response polytope $\BRL(x)$, simultaneously lower-bounding the optimizer's utility and upper-bounding the optimizer's cost on $\mathcal{I}$.
For a fixed $x \in \Delta_{\nO}$, let $\mathcal{F}(x) \coloneq \{y \in \Delta_{\nL} : x^\top \CL y \le \rL\}$ denote the learner's budget-feasible polytope, $V_{\mathtt{L}}(x) \coloneq \max_{y \in \mathcal{F}(x)} x^\top \UL y$ the learner's optimal utility, $\mathcal{V}(x)$ the finite set of vertices of $\mathcal{F}(x)$, and $\mathcal{V}^{\mathrm{sub}}(x) \coloneq \mathcal{V}(x) \setminus \BRL(x)$ the suboptimal vertices of $\mathcal{F}(x)$.
We denote the vertex suboptimality gap of $\mathcal{F}(x)$ by
\begin{equation} \label{eq:app:budgeted_optimizer:vertex_gap}
    \Delta(x)
    \coloneq
    \begin{cases}
        \min_{v \in \mathcal{V}^{\mathrm{sub}}(x)} \qty\big( V_{\mathtt{L}}(x) - x^\top \UL v ) > 0 & \text{if } \mathcal{V}^{\mathrm{sub}}(x) \ne \emptyset, \\
        1 & \text{if } \mathcal{V}^{\mathrm{sub}}(x) = \emptyset.
    \end{cases}
\end{equation}

\begin{lemma} \label{lem:app:budgeted_optimizers:interval_br}
    Let $\mathcal{I} = \{t_1 + 1, \dots, t_2\} \subseteq [T]$ be any contiguous interval in which the optimizer draws $i_t \sim x$ independently in each round $t \in \mathcal{I}$ for some $x \in \Delta_{\nO}$.
    Then against any learning algorithm that on $\mathcal{I}$ has expected stochastic regret $|\mathcal{I}| V_{\mathtt{L}}(x) - \Ex{\sum_{t \in \mathcal{I}} \UL[i_t, j_t]} \le \mathrm{Reg}_{\mathcal{I}}$ and expected overspending $\Ex{\sum_{t \in \mathcal{I}} \CL[i_t, j_t]} - \rL |\mathcal{I}| \le \mathrm{Over}_{\mathcal{I}}$ for some $\mathrm{Reg}_{\mathcal{I}}, \mathrm{Over}_{\mathcal{I}} \ge 0$, the optimizer's expected utility and cost in $\mathcal{I}$ satisfy
    \begin{align*}
        \Ex{\sum_{t \in \mathcal{I}} \UO[i_t, j_t]}
        &\ge
        |\mathcal{I}| \min_{y \in \BRL(x)} x^\top \UO y
        -
        \frac{\mathrm{Reg}_{\mathcal{I}} + 2\,\mathrm{Over}_{\mathcal{I}} / \rL}{\Delta(x)}
        , \\
        \Ex{\sum_{t \in \mathcal{I}} \CO[i_t, j_t]}
        &\le
        |\mathcal{I}| \max_{y \in \BRL(x)} x^\top \CO y
        +
        \frac{\mathrm{Reg}_{\mathcal{I}} + 2\,\mathrm{Over}_{\mathcal{I}} / \rL}{\Delta(x)}
        .
    \end{align*}
\end{lemma}

\begin{proof}
    Let $\bar{y} \coloneq \frac{1}{|\mathcal{I}|} \sum_{t \in \mathcal{I}} \Ex{e_{j_t}} \in \Delta_{\nL}$ be the learner's expected empirical strategy on $\mathcal{I}$.
    Since $i_t \sim x$ is drawn independently of $j_t$ conditioned on the history prior to round $t$, linearity of expectation gives $\Ex{\sum_{t \in \mathcal{I}} M[i_t, j_t]} = |\mathcal{I}| \, x^\top M \bar{y}$ for every matrix $M \in \{\UO, \CO, \UL, \CL\}$.

    We now project $\bar{y}$ onto the feasible polytope $\mathcal{F}(x)$.
    Let $\delta \coloneq \max\qty{0,\, x^\top \CL \bar{y} - \rL}$ and $\gamma \coloneq \frac{\delta}{\rL + \delta} \le \frac{\delta}{\rL}$.
    Since the learner's null action $j_0$ satisfies $x^\top \CL e_{j_0} = 0$, the convex combination $\tilde{y} \coloneq (1 - \gamma) \bar{y} + \gamma e_{j_0} \in \Delta_{\nL}$ satisfies $x^\top \CL \tilde{y} = (1 - \gamma) x^\top \CL \bar{y} \le (1 - \gamma)(\rL + \delta) = \rL$, so $\tilde{y} \in \mathcal{F}(x)$.
    Moreover, since $\UL \in [0, 1]^{\nO \times \nL}$, we have $x^\top \UL \tilde{y} \ge (1 - \gamma) x^\top \UL \bar{y} \ge x^\top \UL \bar{y} - \gamma$.

    Because $\mathcal{F}(x) = \operatorname{conv}(\mathcal{V}(x))$ and $\BRL(x) = \operatorname{conv}(\mathcal{V}(x) \setminus \mathcal{V}^{\mathrm{sub}}(x))$, we can decompose $\tilde{y} = (1 - \beta) y^\star + \beta y^{\mathrm{sub}}$ for some $\beta \in [0, 1]$, $y^\star \in \BRL(x)$, and $y^{\mathrm{sub}} \in \operatorname{conv}(\mathcal{V}^{\mathrm{sub}}(x))$ (with $\beta = 0$ if $\mathcal{V}^{\mathrm{sub}}(x) = \emptyset$).
    By definition of $\Delta(x)$ in \eqref{eq:app:budgeted_optimizer:vertex_gap}, every $y^{\mathrm{sub}} \in \operatorname{conv}(\mathcal{V}^{\mathrm{sub}}(x))$ satisfies $x^\top \UL y^{\mathrm{sub}} \le V_{\mathtt{L}}(x) - \Delta(x)$, and hence
    \begin{equation*}
        V_{\mathtt{L}}(x) - \beta \Delta(x)
        \ge
        (1 - \beta) x^\top \UL y^\star + \beta x^\top \UL y^{\mathrm{sub}}
        =
        x^\top \UL \tilde{y}
        \ge
        x^\top \UL \bar{y} - \gamma
        .
    \end{equation*}

    Rearranging gives $\beta \le \frac{1}{\Delta(x)}\qty\big(V_{\mathtt{L}}(x) - x^\top \UL \bar{y} + \gamma)$.
    For any matrix $M \in [0, 1]^{\nO \times \nL}$, since $x^\top M (\bar{y} - \tilde{y}) = \gamma x^\top M (\bar{y} - e_{j_0}) \in [-\gamma, \gamma]$ and $x^\top M (\tilde{y} - y^\star) = \beta x^\top M (y^{\mathrm{sub}} - y^\star) \in [-\beta, \beta]$, we obtain
    \begin{equation*}
        |\mathcal{I}| \, \abs{ x^\top M \bar{y} - x^\top M y^\star }
        \le
        |\mathcal{I}| (\gamma + \beta)
        \le
        \frac{|\mathcal{I}|}{\Delta(x)} \qty\big( V_{\mathtt{L}}(x) - x^\top \UL \bar{y} + 2\gamma )
        \le
        \frac{\mathrm{Reg}_{\mathcal{I}} + 2\,\mathrm{Over}_{\mathcal{I}} / \rL}{\Delta(x)}
        ,
    \end{equation*}
    $|\mathcal{I}|(V_{\mathtt{L}}(x) - x^\top \UL \bar{y}) \le \mathrm{Reg}_{\mathcal{I}}$ and $|\mathcal{I}|\gamma \le |\mathcal{I}|\delta / \rL \le \mathrm{Over}_{\mathcal{I}} / \rL$.
    Taking $M = \UO$ and $M = \CO$ and using $y^\star \in \BRL(x)$ completes the proof.
\end{proof}

Applying \cref{lem:app:budgeted_optimizers:interval_br} across $S \le 2$ intervals shows that a budgeted optimizer can achieve $\WBSE \cdot T - o(T)$ against any such learner (and in particular against \cref{alg:learner,alg:forecaster}).

\begin{theorem} \label{thm:app:budgeted_optimizers:wbse}
    Fix any feasible solution $(x_1, x_2, q)$ to \eqref{opt:app:budgeted_optimizers:wbse} and let $\Delta \coloneq \min\{\Delta(x_1), \Delta(x_2)\} > 0$.
    If the optimizer plays $i_t \sim x_1$ for the first $T_1 = \floor{q T}$ rounds ($\mathcal{I}_1 = \{1, \dots, T_1\}$) and $i_t \sim x_2$ for the remaining $T_2 = T - T_1$ rounds ($\mathcal{I}_2 = \{T_1 + 1, \dots, T\}$) against 
    any learning algorithm with stochastic regret $\mathrm{Reg}_{\mathcal{I}_s}$ and overspending $\mathrm{Over}_{\mathcal{I}_s}$ on each phase $s \in \{1, 2\}$ (letting $\mathrm{Reg} \coloneq \mathrm{Reg}_{\mathcal{I}_1} + \mathrm{Reg}_{\mathcal{I}_2}$ and $\mathrm{Over} \coloneq \mathrm{Over}_{\mathcal{I}_1} + \mathrm{Over}_{\mathcal{I}_2}$), then
    \begin{align*}
        \Ex{\sum_{t=1}^T \UO[i_t, j_t]}
        &\ge
        \qty( q \min_{y_1 \in \BRL(x_1)} x_1^\top \UO y_1 + (1 - q) \min_{y_2 \in \BRL(x_2)} x_2^\top \UO y_2 ) T
        -
        \frac{\mathrm{Reg} + 2\,\mathrm{Over} / \rL}{\Delta} - 1
        , \\
        \Ex{\sum_{t=1}^T \CO[i_t, j_t]}
        &\le
        \rO T
        +
        \frac{\mathrm{Reg} + 2\,\mathrm{Over} / \rL}{\Delta} + 1
        .
    \end{align*}
    
    Consequently, whenever $\mathrm{Reg}, \mathrm{Over} = o(T)$, by taking $(x_1, x_2, q)$ approaching the supremum in \eqref{opt:app:budgeted_optimizers:wbse} (and shortening the active phases by $\order*{\frac{\mathrm{Reg} + \mathrm{Over}/\rL + 1}{\rO \Delta}}$ rounds of the optimizer's null action), the optimizer achieves expected utility at least $\WBSE \cdot T - o(T)$.
\end{theorem}

\begin{proof}
    Applying \cref{lem:app:budgeted_optimizers:interval_br} to $\mathcal{I}_1$ and $\mathcal{I}_2$, summing the two bounds, and noting that $\abs{T_1 - qT} \le 1$ and $\abs{T_2 - (1-q)T} \le 1$ immediately yields the stated utility and cost guarantees.
\end{proof}

\subsection{Equivalence of \texorpdfstring{$\WBSE$}{WBSE} and \texorpdfstring{$\BSE$}{BSE} for Generic Games}
\label{ssec:budgeted_optimizer:generic}

Finally, we show that when the learner's utilities and costs are generic, i.e., for all but a measure 0 subset of possible cost and utility values, the learner's best response $\BRL(x)$ is unique for any fixed (or generic) optimizer strategy $x \in \Delta_{\nO}$ even when the budget constraint forces the learner to randomize. Moreover, for any $x \in \Delta_{\nO}$, every optimal best-response vertex of the learner can be uniquely induced by an arbitrarily small perturbation of $x$.
As a consequence, the weak and strong Budgeted Stackelberg Equilibrium values coincide ($\WBSE = \BSE$).

\begin{theorem} \label{thm:app:budgeted_optimizers:wbse_eq_bse}
    For generic learner utilities and costs $(\UL, \CL)$, it holds that $\WBSE = \BSE$.
\end{theorem}
Here we outline the idea of the proof; full details are deferred to \cref{sec:app:WBSE=BSE}.
For a given $x \in \Delta_{\nO}$, associate each pure action $j \in [\nL]$ of the learner with its cost-utility pair $p_j(x) \coloneq ((x^\top \CL)_j, (x^\top \UL)_j) \in [0, 1]^2$.
By linear programming duality (which holds strongly since $p_{j_0}(x) = (0, 0)$ and $\rL > 0$), we have
\begin{equation*}
    V_{\mathtt{L}}(x) = \min_{\lambda \ge 0} \max_{j \in [\nL]} \qty\big( (x^\top \UL)_j - \lambda \qty((x^\top \CL)_j - \rL) ),
\end{equation*}
and any $y \in \BRL(x)$ is supported on the set of tight pure actions $S(x, \lambda) \coloneq \arg\max_{j \in [\nL]} \qty\big((x^\top \UL)_j - \lambda (x^\top \CL)_j)$ for an optimal dual multiplier $\lambda \ge 0$, with $x^\top \CL y = \rL$ whenever $\lambda > 0$.
When the best response is a pure action ($\lambda = 0$), two different best responses $j_1$ and $j_2$ must satisfy $(x^\top \UL)_{j_1} = (x^\top \UL)_{j_2}$, which occurs with $0$ probability for a fixed or generic $x$.
When the budget constraint is binding ($\lambda > 0$), any two actions $j_1, j_2 \in S(x, \lambda)$ with $(x^\top \CL)_{j_1} < \rL < (x^\top \CL)_{j_2}$ determine a unique budget-feasible mixture $q e_{j_1} + (1 - q) e_{j_2}$ with $q (x^\top \CL)_{j_1} + (1 - q)(x^\top \CL)_{j_2} = \rL$; thus, unless an action has cost $(x^\top \CL)_j = \rL$, non-uniqueness implies that there are at least three actions $j_1, j_2, j_3 \in S(x, \lambda)$ lying on the same supporting line of slope $\lambda$ in the cost-utility plane, meaning that $p_{j_1}(x), p_{j_2}(x), p_{j_3}(x)$ are collinear and the following determinant is $0$:
\begin{equation*}
    \det \begin{pmatrix}
        1 & (x^\top \CL)_{j_1} & (x^\top \UL)_{j_1} \\
        1 & (x^\top \CL)_{j_2} & (x^\top \UL)_{j_2} \\
        1 & (x^\top \CL)_{j_3} & (x^\top \UL)_{j_3}
    \end{pmatrix} = 0.
\end{equation*}

This defines a non-trivial polynomial equation in the entries of $(\UL, \CL)$, and hence a hypersurface of Lebesgue measure zero.

To show the equivalence of $\WBSE$ and $\BSE$, we also need to consider the case where the optimizer chooses $x \in \Delta_{\nO}$ on a boundary between best-response regions where the learner is indifferent between multiple vertices of the budget-feasible polytope $\mathcal{F}(x)$.
We show that for generic $(\UL, \CL)$, any optimal vertex $v$ of the learner's best-response set can be selected as the unique best response by an arbitrarily small perturbation of $x$.
For a given $x \in \Delta_{\nO}$, let its support be $S_x = \operatorname{supp}(x) \subseteq [\nO]$ of size $s = |S_x| \ge 1$, so that $x$ lies in the $(s-1)$-dimensional open simplex $\Delta_{S_x}^{\circ} \coloneq \{x' \in \Delta_{\nO} : \operatorname{supp}(x') = S_x\}$ with tangent space $T_{S_x} \coloneq \{d \in \mathbb{R}^{\nO} : \operatorname{supp}(d) \subseteq S_x, \, \mathbf{1}^\top d = 0\}$.
Because $x \in \Delta_{S_x}^{\circ}$ has $s - 1$ degrees of freedom, for generic $(\UL, \CL)$ at most $s + 1$ actions can simultaneously lie on the supporting line (or at most $s$ can tie in utility), and in \cref{sec:app:WBSE=BSE} we show that the resulting tie-breaking conditions are linearly independent, allowing any optimal vertex $v$ to be made the unique best response by a small perturbation in a direction $d \in T_{S_x}$.
The theorem then follows by taking the limit as the perturbation goes to $0$.

\section*{AI Disclosure}

The problem formulation, equilibrium concepts, and desiderata for the learning algorithm were developed entirely by the authors.
ChatGPT Sol 5.6 and Astra 6 were helpful in suggesting the initial approach behind the forecasting algorithm and calibration errors.
Both models and Gemini 4 Argon were used interactively as technical sounding boards while refining these ideas and checking auxiliary questions.

Reaching the paper's algorithm, guarantees, and proofs required substantial redesign and technical work by the authors.
Earlier AI-generated versions contained errors or relied on assumptions that would violate standard online-learning models.
For example, they assumed that the learner knows the game matrices or $\nO$ in advance, sees the optimizer's action before choosing its own action, or observes the optimizer's realized action or mixed strategy after each round.
While correct, initial proofs were also heavily over-engineered, relying on optimal transport, Sobolev norm spaces, resetting calibration errors when observing a new environment, and blocks of forced null actions with suboptimal $1/\rL^2$ dependence.
Identifying the core mechanisms, stripping away this machinery, and developing the full scope of results (including interval stochastic regret) required extensive iteration and analysis by the authors.

Furthermore, even when technical details were present, AI-generated write-ups bundled all components and were inaccessible even to an expert, requiring significant effort to decipher and verify.
The authors therefore designed the paper's modular structure, decoupling the calibration notions, and wrote almost all of the text and proofs from scratch.
Gemini 4 Argon was also used to check related-work references and to audit the human-written proofs for typos and minor inaccuracies.

\printbibliography{}

@inproceedings{ACMMSS25,
  author    = {Eshwar Ram Arunachaleswaran and
               Natalie Collina and
               Yishay Mansour and
               Mehryar Mohri and
               Jon Schneider and
               Balasubramanian Sivan},
  title     = {Swap Regret and Correlated Equilibria Beyond Normal-Form Games},
  booktitle = {Proceedings of the 26th {ACM} Conference on Economics and Computation,
               {EC} 2025, Stanford University, Stanford, CA, USA, July 7-10, 2025},
  pages     = {130--157},
  publisher = {{ACM}},
  year      = {2025}
}

@inproceedings{ACS24,
  author    = {Eshwar Ram Arunachaleswaran and
               Natalie Collina and
               Jon Schneider},
  title     = {Pareto-Optimal Algorithms for Learning in Games},
  booktitle = {Proceedings of the 25th {ACM} Conference on Economics and Computation,
               {EC} 2024, New Haven, CT, USA, July 8-11, 2024},
  pages     = {490--510},
  publisher = {{ACM}},
  year      = {2024}
}

@inproceedings{AD16,
  author    = {Shipra Agrawal and
               Nikhil R. Devanur},
  title     = {Linear Contextual Bandits with Knapsacks},
  booktitle = {Advances in Neural Information Processing Systems 29: Annual Conference on Neural Information Processing Systems 2016, December 5-10, 2016,
               Barcelona, Spain},
  pages     = {3450--3458},
  year      = {2016}
}

@inproceedings{AFZ25,
  author    = {Gagan Aggarwal and
               Giannis Fikioris and
               Mingfei Zhao},
  title     = {No-Regret Algorithms in non-Truthful Auctions with Budget and {ROI}
               Constraints},
  booktitle = {Proceedings of the {ACM} on Web Conference 2025, {WWW} 2025, Sydney,
               NSW, Australia, 28 April 2025- 2 May 2025},
  pages     = {1398--1415},
  publisher = {{ACM}},
  year      = {2025}
}

@article{AggarwalBBBDFGLLMMMMLPP24,
  author  = {Gagan Aggarwal and
             Ashwinkumar Badanidiyuru and
             Santiago R. Balseiro and
             Kshipra Bhawalkar and
             Yuan Deng and
             Zhe Feng and
             Gagan Goel and
             Christopher Liaw and
             Haihao Lu and
             Mohammad Mahdian and
             Jieming Mao and
             Aranyak Mehta and
             Vahab Mirrokni and
             {Paes Leme}, Renato and
             Andr{\'{e}}s Perlroth and
             Georgios Piliouras and
             Jon Schneider and
             Ariel Schvartzman and
             Balasubramanian Sivan and
             Kelly Spendlove and
             Yifeng Teng and
             Di Wang and
             Hanrui Zhang and
             Mingfei Zhao and
             Wennan Zhu and
             Song Zuo},
  title   = {Auto-Bidding and Auctions in Online Advertising: {A} Survey},
  journal = {SIGecom Exch.},
  volume  = {22},
  number  = {1},
  pages   = {159--183},
  year    = {2024}
}

@inproceedings{anagnostides-et-al-2026,
  author    = {Ioannis Anagnostides and
               Gabriele Farina and
               Maxwell Fishelson and
               Haipeng Luo and
               Jon Schneider},
  title     = {Swap Regret Minimization through Response-Based Approachability},
  booktitle = {Proceedings of the Thirty-Ninth Conference on Learning Theory},
  series    = {Proceedings of Machine Learning Research},
  volume    = {336},
  pages     = {195--223},
  year      = {2026}
}

@inproceedings{BBFLMSW24,
  author    = {Santiago R. Balseiro and
               Kshipra Bhawalkar and
               Zhe Feng and
               Haihao Lu and
               Vahab Mirrokni and
               Balasubramanian Sivan and
               Di Wang},
  title     = {A Field Guide for Pacing Budget and {ROS} Constraints},
  booktitle = {Forty-first International Conference on Machine Learning, {ICML} 2024,
               Vienna, Austria, July 21-27, 2024},
  series    = {Proceedings of Machine Learning Research},
  pages     = {2607--2638},
  publisher = {{PMLR} / OpenReview.net},
  year      = {2024},
  url       = {https://proceedings.mlr.press/v235/balseiro24a.html}
}

@article{BBW15,
  author  = {Santiago R. Balseiro and
             Omar Besbes and
             Gabriel Y. Weintraub},
  title   = {Repeated Auctions with Budgets in Ad Exchanges: Approximations and
             Design},
  journal = {Manag. Sci.},
  volume  = {61},
  number  = {4},
  pages   = {864--884},
  year    = {2015}
}

@inproceedings{BCCF24,
  author    = {Martino Bernasconi and
               Matteo Castiglioni and
               Andrea Celli and
               Federico Fusco},
  title     = {Bandits with Replenishable Knapsacks: the Best of both Worlds},
  booktitle = {The Twelfth International Conference on Learning Representations,
               {ICLR} 2024, Vienna, Austria, May 7-11, 2024},
  publisher = {OpenReview.net},
  year      = {2024}
}

@article{bernstein-shimkin-2015,
  author  = {Andrey Bernstein and Nahum Shimkin},
  title   = {Response-Based Approachability with Applications to Generalized
             No-Regret Problems},
  journal = {Journal of Machine Learning Research},
  volume  = {16},
  number  = {24},
  pages   = {747--773},
  year    = {2015}
}

@article{BG19,
  author  = {Santiago R. Balseiro and
             Yonatan Gur},
  title   = {Learning in Repeated Auctions with Budgets: Regret Minimization and
             Equilibrium},
  journal = {Management Science},
  volume  = {65},
  number  = {9},
  pages   = {3952--3968},
  year    = {2019}
}

@article{BKK23,
  author  = {Santiago R. Balseiro and
             Christian Kroer and
             Rachitesh Kumar},
  title   = {Contextual Standard Auctions with Budgets: Revenue Equivalence and
             Efficiency Guarantees},
  journal = {Manag. Sci.},
  volume  = {69},
  number  = {11},
  pages   = {6837--6854},
  year    = {2023}
}

@inproceedings{BKMSW23,
  author    = {Santiago R. Balseiro and
               Rachitesh Kumar and
               Vahab Mirrokni and
               Balasubramanian Sivan and
               Di Wang},
  title     = {Robust Budget Pacing with a Single Sample},
  booktitle = {International Conference on Machine Learning, {ICML} 2023, 23-29 July
               2023, Honolulu, Hawaii, {USA}},
  series    = {Proceedings of Machine Learning Research},
  pages     = {1636--1659},
  publisher = {{PMLR}},
  year      = {2023},
  url       = {https://proceedings.mlr.press/v202/balseiro23a.html}
}

@article{BKS18,
  author  = {Ashwinkumar Badanidiyuru and
             Robert Kleinberg and
             Aleksandrs Slivkins},
  title   = {Bandits with Knapsacks},
  journal = {J. {ACM}},
  volume  = {65},
  number  = {3},
  pages   = {13:1--13:55},
  year    = {2018}
}

@article{Blackwell56,
  author  = {David Blackwell},
  title   = {An Analog of the Minimax Theorem for Vector Payoffs},
  journal = {Pacific Journal of Mathematics},
  volume  = {6},
  number  = {1},
  pages   = {1--8},
  year    = {1956},
}

@article{BLM23,
  author    = {Santiago R. Balseiro and
               Haihao Lu and
               Vahab Mirrokni},
  title     = {The Best of Many Worlds: Dual Mirror Descent for Online Allocation
               Problems},
  journal   = {Oper. Res.},
  volume    = {71},
  number    = {1},
  pages     = {101--119},
  year      = {2023},
  url       = {https://doi.org/10.1287/opre.2021.2242},
  doi       = {10.1287/OPRE.2021.2242},
  bibsource = {dblp computer science bibliography, https://dblp.org}
}

@article{BLMS23,
  author     = {Santiago R. Balseiro and
                Haihao Lu and
                Vahab S. Mirrokni and
                Balasubramanian Sivan},
  title      = {From Online Optimization to {PID} Controllers: Mirror Descent with
                Momentum},
  journal    = {CoRR},
  volume     = {abs/2202.06152},
  year       = {2022},
  url        = {https://arxiv.org/abs/2202.06152},
  eprinttype = {arXiv},
  eprint     = {2202.06152}
}

@article{BLMSX25,
  author     = {Mark Braverman and
                Jingyi Liu and
                Jieming Mao and
                Jon Schneider and
                Eric Xue},
  title      = {A New Benchmark for Online Learning with Budget-Balancing Constraints},
  journal    = {CoRR},
  volume     = {abs/2503.14796},
  year       = {2025},
  url        = {https://doi.org/10.48550/arXiv.2503.14796},
  doi        = {10.48550/ARXIV.2503.14796},
  eprinttype = {arXiv},
  eprint     = {2503.14796},
  bibsource  = {dblp computer science bibliography, https://dblp.org}
}

@inproceedings{HW24,
  author    = {Lunjia Hu and Yifan Wu},
  title     = {Predict to Minimize Swap Regret for All Payoff-Bounded Tasks},
  booktitle = {65th {IEEE} Annual Symposium on Foundations of Computer Science, {FOCS} 2024},
  pages     = {244--263},
  publisher = {{IEEE}},
  year      = {2024},
}

@inproceedings{BMSW18,
  author    = {Mark Braverman and
               Jieming Mao and
               Jon Schneider and
               S. Matthew Weinberg},
  title     = {Selling to a No-Regret Buyer},
  booktitle = {Proceedings of the 2018 {ACM} Conference on Economics and Computation,
               Ithaca, NY, USA, June 18-22, 2018},
  pages     = {523--538},
  publisher = {{ACM}},
  year      = {2018}
}

@inproceedings{BSV23,
  author    = {William Brown and
               Jon Schneider and
               Kiran Vodrahalli},
  title     = {Is Learning in Games Good for the Learners?},
  booktitle = {Advances in Neural Information Processing Systems 36: Annual Conference
               on Neural Information Processing Systems 2023, NeurIPS 2023, New Orleans,
               LA, USA, December 10 - 16, 2023},
  year      = {2023},
}

@inproceedings{cai2026onlinelinearoptimizationsufficient,
  title     = {Is Online Linear Optimization Sufficient for Strategic Robustness?},
  author    = {Yang Cai and Haipeng Luo and Chen-Yu Wei and Weiqiang Zheng},
  year      = {2026},
  booktitle = {Proceedings of the 27th {ACM} Conference on Economics and Computation,
               {EC} 2026},
  publisher = {{ACM}}
}

@inproceedings{CCK22,
  author    = {Matteo Castiglioni and
               Andrea Celli and
               Christian Kroer},
  title     = {Online Learning with Knapsacks: the Best of Both Worlds},
  booktitle = {International Conference on Machine Learning, {ICML} 2022, 17-23 July
               2022, Baltimore, Maryland, {USA}},
  series    = {Proceedings of Machine Learning Research},
  pages     = {2767--2783},
  publisher = {{PMLR}},
  year      = {2022}
}

@inproceedings{CCK24,
  author    = {Matteo Castiglioni and
               Andrea Celli and
               Christian Kroer},
  title     = {Online Learning under Budget and {ROI} Constraints via Weak Adaptivity},
  booktitle = {Forty-first International Conference on Machine Learning, {ICML} 2024, Vienna, Austria, July 21-27, 2024},
  series    = {Proceedings of Machine Learning Research},
  pages     = {5792--5816},
  publisher = {{PMLR} / OpenReview.net},
  year      = {2024}
}

@inproceedings{CHJ20,
  author    = {Modibo K. Camara and Jason D. Hartline and Aleck Johnsen},
  title     = {Mechanisms for a No-Regret Agent: Beyond the Common Prior},
  booktitle = {61st {IEEE} Annual Symposium on Foundations of Computer Science, {FOCS} 2020},
  year      = {2020},
  pages     = {259--270},
  publisher = {{IEEE}}
}

@article{CKK24,
  author  = {Xi Chen and
             Christian Kroer and
             Rachitesh Kumar},
  title   = {The Complexity of Pacing for Second-Price Auctions},
  journal = {Math. Oper. Res.},
  volume  = {49},
  number  = {4},
  pages   = {2109--2135},
  year    = {2024}
}

@inproceedings{DBLP:conf/colt/LucierPSZ24,
  author    = {Brendan Lucier and
               Sarath Pattathil and
               Aleksandrs Slivkins and
               Mengxiao Zhang},
  title     = {Autobidders with Budget and {ROI} Constraints: Efficiency, Regret, and Pacing Dynamics},
  booktitle = {Conference on Learning Theory, COLT 2024},
  series    = {Proceedings of Machine Learning Research},
  pages     = {3642--3643},
  publisher = {{PMLR}},
  year      = {2024}
}

@inproceedings{DBLP:conf/innovations/GaitondeLLLS23,
  author    = {Jason Gaitonde and
               Yingkai Li and
               Bar Light and
               Brendan Lucier and
               Aleksandrs Slivkins},
  title     = {Budget Pacing in Repeated Auctions: Regret and Efficiency Without
               Convergence},
  booktitle = {14th Innovations in Theoretical Computer Science Conference, {ITCS}
               2023, MIT, Cambridge, Massachusetts, USA, January 10-13, 2023},
  series    = {LIPIcs},
  pages     = {52:1--52:1},
  publisher = {Schloss Dagstuhl - Leibniz-Zentrum f{\"{u}}r Informatik},
  year      = {2023}
}

@inproceedings{DBLP:conf/sigecom/FikiorisKKKMT25,
  author    = {Giannis Fikioris and
               Robert Kleinberg and
               Yoav Kolumbus and
               Raunak Kumar and
               Yishay Mansour and
               {\'{E}}va Tardos},
  title     = {Learning in Budgeted Auctions with Spacing Objectives},
  booktitle = {Proceedings of the 26th {ACM} Conference on Economics and Computation, {EC} 2025},
  pages     = {158},
  publisher = {{ACM}},
  year      = {2025}
}

@inproceedings{DBLP:journals/corr/abs-2601-03853,
  author    = {Junyao Zhao},
  title     = {From No-Regret to Strategically Robust Learning in Repeated Auctions},
  year      = {2026},
  booktitle = {Proceedings of the 27th {ACM} Conference on Economics and Computation,
               {EC} 2026},
  publisher = {{ACM}}
}

@article{DBLP:journals/ior/ConitzerKSM22,
  author  = {Vincent Conitzer and
             Christian Kroer and
             Eric Sodomka and
             Nicol{\'{a}}s E. Stier-Moses},
  title   = {Multiplicative Pacing Equilibria in Auction Markets},
  journal = {Operations Research},
  volume  = {70},
  number  = {2},
  pages   = {963--989},
  year    = {2022}
}

@article{DBLP:journals/mansci/ConitzerKPSMSW22,
  author  = {Vincent Conitzer and
             Christian Kroer and
             Debmalya Panigrahi and
             Okke Schrijvers and
             Nicol{\'{a}}s E. Stier-Moses and
             Eric Sodomka and
             Christopher A. Wilkens},
  title   = {Pacing Equilibrium in First Price Auction Markets},
  journal = {Management Science},
  volume  = {68},
  number  = {12},
  pages   = {8515--8535},
  year    = {2022}
}

@article{DBLP:journals/mor/FikiorisT25,
  author  = {Giannis Fikioris and
             {\'{E}}va Tardos},
  title   = {Liquid Welfare Guarantees for No-Regret Learning in Sequential Budgeted
             Auctions},
  journal = {Mathematics of Operations Research},
  volume  = {50},
  number  = {2},
  pages   = {1233--1249},
  year    = {2025}
}

@inproceedings{DMMSS23,
  author    = {Christoph Dann and Yishay Mansour and Mehryar Mohri and Jon Schneider and Balasubramanian Sivan},
  title     = {Pseudonorm approachability and applications to regret minimization},
  booktitle = {International Conference on Algorithmic Learning Theory},
  year      = {2023},
  pages     = {471--509},
  publisher = {PMLR}
}

@inproceedings{DSS19,
  author    = {Yuan Deng and
               Jon Schneider and
               Balasubramanian Sivan},
  title     = {Strategizing against No-regret Learners},
  booktitle = {Advances in Neural Information Processing Systems 32: Annual Conference on Neural Information Processing Systems 2019},
  pages     = {1577--1585},
  year      = {2019}
}

@inproceedings{DSS19a,
  author    = {Yuan Deng and
               Jon Schneider and
               Balasubramanian Sivan},
  title     = {Prior-Free Dynamic Auctions with Low Regret Buyers},
  booktitle = {Advances in Neural Information Processing Systems 32: Annual Conference
               on Neural Information Processing Systems 2019, NeurIPS 2019, December
               8-14, 2019, Vancouver, BC, Canada},
  pages     = {4804--4814},
  year      = {2019},
  url       = {https://proceedings.neurips.cc/paper/2019/hash/f15eda31a2da646eea513b0f81a5414d-Abstract.html}
}

@book{durrett2019probability,
  title     = {Probability: Theory and Examples},
  author    = {Durrett, Rick},
  volume    = {49},
  year      = {2019},
  publisher = {Cambridge University Press}
}

@misc{fikioris2026robust,
  title         = {Robust Temporal Guarantees in Budgeted Sequential Auctions},
  author        = {Giannis Fikioris and Robert Kleinberg and Yoav Kolumbus and Yishay Mansour and {\'E}va Tardos},
  year          = {2026},
  eprint        = {2602.17916},
  archiveprefix = {arXiv},
  url           = {https://arxiv.org/abs/2602.17916}
}

@inproceedings{FPW23,
  author    = {Zhe Feng and
               Swati Padmanabhan and
               Di Wang},
  title     = {Online Bidding Algorithms for Return-on-Spend Constrained Advertisers},
  booktitle = {Proceedings of the {ACM} Web Conference 2023, {WWW} 2023, Austin,
               TX, USA, 30 April 2023 - 4 May 2023},
  pages     = {3550--3560},
  publisher = {{ACM}},
  year      = {2023}
}

@article{FST26a,
  author     = {Giannis Fikioris and
                Balasubramanian Sivan and
                {\'{E}}va Tardos},
  title      = {Learning vs. Optimizing Bidders in Budgeted Auctions},
  journal    = {CoRR},
  volume     = {abs/2604.08517},
  year       = {2026},
  eprinttype = {arXiv},
  eprint     = {2604.08517}
}

@inproceedings{FT23,
  author    = {Giannis Fikioris and
               {\'{E}}va Tardos},
  title     = {Approximately Stationary Bandits with Knapsacks},
  booktitle = {The Thirty Sixth Annual Conference on Learning Theory, {COLT} 2023,
               12-15 July 2023, Bangalore, India},
  series    = {Proceedings of Machine Learning Research},
  pages     = {3758--3782},
  publisher = {{PMLR}},
  year      = {2023}
}

@article{FV97,
  title   = {Calibrated Learning and Correlated Equilibrium},
  author  = {Dean P. Foster and Rakesh V. Vohra},
  journal = {Games and Economic Behavior},
  volume  = {21},
  number  = {1--2},
  pages   = {40--55},
  year    = {1997}
}

@article{MTY09,
  author  = {Shie Mannor and John N. Tsitsiklis and Jia Yuan Yu},
  title   = {Online Learning with Sample Path Constraints},
  journal = {Journal of Machine Learning Research},
  volume  = {10},
  number  = {20},
  pages   = {569--590},
  year    = {2009}
}

@article{FH21,
  author  = {Dean P. Foster and Sergiu Hart},
  title   = {Forecast Hedging and Calibration},
  journal = {Journal of Political Economy},
  volume  = {129},
  number  = {12},
  pages   = {3447--3490},
  year    = {2021},
}

@inproceedings{GJLM23,
  author    = {Negin Golrezaei and
               Patrick Jaillet and
               Jason Cheuk Nam Liang and
               Vahab Mirrokni},
  title     = {Pricing against a Budget and {ROI} Constrained Buyer},
  booktitle = {International Conference on Artificial Intelligence and Statistics, 25-27 April 2023, Palau de Congressos, Valencia, Spain},
  series    = {Proceedings of Machine Learning Research},
  volume    = {206},
  pages     = {9282--9307},
  publisher = {{PMLR}},
  year      = {2023}
}

@inproceedings{GKSTVWW24,
  author    = {Guru Guruganesh and
               Yoav Kolumbus and
               Jon Schneider and
               Inbal Talgam{-}Cohen and
               Emmanouil{-}Vasileios Vlatakis{-}Gkaragkounis and
               Joshua R. Wang and
               S. Matthew Weinberg},
  title     = {Contracting with a Learning Agent},
  booktitle = {Neural Information Processing Systems 2024, NeurIPS 2024},
  year      = {2024},
}

@misc{GooglePacing,
  title  = {Power your business by taking control of your budget and advertising cost},
  author = {Google},
  url    = {https://business.google.com/en-all/google-ads/campaign-budget/},
  year   = 2021
}

@article{Hazan16,
  author  = {Elad Hazan},
  title   = {Introduction to Online Convex Optimization},
  journal = {Foundations and Trends in Optimization},
  volume  = {2},
  number  = {3-4},
  pages   = {157--325},
  year    = {2016}
}

@article{HM00,
  title     = {A simple adaptive procedure leading to correlated equilibrium},
  author    = {Hart, Sergiu and Mas-Colell, Andreu},
  journal   = {Econometrica},
  volume    = {68},
  number    = {5},
  pages     = {1127--1150},
  year      = {2000},
  publisher = {Wiley Online Library}
}

@article{HPY24,
  author  = {Nika Haghtalab and Chara Podimata and Kunhe Yang},
  title   = {Calibrated {S}tackelberg Games: Learning Optimal Commitments Against Calibrated Agents},
  journal = {Advances in Neural Information Processing Systems},
  volume  = {36},
  year    = {2023}
}

@article{ISSS22,
  author  = {Nicole Immorlica and
             Karthik Abinav Sankararaman and
             Robert E. Schapire and
             Aleksandrs Slivkins},
  title   = {Adversarial Bandits with Knapsacks},
  journal = {J. {ACM}},
  volume  = {69},
  number  = {6},
  pages   = {40:1--40:47},
  year    = {2022}
}

@inproceedings{KLPST23,
  author    = {Bobby Kleinberg and
               {Paes Leme}, Renato and
               Jon Schneider and
               Yifeng Teng},
  title     = {U-Calibration: Forecasting for an Unknown Agent},
  booktitle = {Conference on Learning Theory, {COLT} 2023},
  series    = {Proceedings of Machine Learning Research},
  volume    = {195},
  publisher = {{PMLR}},
  year      = {2023}
}

@inproceedings{KS20,
  author    = {Thomas Kesselheim and
               Sahil Singla},
  title     = {Online Learning with Vector Costs and Bandits with Knapsacks},
  booktitle = {Conference on Learning Theory, {COLT} 2020, 9-12 July 2020, Virtual
               Event [Graz, Austria]},
  series    = {Proceedings of Machine Learning Research},
  pages     = {2286--2305},
  publisher = {{PMLR}},
  year      = {2020}
}

@inproceedings{KSS24,
  author    = {Rachitesh Kumar and
               Jon Schneider and
               Balasubramanian Sivan},
  title     = {Strategically-Robust Learning Algorithms for Bidding in First-Price
               Auctions},
  booktitle = {Proceedings of the 25th {ACM} Conference on Economics and Computation,
               {EC} 2024, New Haven, CT, USA, July 8-11, 2024},
  pages     = {893},
  publisher = {{ACM}},
  year      = {2024}
}

@inproceedings{BM05,
  author    = {Avrim Blum and Yishay Mansour},
  title     = {From External to Internal Regret},
  booktitle = {Learning Theory, 18th Annual Conference on Learning Theory, {COLT} 2005},
  series    = {Lecture Notes in Computer Science},
  pages     = {621--636},
  publisher = {Springer},
  year      = {2005},
}

@inproceedings{MMSS22,
  author    = {Yishay Mansour and
               Mehryar Mohri and
               Jon Schneider and
               Balasubramanian Sivan},
  title     = {Strategizing against Learners in Bayesian Games},
  booktitle = {Conference on Learning Theory, 2-5 July 2022, London, {UK}},
  series    = {Proceedings of Machine Learning Research},
  pages     = {5221--5252},
  publisher = {{PMLR}},
  year      = {2022}
}

@inproceedings{NRRX25,
  author    = {Georgy Noarov and
               Ramya Ramalingam and
               Aaron Roth and
               Stephan Xie},
  title     = {High-Dimensional Prediction for Sequential Decision Making},
  booktitle = {International Conference on Machine Learning, {ICML} 2025},
  series    = {Proceedings of Machine Learning Research},
  volume    = {267},
  year      = {2025}
}

@inproceedings{rivkin-et-al-2026,
  author    = {Joey Rivkin and {Deo-Campo Vuong}, Ramiro N. and Robert Kleinberg and Chido Onyeze and Erald Sinanaj and {\'E}va Tardos},
  title     = {Improved Multi-Dimensional Forecasting for Swap Regret},
  booktitle = {Proceedings of the 27th {ACM} Conference on Economics and Computation, {EC} 2026},
  publisher = {{ACM}},
  year      = {2026},
}

@inproceedings{RS24,
  author    = {Aaron Roth and Mirah Shi},
  title     = {Forecasting for Swap Regret for All Downstream Agents},
  booktitle = {Proceedings of the 25th {ACM} Conference on Economics and Computation, {EC} 2024},
  pages     = {466--488},
  publisher = {{ACM}},
  year      = {2024},
}

@inproceedings{RZ24,
  author    = {Aviad Rubinstein and Junyao Zhao},
  title     = {Strategizing against no-regret learners in first-price auctions},
  booktitle = {Proceedings of the 25th ACM Conference on Economics and Computation},
  year      = {2024},
  pages     = {894--921}
}

@misc{TwitterPacing,
  title  = {How we built Twitter’s highly reliable ads pacing service},
  author = {Twitter},
  url    = {https://blog.x.com/engineering/en_us/topics/infrastructure/2021/how-we-built-twitter-s-highly-reliable-ads-pacing-service},
  year   = 2021
}

@inproceedings{WYDK23,
  author    = {Qian Wang and
               Zongjun Yang and
               Xiaotie Deng and
               Yuqing Kong},
  title     = {Learning to Bid in Repeated First-Price Auctions with Budgets},
  booktitle = {International Conference on Machine Learning, {ICML} 2023, 23-29 July
               2023, Honolulu, Hawaii, {USA}},
  series    = {Proceedings of Machine Learning Research},
  pages     = {36494--36513},
  publisher = {{PMLR}},
  year      = {2023}
}

@inproceedings{ZKSME21,
  author    = {Shengjia Zhao and
               Michael P. Kim and
               Roshni Sahoo and
               Tengyu Ma and
               Stefano Ermon},
  title     = {Calibrating Predictions to Decisions: A Novel Approach to Multi-Class Calibration},
  booktitle = {Neural Information Processing Systems, NeurIPS 2021},
  year      = {2021}
}

\clearpage
\appendix
\crefalias{section}{appendix}
\crefalias{subsection}{appendix}
\section{Auction Games vs. General Games}
\label{sec:app:comparison}\label{sec:auction_vs_general}

\Cref{tab:comparison} summarizes the main differences between our work on repeated budgeted auctions~\cite{FST26a} and the general budgeted games studied in this paper.
While both papers evaluate non-manipulability against the Budgeted Stackelberg Equilibrium ($\BSE$) and prove low interval stochastic regret, neither result subsumes the other.
In~\cite{FST26a}, the stage game is a Bayesian auction with i.i.d.\ values, where the structure of auctions implies that best responses are parameterized by a single scalar pacing multiplier ($\text{bid} = \lambda \cdot \text{value}$); this allows a simple, deterministic proportional controller to be non-manipulable.
In general games, no deterministic learner can prevent manipulation, and best responses are discontinuous mixtures of up to two actions with no scalar parameterization.
Conversely, our normal-form model has no private values and assumes full-information feedback: casting a Bayesian auction as a normal-form game requires treating entire bidding functions as pure strategies, which makes $\nL$---and with it our running time and regret bounds---exponential even after discretizing values and bids (so the $\order{T^{2/3}}$ rate in~\cite{FST26a} and the $\order{\sqrt{T}}$ rate here are not directly comparable).

\begin{table}[H]
    \centering
    \small
    \renewcommand{\arraystretch}{1.3}
    \begin{tabular}{p{0.12\textwidth}p{0.4\textwidth}p{0.40\textwidth}}
        \hline
        & \textbf{Budgeted auctions~\cite{FST26a}} & \textbf{General budgeted games (this paper)} \\
        \hline
        Stage game & Second-price auctions (and first-price, with pacing bids) & Any game with utility and cost matrices \\
        Environment & Bayesian: values drawn i.i.d.\ from a joint distribution; adaptive optimizer & Adversarial: no distributional assumptions; adaptive optimizer \\
        Learner's strategies & Bids as a function of values (a continuum) & $\nL$ actions \\
        Learner's feedback & Its value and its realized payment & The utility and cost of every action \\
        Structure & Best responses are pacing: $\text{bid}=\lambda\cdot\text{value}$ & None: best responses mix up to two actions \\
        Algorithm & Deterministic proportional controller for $\lambda$ & Randomized best responses to calibrated forecasts \\
        Guarantee & $\BSE\cdot T+O(T^{2/3})$ (under a mild (necessary) assumption on values), plus low interval stochastic regret & $\BSE\cdot T+O(\sqrt{T})$, plus low interval stochastic regret \\
        Analysis & Dynamic programming over a scalar state, with a potential built from Lagrangian duality & Calibration in $L^2([-1,1])$; dual-weighted calibration; forecaster via simulated OGD \\
        \hline
    \end{tabular}
    \caption{This paper and~\cite{FST26a} at a glance.}
    \label{tab:comparison}
\end{table}
\section{Alternative Definitions of Budgeted Stackelberg Equilibrium}
\label{sec:app:bse_defs}

In \cref{sec:defs,sec:budgeted_optimizer}, we defined the BSE under the condition that the learner always tries to best-respond subject to keeping the expected per-round cost at most $\rL$.
This means that in each phase where the optimizer uses a fixed mixed strategy, the learner is meant to best-respond while keeping per-round spending at most $\rL$, similar to what \cref{alg:learner} does.
However, this is not the only behavior that makes sense.
For example, if historical average spending has been substantially less than $\rL$, it is natural to aim for spending higher than $\rL$ per round.
In this section, we formalize two alternative definitions of BSE focusing on an unbudgeted optimizer, corresponding to how a learner might use historical underspending, and show that both alternatives can \emph{only increase} the BSE value---meaning that they can only benefit the optimizer.
Under these definitions it might be natural to define the manipulability of a learning algorithm differently, but such definitions would allow for more utility for the optimizer.

For any optimizer strategy $x \in \Delta_{\nO}$ and any per-round learner budget cap $\tilde\rL > 0$, denote the learner's $\tilde\rL$-budgeted best-response set by
\begin{equation} \label{eq:app:bse_defs:br_rho}
    \BRL(x; \tilde\rL)
    \coloneq
    \argmax_{y \in \Delta_{\nL} \,:\, x^\top \CL y \le \tilde\rL} x^\top \UL y
    .
\end{equation}
Since $\CL \in [0, 1]^{\nO \times \nL}$, setting $\tilde\rL = \rL$ recovers $\BRL(x) = \BRL(x; \rL)$, while setting $\tilde\rL \ge 1$ makes the cost constraint vacuous and recovers the unconstrained best-response set $\BRL(x; 1) = \argmax_{y \in \Delta_{\nL}} x^\top \UL y$.

\paragraph{Three models of learner behavior}
Depending on how the learner adjusts her per-round spending cap after historical underspending, we obtain three equilibrium benchmarks:
\begin{enumerate}
    \item \textbf{Fixed per-round target $\rL$ (our baseline, $V_0 = \BSE$):}
    The learner always enforces her target rate $\rL$ in every phase ($y \in \BRL(x; \rL)$), regardless of past underspending.
    When the optimizer is unbudgeted, time-multiplexing across phases cannot improve upon a single mixed strategy, yielding \eqref{opt:defs:bse}:
    \begin{equation} \label{eq:app:bse_defs:V0}
        V_0
        \coloneq
        \BSE
        =
        \max_{x \in \Delta_{\nO},\; y \in \BRL(x; \rL)} x^\top \UO y
        .
    \end{equation}

    \item \textbf{Unconstrained best response under budget surplus ($V_1$):}
    Consider a learner who is willing to overspend when her historical average spending is less than $\rL$.
    This can happen when using a primal/dual approach (e.g., LagrangeBwK~\cite{ISSS22}), where the primal algorithm is learning to maximize the Lagrangian function, controlled by the dual algorithm that is picking the Lagrange multiplier to minimize the Lagrangian function.
    While historical spending is low, the Lagrange multiplier tends to be low, which makes the primal algorithm use an unrestricted best-response.

    We model such a learner by assuming two different behaviors:
    \begin{itemize}
        \item \emph{restricted phases} (historical spending equal to $\rL$), where a constrained best response $y_1 \in \BRL(x_1; \rL)$ is used,
        \item \emph{unrestricted phases} (historical spending strictly less than $\rL$), where the budget constraint is ignored and an unconstrained best response $y_2 \in \BRL(x_2; 1)$ is used.
    \end{itemize}

    Assuming that the optimizer uses only one restricted and one unrestricted phase, we make the following observations.
    The restricted phase must precede the unrestricted phase and the learner will continue using the unrestricted phase's $y_2$ while the historical average spending is at most $\rL$.
    This introduces our first alternate BSE value $V_1$, where 
    \begin{optimization} \label{opt:app:bse_defs:V1}
        V_1 \coloneq
        \max_{\substack{x_1, x_2 \in \Delta_{\nO},\; y_1, y_2 \in \Delta_{\nL} \\ q_1, q_2 \ge 0,\; q_1 + q_2 = 1}}
        \quad &
        \sum_{i=1}^2 q_i \, x_i^\top \UO y_i
        \\
        \textrm{\textup{such that}} \;\qquad &
        y_1 \in \BRL(x_1; \rL), y_2 \in \BRL(x_2; 1),
        \\
        \textrm{\textup{and}} \;\qquad\quad &
        \sum_{i=1}^2 q_i \, x_i^\top \CL y_i \le \rL
        .
    \end{optimization}

    We now explain how an optimizer can implement a feasible solution.
    The first $q_1 T$ rounds the optimizer uses $x_1$ and the learner best responds with $y_1$.
    Since $y_1 \in \BRL(x_1; \rL)$ the learner's average spending is at most her budget: $x_1^\top \CL y_1 \le \rL$.
    We now distinguish two cases based on the spending $x_1^\top \CL y_1$
    \begin{itemize}
        \item If $x_1^\top \CL y_1 < \rL$, i.e., the learner has been underspending after the first phase, then when the optimizer switches to $x_2$ after round $q_1 T$, the learner has an excess of budget.
        This allows them to overspend and select the unbudgeted best-response $y_2$.
        In fact, the inequality of \cref{opt:app:bse_defs:V1} guarantees that this excess of budget will continue up to the final round $T$, since for any $q' \in (0, q_2)$,
        \begin{equation*}
            q_1 x_1^\top \CL y_1
            +
            q' x_2^\top \CL y_2
            <
            (q_1 + q') \rL
            .
        \end{equation*}

        This guarantees that $y_2$ will continue being a best-response to $x_2$ up to the final round.
        
        \item If $x_1^\top \CL y_1 = \rL$, i.e., the spending matches the budget, then by the inequality of \cref{opt:app:bse_defs:V1} it must hold $x_2^\top \CL y_2 \le \rL$, i.e., the learner does not overspend in the second phase either.
        This implies that $y_2 \in \BRL(x_2; \rL)$, and the learner best-responds with $y_2$ when the optimizer starts using $x_2$ in rounds $q_1 T, \ldots, T$, and never overspends her budget.
        This also implies that the value of such a solution cannot be more than $V_0$:
        Both $(x_1, y_1)$ and $(x_2, y_2)$ are feasible solutions of \cref{eq:app:bse_defs:V0}, so any convex combination of $x_1^\top \UO y_1$ and $x_2^\top \UO y_2$ cannot exceed $V_0$.
    \end{itemize}

    We also note that because restricted and unrestricted phases decouple linearly adding more phases of either kind cannot make the value of \cref{opt:app:bse_defs:V1} larger:
    given a solution with more phases $(x_i, y_i, q_i)_{i=1,2,\ldots}$ we can fix $(x_i, y_i)$ and write an LP with only the $q_i$ as variables; since this has only two constraints ($\sum_i q_i = 1$ and the inequality of \cref{opt:app:bse_defs:V1}) the optimal solution can be supported on at most two non-zero variables. 

    \item \textbf{Best response to the remaining per-round budget ($V_2$):}
    Here we consider a learner who adjusts the target budget to the remaining average budget.
    When there are $\tau$ rounds remaining and the remaining budget is $\tilde\rL \tau$, she best-responds subject to her \emph{remaining per-round budget rate} $\tilde\rL$, choosing $y \in \BRL(x; \tilde\rL)$.
    In a multi-phase plan with phase lengths $q_i T$ and strategies $(x_i, y_i)$, the effective budget rate entering phase $i$ is $\tilde\rL^{(i)} = (\rL - \sum_{s < i} q_s \, x_s^\top \CL y_s) / (1 - \sum_{s < i} q_s)$, and the learner plays $y_i \in \BRL(x_i; \tilde\rL^{(i)})$.
    Without loss of generality, the learner strictly underspends ($x_i^\top \CL y_i < \tilde\rL^{(i)}$) in every phase except the last:
    If $x_i^\top \CL y_i = \tilde\rL^{(i)}$ in an intermediate phase $i$, the remaining budget rate is unchanged after phase $i$, so the optimizer can do at least as well by either extending phase $i$ to the end of the game or removing phase $i$ and proportionally rescaling the remaining phases.
    Since the budget constraint is non-binding in all phases prior to the final phase, each earlier $y_i$ is an unconstrained best response and hence also lies in $\BRL(x_i; \tilde\rL)$, where $\tilde\rL \ge \rL$ denotes the final phase's budget rate (and ordering the phases so that $x_1^\top \CL y_1 \le x_2^\top \CL y_2 \le \dots$ guarantees $x_i^\top \CL y_i \le \tilde\rL^{(i)} \le \tilde\rL$ throughout).
    By Carath\'eodory's theorem, at most two phases again suffice, yielding
    \begin{optimization} \label{opt:app:bse_defs:V2}
        V_2 \coloneq
        \max_{\substack{x_1, x_2 \in \Delta_{\nO},\; y_1, y_2 \in \Delta_{\nL} \\ \tilde\rL \ge \rL,\; q_1, q_2 \ge 0,\; q_1 + q_2 = 1}}
        \quad &
        \sum_{i=1}^2 q_i \, x_i^\top \UO y_i
        \\
        \textrm{\textup{such that}} \;\qquad\quad &
        y_i \in \BRL(x_i; \tilde\rL) \qquad \qquad \forall i \in \{1, 2\},
        \\
        \textrm{\textup{and}} \;\qquad\qquad &
        \sum_{i=1}^2 q_i \, x_i^\top \CL y_i \le \rL
        .
    \end{optimization}
\end{enumerate}

\paragraph{Alternative definitions can only benefit the optimizer}
We next show that allowing the learner to spend accumulated budget surpluses never hurts the optimizer and can strictly increase the optimizer's equilibrium payoff in some games under both alternatives.

\begin{proposition} \label{prop:app:bse_defs:hierarchy}
    For every budgeted game, $\BSE = V_0 \le V_1 \le V_2$.
    Moreover, both inequalities can be strict: there exist games with $V_0 < V_1$ and games with $V_1 < V_2$.
\end{proposition}

\begin{proof}
    Setting $q_1 = 1$ and $q_2 = 0$ in \cref{opt:app:bse_defs:V1} recovers \cref{eq:app:bse_defs:V0}, so $V_0 \le V_1$.
    Setting $\tilde \rL = 1$ in \cref{opt:app:bse_defs:V2} recovers \cref{opt:app:bse_defs:V1}, so $V_1 \le V_2$.
    We establish the two strict separations via the following examples:

    \emph{Strict separation $V_0 < V_1$:}
    Consider a $2 \times 2$ game with budget $\rL = \nicefrac{1}{2}$ and matrices
    \begin{equation*}
        \UO = \m{1 & 0 \\ 0 & 0},
        \qquad
        \UL = \m{1 & 0 \\ 0 & \nicefrac{1}{2}},
        \qquad
        \CL = \m{1 & \nicefrac{1}{2} \\ \nicefrac{1}{2} & 0}
        .
    \end{equation*}
    Under $V_0$, any feasible pair $(x, y)$ satisfies $x^\top \CL y = \frac{x_1 + y_1}{2} \le \rL = \frac{1}{2}$, so by the AM--GM inequality the optimizer's utility is at most $x^\top \UO y = x_1 y_1 \le \qty(\frac{x_1 + y_1}{2})^2 \le \frac{1}{4}$. 
    This is attained at $x^\star = y^\star = (\nicefrac{1}{2}, \nicefrac{1}{2})$, giving $\BSE = V_0 = \nicefrac{1}{4}$.
    
    Under $V_1$, however, the optimizer achieves $V_1 = \frac{1}{2}$.
    First, the optimizer uses $x_1 = (0, 1)$ for the first $T/2$ rounds ($q_1 = \nicefrac{1}{2}$), inducing the best response $y_1 = (0, 1)$ at cost $0$.
    Then, they switch to $x_2 = (1, 0)$ for the remaining $T/2$ rounds ($q_2 = \nicefrac{1}{2}$), inducing the unconstrained best response $y_2 = (1, 0)$ at cost $1$.
    The average cost across the horizon is $\frac{1}{2} \cdot 0 + \frac{1}{2}\cdot 1 = \frac{1}{2} = \rL$, and the optimizer earns $V_1 = \frac{1}{2} \cdot 0 + \frac{1}{2}\cdot 1 = \frac{1}{2} > \frac{1}{4} = V_0$.
    
    \emph{Strict separation $V_1 < V_2$:}
    Consider a $2 \times 3$ game with budget $\rL = \nicefrac{1}{4}$ and matrices
    \begin{equation*}
        \UO = \m{1 & \nicefrac{1}{2} & 0 \\ 1 & \nicefrac{1}{2} & 0},
        \qquad
        \UL = \m{\nicefrac{3}{4} & 1 & 0 \\ \nicefrac{3}{4} & 1 & 0},
        \qquad
        \CL = \m{\nicefrac{1}{2} & 0 & 0 \\ \nicefrac{1}{2} & 1 & 0}
        .
    \end{equation*}

    We start by showing that $V_0 = 1/2$ and then that also $V_1 = V_0 = 1/2$.
    If $x$ is such that $x_2 \le \nicefrac{1}{4}$, then the learner's 2nd action is budget-feasible and the unique best-response, yielding optimizer utility $\nicefrac{1}{2}$.
    If $x_2 > \nicefrac{1}{4}$, then the optimizer's utility is
    \begin{equation*}
        x^\top \UO y
        =
        y_1 + \frac{1}{2} y_2
        \le
        y_1 + 2 x_2 y_2
        =
        2 \, x^\top \CL y
        \le
        \frac{1}{2}
    \end{equation*}
    where in the last inequality we assume that the learner's response is budget-feasible.
    This proves that $V_0 = 1/2$.

    We now prove that under $V_1$ the optimizer cannot get additional utility.
    This is because for every unbudgeted best-response $y \in \BRL(x; 1)$ it holds $y_2 = 1$, which means that $x^\top \UO y = 1/2$.
    Since the objective of $V_1$ in \cref{opt:app:bse_defs:V1} is a convex combination of $(x, y)$ that is feasible in $V_0$ and $(x, y)$ with $y \in \BRL(x; 1)$, its value cannot exceed $1/2$.

    Under $V_2$, however, the optimizer achieves $V_2 = \nicefrac{3}{4}$.
    First, the optimizer uses $x = (1, 0)$ for the first $T/2$ rounds ($q_1 = \nicefrac{1}{2}$), inducing the best response $y = (0, 1, 0)$ at cost $0$.
    This earns the optimizer utility $\nicefrac{1}{2}$ and raises the learner's effective per-round budget for the remaining $T/2$ rounds ($q_2 = \nicefrac{1}{2}$) from $\rL = \nicefrac{1}{4}$ to $\tilde\rL = \nicefrac{1}{2}$.
    Then, the optimizer switches to $x = (0, 1)$, under which the unique budget-adjusted best response is $(1, 0, 0) \in \BRL\qty\big( (0,1) ; \tilde\rL )$:
    the utility to cost ratios of the first two actions are $(3/2, 1)$, so since the first action has cost $\tilde\rL$ it is the unique best response.
    Thus the optimizer earns $V_2 = \nicefrac{1}{2} \cdot \nicefrac{1}{2} + \nicefrac{1}{2} \cdot 1 = \nicefrac{3}{4} > \nicefrac{1}{2} = V_1$.

\end{proof}

\section{Manipulability of Online Gradient Ascent}
\label{sec:app:OGA_manipulable}

The example in this section is just for completeness: this example is directly from~\cite{FST26a}.

In this section, we present a simple $3 \times 3$ unbudgeted game ($\rL = 1$ and $\CL = 0$) in which Projected Online Gradient Ascent (OGA) is manipulable, even though OGA satisfies low \emph{interval regret} (and dynamic regret) on every contiguous interval $\mathcal I \subseteq [T]$---a property similar to, and technically stronger than, the stochastic per-interval regret guarantee we prove in \cref{thm:lagr_cal:stochastic_regret}.
Specifically, by switching actions each time the learner's mixed strategy reaches a corner of the simplex, the optimizer forces OGA to cycle around the boundary of $\Delta_3$ and achieves $\Omega(T)$ more utility than the Stackelberg equilibrium value $\SE$.

\paragraph{Game definition and Stackelberg value}
Consider an unbudgeted game with $\nO = \nL = 3$ actions and payoff matrices
\begin{equation}
    \label{eq:app:OGA:payoffs}
    \UO = \m{
        1 & 0 & 0 \\
        0 & 1 & 0 \\
        0 & 0 & 1
    },
    \qquad
    \UL = \m{
        0 & 1 & \nicefrac{1}{2} \\
        \nicefrac{1}{2} & 0 & 1 \\
        1 & \nicefrac{1}{2} & 0
    }.
\end{equation}

When the optimizer commits to a mixed strategy $x = (x_1, x_2, x_3) \in \Delta_3$, the learner's expected utilities for her three actions are $x^\top \UL = (\frac{1}{2}x_2 + x_3,\; x_1 + \frac{1}{2}x_3,\; \frac{1}{2}x_1 + x_2)$.
For pure action $1$ to be a best response for the learner, it must beat action $2$:
\begin{equation*}
    \frac{1}{2}x_2 + x_3 \ge x_1 + \frac{1}{2}x_3
    \quad\Longleftrightarrow\quad
    x_1 \le \frac{1}{2}(x_2 + x_3) = \frac{1}{2}(1 - x_1)
    \quad\Longleftrightarrow\quad
    x_1 \le \frac{1}{3}.
\end{equation*}

By cyclic symmetry, any pure action $j \in \operatorname{supp}(y)$ in a learner best response $y \in \BRL(x)$ satisfies $x_j \le \nicefrac{1}{3}$, so $x^\top \UO y = \sum_{j} y_j x_j \le \nicefrac{1}{3}$.
Since the uniform strategy $x^\star = (\nicefrac{1}{3}, \nicefrac{1}{3}, \nicefrac{1}{3})$ achieves $\nicefrac{1}{3}$, the Stackelberg value is $\SE = \nicefrac{1}{3}$.

\paragraph{How OGA is manipulated by corner switching}
Suppose the learner updates her mixed strategy $y_t \in \Delta_3$ via Projected OGA with step size $\eta > 0$ starting from $y_1 = (\nicefrac{1}{3}, \nicefrac{1}{3}, \nicefrac{1}{3})$:
\begin{equation}
    \label{eq:app:OGA:update}
    y_{t+1} = \Pi_{\Delta_3}\qty(y_t + \eta \, \UL[i_t, :]^\top),
    \qquad\text{where}\qquad
    \Pi_{\Delta_3}(z) = \argmin_{y \in \Delta_3} \norm{y - z}_2^2.
\end{equation}
By standard online convex optimization analysis~\cite{Hazan16}, for $\eta = \Theta(1/\sqrt{T})$, Projected OGA guarantees low dynamic regret: $\sum_{t \in \mathcal I} \UL[i_t, :]\qty(y^\star - y_t) \le \frac{1}{\eta} + \frac{3\eta}{2}|\mathcal I| = O(\sqrt{T})$ regret on \emph{every} contiguous interval $\mathcal I \subseteq [T]$ against \emph{any} comparator $y^\star \in \Delta_3$ and any adversarial sequence of actions (which directly implies the stochastic interval regret guarantee of \cref{thm:lagr_cal:stochastic_regret}).
Nevertheless, the optimizer can manipulate OGA using a simple state-based switching rule (note this is known by~\cite{MMSS22} since OGA can have linear swap regret).

The optimizer's strategy goes as follows.
Start by playing action $1$ until $y_t$ reaches the corner $e_2 = (0, 1, 0)$ (the learner's best response to action $1$); as soon as $y_t = e_2$, switch to action $2$ until $y_t$ reaches $e_3 = (0, 0, 1)$; as soon as $y_t = e_3$, switch to action $3$ until $y_t$ reaches $e_1 = (1, 0, 0)$; and repeat cyclically ($e_i \to e_{(i \bmod 3) + 1}$) whenever $y_t$ reaches a corner.

To see why this yields $\nicefrac{1}{2}$ average utility per round for the optimizer, consider the phase starting at $y_t = e_1 = (1, 0, 0)$, where the optimizer switches to action $1$ and induces learner utility vector $u^{(1)} = (0, 1, \nicefrac{1}{2})^\top$.
For any point $y_t = (y_{t,1}, 1 - y_{t,1}, 0)$ along the edge $e_1 e_2$, the Euclidean projection $\Pi_{\Delta_3}(y_t + \eta u^{(1)})$ subtracts the threshold $\theta = \frac{0 + 1}{2}\eta = \frac{\eta}{2}$ from each coordinate and clips at zero.
Because action $3$'s unprojected increment is $\frac{\eta}{2} = \theta$, the third coordinate remains pinned at zero ($y_{t+1,3} = 0$), while $y_{t,1}$ decreases linearly from $1$ to $0$ in steps of $\frac{\eta}{2}$ over $\ceil{2/\eta}$ rounds until $y_t$ reaches the next corner $e_2 = (0, 1, 0)$.
Throughout this traversal of $e_1 \to e_2$, the optimizer plays action $1$ and earns utility $y_{t,1}$ in round $t$, which averages $\nicefrac{1}{2}$ per round as $y_{t,1}$ drops linearly from $1$ to $0$.
By cyclic symmetry, every subsequent edge traversal $e_2 \to e_3$, $e_3 \to e_1$, and $e_1 \to e_2$ likewise keeps the third coordinate pinned at $0$ and yields average optimizer utility $\nicefrac{1}{2}$, giving total optimizer utility $\frac{1}{2}T - O(1/\eta) = \SE \cdot T + \frac{1}{6}T - O(\sqrt{T})$ for $\eta = 1/\sqrt{T}$.
Notice that this does not contradict OGA's $O(\sqrt{T})$ interval regret: each single-edge phase has length $\ceil{2/\eta} = \Theta(\sqrt{T})$ (within the $O(1/\eta) = O(\sqrt{T})$ interval-regret slack), while over any longer interval spanning multiple cycles the optimizer's empirical play is uniform, against which every fixed action earns $\nicefrac{1}{2}$ per round---matching OGA's own average utility of $\nicefrac{1}{2}$.

\begin{figure}[t!]
    \centering
    \includegraphics[width=\linewidth]{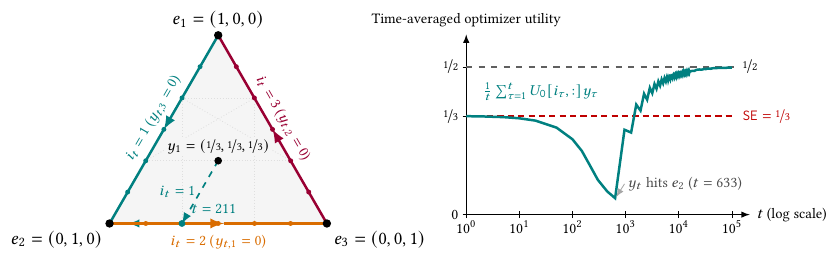}
    \caption{Simulation of Projected OGA in the $3 \times 3$ game of \cref{eq:app:OGA:payoffs} over $T = 100{,}000$ rounds with step size $\eta = 1/\sqrt{T}$. \textbf{Left:} Trajectory of the learner's mixed strategy $y_t \in \Delta_3$ from $y_1 = (\nicefrac{1}{3}, \nicefrac{1}{3}, \nicefrac{1}{3})$ to the boundary (dashed) and around the boundary cycle $e_2 \to e_3 \to e_1 \to e_2$ (solid). \textbf{Right:} Optimizer's running average utility $\frac{1}{t}\sum_{\tau=1}^t \UO[i_\tau, :] y_\tau$ converging to $\nicefrac{1}{2} > \SE = \nicefrac{1}{3}$.}
    \label{fig:app:OGA_manipulable}
\end{figure}

\paragraph{Experimental results}
We simulate Projected OGA (\cref{eq:app:OGA:update}) over $T = 100{,}000$ rounds with $\eta = 1/\sqrt{T}$ starting from $y_1 = (\nicefrac{1}{3}, \nicefrac{1}{3}, \nicefrac{1}{3})$.
As shown in \cref{fig:app:OGA_manipulable} (left), during the initial phase ($i_t = 1$), $y_t$ moves parallel to $e_1 e_2$ until hitting the boundary face $y_{t,1} = 0$ at $t = 211$ and then slides along $e_2 e_3$ into the corner $e_2 = (0, 1, 0)$ at $t = 633$; thereafter, $y_t$ cycles counterclockwise along the boundary edges $e_2 \to e_3 \to e_1 \to e_2$, taking $\ceil{2/\eta} = 633$ rounds per edge ($157$ corner switches in total).
As shown in \cref{fig:app:OGA_manipulable} (right), the optimizer's time-averaged utility converges to $0.498 \approx \nicefrac{1}{2}$, exceeding $\SE = \nicefrac{1}{3}$ by $\Omega(1)$ per round, while the learner still achieves $0.502$ average utility per round (matching the best fixed action's $0.503$ in hindsight).

\section{Why Naive Forecast Error Fails for Non-Manipulability}
\label{sec:app:resp_cal_example}

In this section, we give a simple $2 \times 2$ unbudgeted example ($\rL = 1$ and $\CL = 0$) illustrating why tracking only the error between the realized vector $uc_t$ and the predicted vector $uc(f^{(t)})$---without conditioning on the learner's induced best response---fails to prevent manipulation.

Consider the unbudgeted matching pennies game with $\nO = \nL = 2$ actions and utility matrices
\begin{equation*}
    \UO = \m{
        0 & 1 \\
        1 & 0
    }
    \qquad\text{and}\qquad
    \UL = \m{
        1 & 0 \\
        0 & 1
    },
\end{equation*}
where the learner wants to match the optimizer's action ($j = i$) and the optimizer wants the learner to mismatch ($j \ne i$).
Here the two rows of $\UL$ are $u^{(1)} = (1, 0)$ and $u^{(2)} = (0, 1)$, so a forecast $f = (f_1, 1 - f_1) \in \Delta_2$ predicts $u(f) = (f_1, 1 - f_1)$.
The learner's best response $\BRL\qty\big(u(f))$ is discontinuous at $f_1 = \nicefrac{1}{2}$: it uniquely selects pure action $1$ when $f_1 > \nicefrac{1}{2}$ and pure action $2$ when $f_1 < \nicefrac{1}{2}$.
Consequently, against any committed mixed strategy $x = (x_1, 1 - x_1) \in \Delta_2$, the optimizer obtains utility $1 - x_1 < \nicefrac{1}{2}$ if $x_1 > \nicefrac{1}{2}$, utility $x_1 < \nicefrac{1}{2}$ if $x_1 < \nicefrac{1}{2}$, and utility $\nicefrac{1}{2}$ if $x_1 = \nicefrac{1}{2}$; thus $\BSE = \nicefrac{1}{2}$.

Now fix any small $\varepsilon \in (0, \nicefrac{1}{2})$ and consider how the naive forecast error fails around the discontinuity at $f_1 = \nicefrac{1}{2}$.
Suppose that in odd rounds the forecaster predicts $f^{(t)} = (\nicefrac{1}{2} + \varepsilon, \nicefrac{1}{2} - \varepsilon)$ while the optimizer plays $i_t = 2$ (realizing $u^{(2)} = (0, 1)$), and in even rounds the forecaster predicts $f^{(t)} = (\nicefrac{1}{2} - \varepsilon, \nicefrac{1}{2} + \varepsilon)$ while the optimizer plays $i_t = 1$ (realizing $u^{(1)} = (1, 0)$).
Over any even horizon $T$, the cumulative error between the realized and predicted utility vectors is identically zero:
\begin{equation*}
    \sum_{t=1}^T \qty\big(u_t - u(f^{(t)}))
    =
    \frac{T}{2}\qty\Big[(0, 1) - \qty(\tfrac{1}{2}+\varepsilon, \tfrac{1}{2}-\varepsilon)]
    +
    \frac{T}{2}\qty\Big[(1, 0) - \qty(\tfrac{1}{2}-\varepsilon, \tfrac{1}{2}+\varepsilon)]
    =
    (0, 0).
\end{equation*}
Yet because the two forecasts lie on opposite sides of the discontinuity at $\nicefrac{1}{2}$, the learner plays $j_t = 1$ in every odd round (when $i_t = 2$) and $j_t = 2$ in every even round (when $i_t = 1$).
The optimizer therefore earns utility $1$ in every round, achieving $T = \BSE \cdot T + \frac{1}{2}T$.
The response-calibration error $R_T$ in \eqref{eq:regret_function} prevents this by conditioning the forecast error on the learner's induced best response $\vec J(f^{(t)})$, separating odd rounds ($\vec J = (1, 1)$) from even rounds ($\vec J = (2, 2)$) so their errors cannot cancel.

\section{Why Indexing Response-Calibration Error by \texorpdfstring{$(k, j)$}{(k, j)} Pairs Fails}
\label{sec:app:resp_cal_pairs}

Instead of indexing $\d^R$ and $R_T$ in \cref{sec:resp_cal} by triplets $(k, j_1, j_2) \in [\hatnO] \times [\nL]^2$, a natural alternative is to index by pairs $(k, j) \in [\hatnO] \times [\nL]$, recording the step function $\One{q \le Q(f)}$ on coordinate $(k, J_1(f))$ and $\One{q \le 1 - Q(f)}$ on coordinate $(k, J_2(f))$:
\begin{equation*}
    \tilde{R}_{T, (k, j)}(q)
    \coloneq
    \sum_{t \in \mathcal T} \Ex[f \sim F_t]{
        \qty\big(\One{k_t = k} - f_k)
        \qty\Big(
            \One{j = J_1(f)} \One{q \le Q(f)}
            +
            \One{j = J_2(f)} \One{q \le 1 - Q(f)}
        )
    }
    .
\end{equation*}

When each environment $uc^{(k)}$ corresponds to a unique optimizer payoff row (that is, if $\hatnO = \nO$), $g_{k, j_1, j_2}(q) = q \UO[k, j_1] + (1 - q) \UO[k, j_2]$ decomposes additively across $j_1$ and $j_2$, so bounding $\norm{\tilde{R}_T}_{L^2}$ suffices and reduces the dimension from $\hatnO \nL^2$ to $\hatnO \nL$ (saving a $\sqrt{\nL}$ factor in \cref{thm:resp_cal:manipulability}).
The same argument also works if $i_t$ is observed, and we can collect calibration error $\tilde{R}_{T, (i, j)}(q)$ for $(i,j)$ pairs instead of $(k, j)$ pairs.

However, when multiple optimizer actions $i_1, i_2 \in [\nO]$ induce the same learner environment $uc^{(k)}$ ($\hatnO < \nO$), $\tilde{R}_T$ fails.
To see why, consider a learner's mixed best response $y(f) = Q(f) e_{j_1} + (1 - Q(f)) e_{j_2}$ to a forecast $f$.
We estimated the optimizer's payoff under $uc^{(k)}$ as
\begin{equation*}
    g_{k, j_1, j_2}(q)
    =
    \max_{i : (\UL[i, :], \CL[i, :]) = uc^{(k)}} \qty\big(q \UO[i, j_1] + (1 - q) \UO[i, j_2])    
\end{equation*}
which depends on the \emph{pair} $(j_1, j_2)$ jointly.
Unbundling into $(k, j)$ pairs forgets which other action $j$ was mixed with; bounding $(k, j_1)$ and $(k, j_2)$ separately would replace the maximum of the sum by the sum of the maxima over $i$, allowing the optimizer to play $i_1$ when $j_t = j_1$ and $i_2$ when $j_t = j_2$ within the same coin flip.
The example below shows how this can cause higher value for the optimizer despite $ \tilde{R}_{T, (k, j)}(q)=0$.

\paragraph{Counterexample}
Consider a game with $\nO = \nL = 4$, budget $\rL = \nicefrac{1}{2}$, and matrices
\begin{equation*}
    \UO = \m{
        0 & 0 & 0 & 1 \\
        1 & 0 & 0 & 0 \\
        1 & 0 & 1 & 0 \\
        0 & 1 & 0 & 1
    }, \qquad
    \UL = \frac{1}{3}\m{
        1 & 0 & 2 & 2 \\
        0 & 0 & 2 & 3 \\
        0 & 1 & 3 & 2 \\
        0 & 1 & 3 & 2
    }, \qquad
    \CL = \m{
        0 & 0 & 1 & 1 \\
        0 & 0 & 1 & 1 \\
        0 & 0 & 1 & 1 \\
        0 & 0 & 1 & 1
    }.
\end{equation*}
Here optimizer actions $i = 1$ and $i = 2$ induce $uc^{(1)}$ and $uc^{(2)}$, respectively, while $i = 3, 4$ induce the same environment $uc^{(3)}$ ($\hatnO = 3$).
In every row of $\UL$, the free actions $\{1, 2\}$ (cost $0$) have utility at most $\nicefrac{1}{3}$ and the expensive actions $\{3, 4\}$ (cost $1$) have utility at least $\nicefrac{2}{3}$.
Thus, for any forecast $f \in \Delta_3$, the learner's best response mixes one free action $j_1 \in \argmax_{j \in \{1, 2\}} u(f)[j]$ and one expensive action $j_2 \in \argmax_{j \in \{3, 4\}} u(f)[j]$ with weight $Q(f) = \nicefrac{1}{2}$ each.
Moreover, since $u(f)[1] - u(f)[2] = \frac{1}{3}(f_1 - f_3)$ and $u(f)[3] - u(f)[4] = \frac{1}{3}(f_3 - f_2)$, the learner's best response is determined simply by comparing $f_3$ to $f_1$ and $f_2$:
\begin{itemize}
    \item \emph{Aligned} pairs: $(1, 3)$ when $f_1 > f_3 > f_2$, and $(2, 4)$ when $f_2 > f_3 > f_1$.
    Under $uc^{(3)}$ and when $(j_1, j_2)$ is an aligned pair, the optimizer gets $g_{3, j_1, j_2}(\nicefrac{1}{2}) = 1$ by playing $i = 3$ on $(1, 3)$ and $i = 4$ on $(2, 4)$.
    
    \item \emph{Crossed} pairs: $(2, 3)$ when $f_3 > f_1, f_2$, and $(1, 4)$ when $f_3 < f_1, f_2$. 
    Under $uc^{(3)}$ and when $(j_1, j_2)$ is a crossed pair, the optimizer gets $g_{3, j_1, j_2}(\nicefrac{1}{2}) = \nicefrac{1}{2}$ for either $i = 3, 4$.
\end{itemize}

Under $uc^{(1)}$ ($i = 1$) and $uc^{(2)}$ ($i = 2$), each row of $\UO$ contains a single $1$, so the optimizer gets at most $\nicefrac{1}{2}$ on every pair, and gets $0$ under $uc^{(1)}$ on $(1, 3)$ and under $uc^{(2)}$ on $(2, 4)$.
Thus, for any forecast $f \in \Delta_3$, the optimizer's expected utility is at most $\nicefrac{1}{2}$ on the crossed pairs; on the aligned pair $(1, 3)$ (where $f_1 \ge f_3 \ge f_2$), it is $0 \cdot f_1 + \frac{1}{2} f_2 + 1 \cdot f_3 \le \frac{1}{2}(f_1 + f_2 + f_3) = \nicefrac{1}{2}$ (and symmetrically on $(2, 4)$).
Hence $\BSE = \nicefrac{1}{2}$ (achieved, e.g., at $f = (0, 0, 1)$).

Now divide $T$ into four equal phases of $T/4$ rounds.
In each phase a deterministic forecast $f$ is used that uniquely induces one of the four pairs $(j_1, j_2) \in \{1, 2\} \times \{3, 4\}$ with $Q(f) = \nicefrac{1}{2}$.
Let $d \coloneqq \frac{1}{8}(-1, -1, 2)$, which will be the forecasting error.
Specifically, for phases with aligned pairs, the optimizer's mixed strategy induces $\hat f$ such that $\hat f - f = d$ and for crossed pairs $\hat f - f = -d$, making the average forecast correct.
In particular, even if we condition on a certain learner action, the average forecast is correct, making $\tilde R_T = 0$.
This cannot happen when conditioning on pairs of learner actions (since for each pair $\hat f - f$ is $d$ or $-d$) which makes $\norm{R_T} = \Omega(T)$.

The phases follow.
\begin{itemize}
    \item \textbf{Aligned phase $(1, 3)$:}
    The forecast is $f = \frac{1}{8}(4, 1, 3)$ ($f_1 > f_3 > f_2$), which does induce $(1, 3)$.
    The optimizer plays $i = 1$ ($uc^{(1)}$) on $\nicefrac{3}{8}$ of the rounds and $i = 3$ ($uc^{(3)}$) on $\nicefrac{5}{8}$, giving empirical environment distribution $\hat{f} = f + d = \frac{1}{8}(3, 0, 5)$.
    The optimizer's expected utility per round is $\nicefrac{5}{8}$.

    \item \textbf{Aligned phase $(2, 4)$:}
    The forecast is $f = \frac{1}{8}(1, 4, 3)$ ($f_2 > f_3 > f_1$), inducing $(2, 4)$.
    The optimizer plays $i = 2$ ($uc^{(2)}$) on $\nicefrac{3}{8}$ of the rounds and $i = 4$ ($uc^{(3)}$) on $\nicefrac{5}{8}$, giving $\hat{f} = f + d = \frac{1}{8}(0, 3, 5)$ and optimizer utility $\nicefrac{5}{8}$ per round.

    \item \textbf{Crossed phase $(2, 3)$:}
    The forecast is $f = (0, 0, 1)$ ($f_3 > f_1, f_2$), inducing $(2, 3)$.
    The optimizer plays $i = 1, 2$ on $\nicefrac{1}{8}$ of the rounds each and $i = 3$ ($uc^{(3)}$) on $\nicefrac{6}{8}$, giving $\hat{f} = f - d = \frac{1}{8}(1, 1, 6)$ and optimizer utility $\frac{6}{8} \cdot \frac{1}{2} = \nicefrac{3}{8}$ per round.

    \item \textbf{Crossed phase $(1, 4)$:}
    The forecast is $f = \frac{1}{8}(3, 3, 2)$ ($f_3 < f_1, f_2$), inducing $(1, 4)$.
    The optimizer plays $i = 1, 2$ on $\nicefrac{1}{2}$ of the rounds each (never $uc^{(3)}$), giving $\hat{f} = f - d = (\nicefrac{1}{2}, \nicefrac{1}{2}, 0)$ and optimizer utility $\nicefrac{1}{2}$ per round.
\end{itemize}
Because every pure learner action $j \in \{1, 2, 3, 4\}$ is played with weight $\nicefrac{1}{2}$ in exactly one aligned phase (error $+d$) and one crossed phase (error $-d$), the pairwise calibration error vanishes identically for every $(k, j) \in [3] \times [4]$:
\begin{equation*}
    \tilde{R}_{T, (k, j)}(q)
    =
    \frac{T}{4}\qty\big(d_k - d_k) \One{q \le \nicefrac{1}{2}}
    =
    0
    .
\end{equation*}
Yet by inducing $uc^{(3)}$ more often than forecast on aligned pairs (where $uc^{(3)}$ pays $1$) and less often on crossed pairs (where $uc^{(3)}$ pays $\nicefrac{1}{2}$), the optimizer achieves total utility $\frac{T}{4}\qty(\frac{5}{8} + \frac{5}{8} + \frac{3}{8} + \frac{4}{8}) = \frac{17}{32}T = \BSE \cdot T + \frac{1}{32}T$.

Tracking triplets $(k, j_1, j_2)$ in $R_T$ prevents this cancellation by keeping the four pairs in separate coordinates.
Specifically, for each $k \in [3]$, the aligned triplets $(k, 1, 3)$ and $(k, 2, 4)$ each accumulate error $\frac{T}{4} d_k \One{q \le \nicefrac{1}{2}}$ in $R_T$, whereas the crossed triplets $(k, 2, 3)$ and $(k, 1, 4)$ each accumulate $-\frac{T}{4} d_k \One{q \le \nicefrac{1}{2}}$.
Because the $+d_k$ error from $(1, 3)$ and the $-d_k$ error from $(1, 4)$ are no longer summed into a single $(k, 1)$ coordinate (and similarly for $j = 2, 3, 4$), they cannot cancel across distinct pairs, yielding $\norm{R_T}_{L^2} = \Omega(T)$.

\section{Why Response-calibration is Insufficient for Stochastic Regret}
\label{sec:app:resp_cal:not_enough}

In this section we show that even if $R_T = 0$, the learner may suffer $\Omega(T)$ regret in a stochastic environment by best-responding to each of the optimizer's pure actions instead of the mixed ones.
Our example is as follows.
Let $\rL = 1/2$, $\nO = \hatnO = 2$, $\nL = 3$ (where the third action is the null action), and
\begin{equation*}
    \UL
    =
    \m{1/2 & 4/5 & 0 \\ 1/2 & 2/5 & 0}
    \qquad\text{ and }\qquad
    \CL
    =
    \m{1/2 & 1 & 0 \\ 1/2 & 0 & 0}
    ,
\end{equation*}
where the optimizer draws $uc^{(1)}$ (row $1$) and $uc^{(2)}$ (row $2$) with probability $1/2$ each.
This makes the ``correct'' forecast $f^\star = (\frac{1}{2}, \frac{1}{2})$ whose best response is action $2$, achieving expected utility $3/5$ and cost $1/2$.

Under forecast $f = (1, 0)$ (predicting $uc^{(1)}$), pure action $1$ is the unique best response, achieving utility $1/2$ at cost $1/2$.
Under $f = (0, 1)$ (predicting $uc^{(2)}$), pure action $1$ is again the unique best response, achieving utility $1/2$ at cost $1/2$.
Since both induce pure action $1$ (so $\vec J = (1, 1)$ and $Q = 0$), a forecaster that predicts each half the time has calibration error for the triplet $(k, 1, 1)$ for each of $k \in\{1,2\}$
\begin{equation*}
    \frac{1}{2} \d^R_{k, 1, 1}\qty\big((1, 0), k_t)
    +
    \frac{1}{2} \d^R_{k, 1, 1}\qty\big((0, 1), k_t)
    =
    \qty( \One{k_t = k} - \frac{1}{2})\One{\cdot \le 0}
\end{equation*}
which makes it $0$ in expectation if $k_t$ is sampled uniformly from $\{1 ,2\}$.
Thus, the learner always plays action $1$ and earns $1/2$ per round which is suboptimal compared to the true optimal of $3/5$.
Notice that this issue does not benefit the optimizer: each learner behavior is best-responding, so the definition of the BSE (\cref{opt:defs:bse}) already considers this behavior as feasible actions.

For the learner's stochastic regret, however, the problem is that because both forecasts induce the same action, $R_T$ only tracks their average $(1/2, 1/2)$, achieving $R_T = 0$ without ever forcing the forecaster to predict the actual mixture $f = (1/2, 1/2)$.
In \cref{sec:lagr_cal}, we introduce a calibration error that prevents such errors from canceling by adding conditions that distinguish forecasts where the budget constraint is binding from those where it is not.

We note that this cannot happen if the learner was unbudgeted.
In that case, if some action is best-responding to both $f = (1, 0)$ and $f = (0, 1)$, then it is also best-responding to $f = (\frac{1}{2}, \frac{1}{2})$.
In fact, for an unbudgeted learner our response-calibration error reduces to a calibration error that minimizes swap regret; see \cref{sec:app:resp_cal_unbudgeted} for details.

\section{Response-calibration Error in Unbudgeted Games}
\label{sec:app:resp_cal_unbudgeted}

In this section, we show how the response-calibration error $R_T$ in \eqref{eq:regret_function} simplifies in unbudgeted games ($\rL = 1$ or $\CL = 0$), why bounding $\norm{R_T}_{L^2}$ directly implies low swap regret for the learner in addition to non-manipulability, and how $R_T$ relates to \citet{bernstein-shimkin-2015, rivkin-et-al-2026, anagnostides-et-al-2026}.

\paragraph{Simplification in the unbudgeted case}
When $\rL = 1$ or $\CL = 0$, every round is active ($\mathcal{T} = [T]$) and every forecast $f \in \Delta_{\hatnO}$ induces a pure best response $j(f) \in \argmax_{j \in [\nL]} u(f)[j]$, represented by $\vec{J}(f) = (j(f), j(f))$ and $Q(f) = 0$.
In \eqref{eq:algo:delta_R_alt}, all off-diagonal coordinates $(k, j_1, j_2)$ with $j_1 \ne j_2$ vanish identically, and each diagonal coordinate $(k, j, j)$ is a scalar multiple of the step function $\One{\cdot \le 0}$:
\begin{equation*}
    R_{T, (k, j, j)}(q)
    =
    r_{k, j} \, \One{q \le 0},
    \qquad\text{where}\qquad
    r_{k, j}
    \coloneq
    \sum_{t=1}^T \Ex[f \sim F_t]{ \One{j(f) = j} \qty\big( \One{k_t = k} - f_k ) }
    \in \R^{\hatnO \times \nL}
    .
\end{equation*}
Since $\norm{\One{\cdot \le 0}}_{L^2([-1, 1])}^2 = \int_{-1}^0 1^2 \, dx = 1$, the function-space norm collapses to the Frobenius norm of the $\hatnO \times \nL$ response-conditioned frequency table $r$:
\begin{equation*}
    \norm{R_T}_{L^2}
    =
    \norm{r}_F
    =
    \sqrt{ \sum_{k=1}^{\hatnO} \sum_{j=1}^{\nL} r_{k, j}^2 }
    .
\end{equation*}
\paragraph{Why $\norm{R_T}_{L^2}$ bounds both swap regret and strategic manipulation}
For any action-modification rule $\pi \colon [\nL] \to [\nL]$, optimality of $j(f) = j$ under the predicted utility vector $u(f) = \sum_{k=1}^{\hatnO} f_k u^{(k)}$ implies $u(f)[\pi(j)] \le u(f)[j]$.
Adding and subtracting $u(f)$ therefore bounds the learner's swap regret directly in terms of the $\nL \times \nL$ projected utility-error matrix $\tilde{r}_{j', j} \coloneq \sum_{k=1}^{\hatnO} u^{(k)}[j']\,r_{k, j}$:
\begin{align*}
    & \sum_{t=1}^T \Ex[f \sim F_t]{ u^{(k_t)}[\pi(j(f))] - u^{(k_t)}[j(f)] }
    \\
    &\quad \le
    \sum_{j=1}^{\nL} \sum_{t=1}^T \Ex[f \sim F_t]{ \One{j(f) = j} \qty\Big( \qty(u^{(k_t)}[\pi(j)] - u(f)[\pi(j)]) - \qty(u^{(k_t)}[j] - u(f)[j]) ) } \\
    &\quad =
    \sum_{j=1}^{\nL} \qty(\tilde{r}_{\pi(j), j} - \tilde{r}_{j, j})
    \le
    \sqrt{\hatnO\,\nL}\,\norm{r}_F
    =
    \sqrt{\hatnO\,\nL}\,\norm{R_T}_{L^2}
    .
\end{align*}

Notice the distinction between learner swap regret and optimizer non-manipulability:
To control the \emph{learner's} swap regret, it suffices to bound the $\nL \times \nL$ matrix $\tilde{r}_{j', j} = \sum_{k=1}^{\hatnO} u^{(k)}[j']\,r_{k, j}$, which projects the type errors $r_{k,j}$ onto the learner's known utility vectors $u^{(k)} \in [0, 1]^{\nL}$.
This is the space in which \citet{rivkin-et-al-2026} and \citet{anagnostides-et-al-2026} operate: both works study arbitrary outcome vectors $u_t \in [0, 1]^{\nL}$ rather than discrete optimizer types $k_t \in [\hatnO]$, and neither model includes an unknown optimizer utility matrix $\UO$.
Specifically, for a fixed best-response map $j(\cdot)$, the partition bias of \citet{rivkin-et-al-2026} is $\sum_{j=1}^{\nL} \norm{\tilde{r}_{\cdot, j}}_\infty$, while \citet{anagnostides-et-al-2026} apply the response-based approachability framework of \citet{bernstein-shimkin-2015} directly to outer products $u_t \otimes p_t$ in $\R^{\nL \times \nL}$ to approach $\operatorname{conv}\{\hat{u} \otimes e_{j(\hat{u})} : \hat{u} \in [0,1]^{\nL}\}$.
Note that this is a special case of the setting of \cite{anagnostides-et-al-2026}, since they examine linear swap regret minimization, i.e., when the action space of the learner is not the simplex.
\section{Equivalence of \texorpdfstring{$\WBSE$}{WBSE} and \texorpdfstring{$\BSE$}{BSE} for Generic Games}
\label{sec:app:WBSE=BSE}

Here we show that when the learner's utilities and costs are generic, the learner's best response $\BRL(x)$ is unique for any fixed (or generic) optimizer strategy $x \in \Delta_{\nO}$ even when the budget constraint forces the learner to randomize, and moreover every optimal best-response vertex at any $x \in \Delta_{\nO}$ can be uniquely induced by an arbitrarily small perturbation of $x$.
As a consequence, the weak and strong Budgeted Stackelberg Equilibrium values coincide ($\WBSE = \BSE$). 

Throughout this section, because the learner's null action $j_0 \in [\nL]$ has zero utility and cost ($\UL[i, j_0] = \CL[i, j_0] = 0$ for all $i \in [\nO]$), we say that a property holds for \textit{generic} learner utilities and costs $(\UL, \CL)$ if it holds for all $(\UL, \CL) \in [0, 1]^{\nO \times \nL} \times [0, 1]^{\nO \times \nL}$ (subject to $\UL[\cdot, j_0] = \CL[\cdot, j_0] = 0$) outside a set of Lebesgue measure zero on the unrestricted entries $[\nO] \times ([\nL] \setminus \{j_0\})$.

We first characterize the vertices $\mathcal{V}(x)$ of the learner's budget-feasible polytope $\mathcal{F}(x) = \{y \in \Delta_{\nL} : x^\top \CL y \le \rL\}$ and explain why randomization due to a binding budget does not create ties in $\BRL(x)$.
Recall that $\mathcal{F}(x)$ is the intersection of the simplex $\Delta_{\nL}$ with a single linear inequality $x^\top \CL y \le \rL$, so every vertex $v \in \mathcal{V}(x)$ has support of size at most two:
\begin{enumerate}
    \item \textit{Pure vertices:} $v = e_j$ for some $j \in [\nL]$ satisfying $(x^\top \CL)_j \le \rL$.
    \item \textit{Two-action budget-binding vertices:} $v = v_{j, j'}(x) \coloneq q_{j, j'}(x) e_j + (1 - q_{j, j'}(x)) e_{j'}$ for a pair of pure actions $j, j' \in [\nL]$ with $(x^\top \CL)_j < \rL < (x^\top \CL)_{j'}$, where the mixing weight
    \begin{equation} \label{eq:app:budgeted_optimizer:binding_alpha}
        q_{j, j'}(x) \coloneq \frac{(x^\top \CL)_{j'} - \rL}{(x^\top \CL)_{j'} - (x^\top \CL)_j} \in (0, 1)
    \end{equation}
    is uniquely determined by the binding budget constraint $x^\top \CL v_{j, j'}(x) = \rL$.
\end{enumerate}

When a two-action vertex $v_{j, j'}(x)$ is optimal for the learner (with $(x^\top \UL)_j < V_{\mathtt{L}}(x) < (x^\top \UL)_{j'}$), both $j$ and $j'$ lie in the support of the optimal response and share the same Lagrangian payoff $(x^\top \UL)_k - \lambda (x^\top \CL)_k$ for the optimal dual multiplier $\lambda = \frac{(x^\top \UL)_{j'} - (x^\top \UL)_j}{(x^\top \CL)_{j'} - (x^\top \CL)_j} > 0$.
Thus $v_{j, j'}(x)$ is a single point in $\Delta_{\nL}$, and $\BRL(x) = \operatorname{conv}(\mathcal{V}(x) \cap \BRL(x))$ fails to be a singleton if and only if two distinct vertices in $\mathcal{V}(x)$ simultaneously achieve $V_{\mathtt{L}}(x)$.

\begin{lemma} \label{lem:app:budgeted_optimizers:generic_fixed_x}
    Fix any optimizer strategy $x \in \Delta_{\nO}$.
    For generic learner utilities and costs $(\UL, \CL)$, the learner's best-response set $\BRL(x)$ is a singleton.
    Consequently, for generic $(\UL, \CL)$, $\BRL(x)$ is a singleton for almost every $x \in \Delta_{\nO}$ (and for every pure optimizer action $x = e_i$, $i \in [\nO]$).
\end{lemma}

\begin{proof}
    For a given $x \in \Delta_{\nO}$, associate each pure action $j \in [\nL]$ of the learner with its cost-utility pair $p_j(x) \coloneq ((x^\top \CL)_j, (x^\top \UL)_j) \in [0, 1]^2$.
    By linear programming duality (which holds strongly since $p_{j_0}(x) = (0, 0)$ and $\rL > 0$), we have
    \begin{equation*}
        V_{\mathtt{L}}(x) = \min_{\lambda \ge 0} \max_{j \in [\nL]} \qty\big( (x^\top \UL)_j - \lambda \qty((x^\top \CL)_j - \rL) ),
    \end{equation*}
    and any $y \in \BRL(x)$ is supported on the set of tight pure actions $S(x, \lambda) \coloneq \arg\max_{j \in [\nL]} \qty\big((x^\top \UL)_j - \lambda (x^\top \CL)_j)$ for an optimal dual multiplier $\lambda \ge 0$, with $x^\top \CL y = \rL$ whenever $\lambda > 0$.

    Suppose $\BRL(x)$ is not a singleton, so that at least two distinct vertices in $\mathcal{V}(x)$ belong to $\BRL(x)$.
    If $\lambda = 0$ is an optimal dual multiplier, then at least two distinct pure actions $j_1 \ne j_2$ must tie for the maximum utility:
    \begin{equation} \label{eq:app:budgeted_optimizer:deg_tie}
        (x^\top \UL)_{j_1} - (x^\top \UL)_{j_2} = 0.
    \end{equation}
    If every optimal dual multiplier satisfies $\lambda > 0$, then every $y \in \BRL(x)$ satisfies $x^\top \CL y = \rL$ and is supported on $S(x, \lambda)$.
    If $|S(x, \lambda)| = 1$, say $S(x, \lambda) = \{j\}$, then $\BRL(x) = \{e_j\}$ is a singleton.
    If $|S(x, \lambda)| = 2$, say $S(x, \lambda) = \{j, j'\}$, then the equation $q (x^\top \CL)_j + (1 - q) (x^\top \CL)_{j'} = \rL$ has a unique solution $q = q_{j, j'}(x) \in [0, 1]$ unless $(x^\top \CL)_j = (x^\top \CL)_{j'} = \rL$, which requires
    \begin{equation} \label{eq:app:budgeted_optimizer:deg_cost}
        (x^\top \CL)_j - \rL = 0.
    \end{equation}
    Finally, if $|S(x, \lambda)| \ge 3$, then there exist three distinct pure actions $j_1, j_2, j_3 \in [\nL]$ lying on the same supporting line of slope $\lambda$ in the cost-utility plane, which means $p_{j_1}(x), p_{j_2}(x), p_{j_3}(x)$ are collinear:
    \begin{equation} \label{eq:app:budgeted_optimizer:deg_collinear}
        \det \begin{pmatrix}
            1 & (x^\top \CL)_{j_1} & (x^\top \UL)_{j_1} \\
            1 & (x^\top \CL)_{j_2} & (x^\top \UL)_{j_2} \\
            1 & (x^\top \CL)_{j_3} & (x^\top \UL)_{j_3}
        \end{pmatrix} = 0.
    \end{equation}
    For any fixed $x \in \Delta_{\nO}$ (having at least one positive coordinate $x_i > 0$), each of the finitely many conditions \eqref{eq:app:budgeted_optimizer:deg_tie}--\eqref{eq:app:budgeted_optimizer:deg_collinear} over distinct tuples of actions in $[\nL]$ is a non-trivial polynomial equation in the unrestricted entries of $(\UL, \CL)$, and hence defines a hypersurface of Lebesgue measure zero.
    Outside the finite union of these measure-zero hypersurfaces, $\BRL(x)$ is a singleton.
    The second claim follows immediately by Fubini's theorem (and by taking a union over the $\nO$ pure strategies $x = e_i$).
\end{proof}

When the optimizer chooses $x \in \Delta_{\nO}$ strategically, as in the definition of the BSE in \eqref{opt:defs:bse} or \eqref{opt:app:budgeted_optimizers:bse}, the optimizer's best choice may be an $x$ on a boundary between best-response regions where the learner is indifferent between multiple vertices of $\mathcal{F}(x)$.
We next show that for generic $(\UL, \CL)$, no optimal vertex of $\mathcal{F}(x)$ is isolated or degenerate: any optimal vertex $v \in \mathcal{V}(x) \cap \BRL(x)$ can be selected as the unique best response by an arbitrarily small perturbation of $x$.

\begin{lemma} \label{lem:app:budgeted_optimizers:generic_perturbation}
    For generic learner utilities and costs $(\UL, \CL)$, the following holds for every $x \in \Delta_{\nO}$ and every optimal vertex $v \in \mathcal{V}(x) \cap \BRL(x)$: there exists a sequence of optimizer strategies $\{x_\varepsilon\}_{\varepsilon > 0} \subseteq \Delta_{\nO}$ with $\lim_{\varepsilon \rightarrow 0} x_\varepsilon = x$ such that $\BRL(x_\varepsilon) = \{v_\varepsilon\}$ is a singleton for every $\varepsilon > 0$ and $\lim_{\varepsilon \rightarrow 0} v_\varepsilon = v$.
\end{lemma}

\begin{proof}
    Fix any $x\in \Delta_{\nO}$, and let support $S_x = \operatorname{supp}(x) \subseteq [\nO]$ of size $s = |S_x| \ge 1$, so that $x$ lies in the $(s-1)$-dimensional open simplex $\Delta_{S_x}^{\circ} \coloneq \{x' \in \Delta_{\nO} : \operatorname{supp}(x') = S_x\}$ with tangent space $T_{S_x} \coloneq \{d \in \mathbb{R}^{\nO} : \operatorname{supp}(d) \subseteq S_x, \, \mathbf{1}^\top d = 0\}$.
    Let $v \in \mathcal{V}(x) \cap \BRL(x)$ be any optimal vertex of $\mathcal{V}(x)$ that is a best response to  $x$, and let $\lambda \ge 0$ be a corresponding optimal dual multiplier with intercept $\tau \coloneq V_{\mathtt{L}}(x) - \lambda \rL$ and tight action set $S \coloneq \{j \in [\nL] : (x^\top \UL)_j - \lambda (x^\top \CL)_j = \tau\}$.
    
    \paragraph{Case 1: Unconstrained utility maximum ($\lambda = 0$ and $v = e_{j^\star}$ with $(x^\top \CL)_{j^\star} \le \rL$)}
    Here $S = \{j \in [\nL] : x_{S_x}^\top u_j = V_{\mathtt{L}}(x)\}$.
    Note that for generic $\UL$, every non-null action has positive utility entries, so $V_{\mathtt{L}}(x) > 0$ and $j_0 \notin S$.
    Every action $j \in S \setminus \{j^\star\}$ is tied with $j^\star$ at $x$, meaning
    \begin{equation*}
        x_{S_x}^\top (u_j - u_{j^\star}) = 0 \qquad \text{for all } j \in S \setminus \{j^\star\}.
    \end{equation*}
    For generic $\UL$, any at most $s$ difference vectors of the form $u_j - u_{j^\star}$ in $\mathbb{R}^s$ are linearly independent.
    Since the nonzero vector $x_{S_x} \in \mathbb{R}^s$ is orthogonal to all $|S| - 1$ vectors $\{u_j - u_{j^\star}\}_{j \in S \setminus \{j^\star\}}$, there can be at most $|S| - 1 \le s - 1$ such vectors (so $|S| \le s$), and because $\mathbf{1}^\top x_{S_x} = 1 \ne 0$, the all-ones vector $\mathbf{1} \in \mathbb{R}^s$ does not lie in their span.
    Consequently, the $|S|$ vectors $\{\mathbf{1}\} \cup \{u_j - u_{j^\star}\}_{j \in S \setminus \{j^\star\}}$ in $\mathbb{R}^s$ are linearly independent (and if $(x^\top \CL)_{j^\star} = \rL$, the vector $c_{j^\star} - \rL \mathbf{1}$, which is also orthogonal to $x_{S_x}$ and has generic entries, is likewise linearly independent of them, with $|S| \le s - 1$).
    Therefore, the linear system
    \begin{equation*}
        \mathbf{1}^\top d = 0, \qquad (u_j - u_{j^\star})^\top d = -1 \quad (\forall j \in S \setminus \{j^\star\})
    \end{equation*}
    (together with $c_{j^\star}^\top d = -1$ if $(x^\top \CL)_{j^\star} = \rL$) has a solution $d \in \mathbb{R}^s$.
    For all sufficiently small $\varepsilon > 0$, $x_\varepsilon = x + \varepsilon d \in \Delta_{\nO}$ satisfies $(x_\varepsilon^\top \CL)_{j^\star} < \rL$ and $(x_\varepsilon^\top \UL)_{j^\star} > (x_\varepsilon^\top \UL)_j$ for all $j \ne j^\star$, so $\BRL(x_\varepsilon) = \{e_{j^\star}\}$.

    \paragraph{Case 2: Budget-binding vertex ($x^\top \CL v = \rL$)}
    First suppose $v = v_{j_1, j_2}(x)$ is a two-action vertex with $j_1, j_2 \in S$ and $x_{S_x}^\top c_{j_1} < \rL < x_{S_x}^\top c_{j_2}$, where the supporting line through $p_{j_1}(x)$ and $p_{j_2}(x)$ has slope
    \begin{equation*}
        \lambda = \frac{x_{S_x}^\top (u_{j_2} - u_{j_1})}{x_{S_x}^\top (c_{j_2} - c_{j_1})}.
    \end{equation*}
    For each additional tight action $j \in S \setminus \{j_1, j_2\}$, let
    \begin{equation*}
        \beta_j(x) \coloneq \frac{x_{S_x}^\top (c_j - c_{j_1})}{x_{S_x}^\top (c_{j_2} - c_{j_1})}
    \end{equation*}
    denote the cost-interpolation weight of $j$ relative to $j_1$ and $j_2$, and define the vector
    \begin{equation*}
        w_j(x) \coloneq \qty\big((u_j - u_{j_1}) - \lambda (c_j - c_{j_1})) - \beta_j(x) \qty\big((u_{j_2} - u_{j_1}) - \lambda (c_{j_2} - c_{j_1})) \in \mathbb{R}^s.
    \end{equation*}
    Notice that for any $x' \in \Delta_{\nO}$ supported on $S_x$, the vertical distance of $p_j(x')$ above the line passing through $p_{j_1}(x')$ and $p_{j_2}(x')$ is $(x'_{S_x})^\top w_j(x')$.
    At $x' = x$, since $j, j_1, j_2 \in S$ all lie on the supporting line of slope $\lambda$, both terms in the definition of $w_j(x)$ are orthogonal to $x_{S_x}$, so
    \begin{equation*}
        x_{S_x}^\top w_j(x) = 0 \qquad \text{for all } j \in S \setminus \{j_1, j_2\},
    \end{equation*}
    and differentiating $(x'_{S_x})^\top w_j(x')$ at $x' = x$ along a perturbation $x_\varepsilon = x + \varepsilon d$ (using $x_{S_x}^\top \qty((u_{j_2} - u_{j_1}) - \lambda (c_{j_2} - c_{j_1})) = 0$) shows that the first-order change in the vertical distance of $p_j(x_\varepsilon)$ above the line through $p_{j_1}(x_\varepsilon)$ and $p_{j_2}(x_\varepsilon)$ is simply $d^\top w_j(x)$.

    Crucially, for each $j \in S \setminus \{j_1, j_2\}$, the vector $w_j(x)$ equals $u_j$ minus a vector that does not depend on $u_j$ (since neither $\lambda$ nor $\beta_j(x)$ depends on $u_j$, and at most one action in $S$ can be the null action $j_0$, which can be taken as $j_1$ if $j_0 \in S$).
    Consequently, for generic $(\UL, \CL)$, the vectors $\{w_j(x)\}_{j \in S \setminus \{j_1, j_2\}}$ in the $(s-1)$-dimensional subspace $\{w \in \mathbb{R}^s : x_{S_x}^\top w = 0\}$ are linearly independent; in particular, there can be at most $|S| - 2 \le s - 1$ such vectors (so $|S| \le s + 1$), and because $\mathbf{1}^\top x_{S_x} = 1 \ne 0$, the vectors $\{\mathbf{1}\} \cup \{w_j(x)\}_{j \in S \setminus \{j_1, j_2\}}$ are linearly independent in $\mathbb{R}^s$.
    Therefore, the linear system
    \begin{equation*}
        \mathbf{1}^\top d = 0, \qquad w_j(x)^\top d = -1 \quad (\forall j \in S \setminus \{j_1, j_2\})
    \end{equation*}
    has a solution $d \in \mathbb{R}^s$ (and if $\lambda = 0$, i.e., $x_{S_x}^\top (u_{j_2} - u_{j_1}) = 0$, the vector $u_{j_2} - u_{j_1}$ is also orthogonal to $x_{S_x}$ and linearly independent of them, so we can additionally require $(u_{j_2} - u_{j_1})^\top d = 1$ to ensure the perturbed slope $\lambda_\varepsilon$ is strictly positive).
    For $\varepsilon > 0$ sufficiently small, let $\lambda_\varepsilon > 0$ and $\tau_\varepsilon \in \mathbb{R}$ be the slope and intercept of the line passing through $p_{j_1}(x_\varepsilon)$ and $p_{j_2}(x_\varepsilon)$.
    By construction, for every $j \in S \setminus \{j_1, j_2\}$, the vertical distance of $p_j(x_\varepsilon)$ above this line is $\varepsilon \, d^\top w_j(x) + \order{\varepsilon^2} = -\varepsilon + \order{\varepsilon^2} < 0$ for all small $\varepsilon > 0$, while all $j \notin S$ remain strictly below the line and $(x_\varepsilon^\top \CL)_{j_1} < \rL < (x_\varepsilon^\top \CL)_{j_2}$ holds by continuity.
    Hence $S(x_\varepsilon, \lambda_\varepsilon) = \{j_1, j_2\}$, so $\BRL(x_\varepsilon) = \{v_{j_1, j_2}(x_\varepsilon)\}$ is a singleton, and by continuity of $q_{j_1, j_2}$ in \eqref{eq:app:budgeted_optimizer:binding_alpha} we have $\lim_{\varepsilon \rightarrow 0} v_{j_1, j_2}(x_\varepsilon) = v_{j_1, j_2}(x) = v$.

    Finally, if $v = e_{j_1}$ is a boundary pure vertex with $j_1 \in S$ and $x_{S_x}^\top c_{j_1} = \rL$, then if $S = \{j_1\}$ we already have $\BRL(x) = \{e_{j_1}\}$; otherwise, picking any $j_2 \in S$ with $x_{S_x}^\top c_{j_2} > \rL$ (or reducing to Case 1 if none exists), the additional vector $c_{j_1} - \rL \mathbf{1}$ is also orthogonal to $x_{S_x}$ and linearly independent of $\{\mathbf{1}\} \cup \{w_j(x)\}_{j \in S \setminus \{j_1, j_2\}}$, so we can additionally require $c_{j_1}^\top d = -1$.
    Then $(x_\varepsilon^\top \CL)_{j_1} = \rL - \varepsilon < \rL$ and $S(x_\varepsilon, \lambda_\varepsilon) = \{j_1, j_2\}$, so $\BRL(x_\varepsilon) = \{v_{j_1, j_2}(x_\varepsilon)\}$ is a singleton with $q_{j_1, j_2}(x_\varepsilon) \to 1$ and $v_{j_1, j_2}(x_\varepsilon) \to e_{j_1} = v$ as $\varepsilon \rightarrow 0$.
\end{proof}

Combining \cref{lem:app:budgeted_optimizers:generic_perturbation} with the two-phase structure of \eqref{opt:app:budgeted_optimizers:bse} and \eqref{opt:app:budgeted_optimizers:wbse} immediately establishes that $\WBSE$ and $\BSE$ coincide for generic games.

\begin{proof}[Proof of \cref{thm:app:budgeted_optimizers:wbse_eq_bse}]
   Clearly, $\WBSE \le \BSE$. To prove the reverse inequality $\WBSE \ge \BSE$, recall from the proof of \cref{lem:app:budgeted_optimizers:lagrangian_equiv} that $\BSE = \max_{(u, c) \in \operatorname{conv}(\mathcal{S}) : c \le \rO} u$, where $\mathcal{S} = \{(x^\top \UO y, x^\top \CO y) : x \in \Delta_{\nO}, \, y \in \BRL(x)\}$.
    For any fixed $x \in \Delta_{\nO}$, the map $y \mapsto (x^\top \UO y, x^\top \CO y)$ is linear and $\BRL(x) = \operatorname{conv}(\mathcal{V}(x) \cap \BRL(x))$, so $\operatorname{conv}(\mathcal{S}) = \operatorname{conv}(\mathcal{S}^{\mathrm{vert}})$ where
    \begin{equation*}
        \mathcal{S}^{\mathrm{vert}} \coloneq \{ (x^\top \UO v, x^\top \CO v) : x \in \Delta_{\nO}, \, v \in \mathcal{V}(x) \cap \BRL(x) \}.
    \end{equation*}
    Therefore, there exist $x_1^\star, x_2^\star \in \Delta_{\nO}$, optimal vertices $v_1^\star \in \mathcal{V}(x_1^\star) \cap \BRL(x_1^\star)$ and $v_2^\star \in \mathcal{V}(x_2^\star) \cap \BRL(x_2^\star)$, and $q^\star \in [0, 1]$ that attain $\BSE$ in \eqref{opt:app:budgeted_optimizers:bse}, with $u_k^\star \coloneq (x_k^\star)^\top \UO v_k^\star$ and $c_k^\star \coloneq (x_k^\star)^\top \CO v_k^\star$ for $k \in \{1, 2\}$.

    By \cref{lem:app:budgeted_optimizers:generic_perturbation}, for each $k \in \{1, 2\}$ there exists a sequence $x_{k, \varepsilon} \in \Delta_{\nO}$ with $x_{k, \varepsilon} \to x_k^\star$ such that $\BRL(x_{k, \varepsilon}) = \{v_{k, \varepsilon}\}$ is a singleton and $v_{k, \varepsilon} \to v_k^\star$ as $\varepsilon \rightarrow 0$.
    Because $\BRL(x_{k, \varepsilon})$ is a singleton, pessimistic and optimistic tie-breaking coincide at $x_{k, \varepsilon}$:
    \begin{equation*}
        \min_{y \in \BRL(x_{k, \varepsilon})} x_{k, \varepsilon}^\top \UO y = x_{k, \varepsilon}^\top \UO v_{k, \varepsilon} \eqcolon u_{k, \varepsilon} \xrightarrow{\varepsilon \rightarrow 0} u_k^\star,
        \qquad
        \max_{y \in \BRL(x_{k, \varepsilon})} x_{k, \varepsilon}^\top \CO y = x_{k, \varepsilon}^\top \CO v_{k, \varepsilon} \eqcolon c_{k, \varepsilon} \xrightarrow{\varepsilon \rightarrow 0} c_k^\star.
    \end{equation*}
    If $\min\{c_1^\star, c_2^\star\} < \rO < \max\{c_1^\star, c_2^\star\}$ (say $c_1^\star < \rO < c_2^\star$), then $c_{1, \varepsilon} < \rO < c_{2, \varepsilon}$ for all sufficiently small $\varepsilon > 0$; setting $q_\varepsilon \coloneq \frac{c_{2, \varepsilon} - \rO}{c_{2, \varepsilon} - c_{1, \varepsilon}} \to q^\star$ gives a feasible solution $(x_{1, \varepsilon}, x_{2, \varepsilon}, q_\varepsilon)$ to \eqref{opt:app:budgeted_optimizers:wbse} whose objective value $q_\varepsilon u_{1, \varepsilon} + (1 - q_\varepsilon) u_{2, \varepsilon}$ converges to $q^\star u_1^\star + (1 - q^\star) u_2^\star = \BSE$.
    Otherwise, at least one of the two points (say $(x_1^\star, v_1^\star)$) satisfies both $c_1^\star \le \rO$ and $u_1^\star = \BSE$.
    Pairing $x_{1, \varepsilon}$ with the optimizer's null action $x_2 = e_{i_0}$ (which has cost $0$ regardless of the learner's response) and weight $q_\varepsilon \coloneq \min\{1, \rO / c_{1, \varepsilon}\} \to 1$ yields a feasible solution $(x_{1, \varepsilon}, e_{i_0}, q_\varepsilon)$ to \eqref{opt:app:budgeted_optimizers:wbse} with objective value at least $q_\varepsilon u_{1, \varepsilon} \to u_1^\star = \BSE$.
    In either case, taking $\varepsilon \rightarrow 0$ shows $\WBSE \ge \BSE$, completing the proof.
\end{proof}

\end{document}